\documentclass[review]{elsarticle}
\usepackage[letterpaper, margin=1.2in, top=1.2in]{geometry}

\usepackage{lineno,hyperref}
\modulolinenumbers[5]

\usepackage{color}
\usepackage{amssymb}
\usepackage{bm}
\usepackage{amsmath}
\usepackage{graphicx}
\usepackage{subfig}
\usepackage[mathscr]{euscript}
\usepackage[makeroom]{cancel} 

\biboptions{numbers,sort&compress}

\usepackage{multicol}

\journal{Journal of \LaTeX\ Templates}

\usepackage{amsthm}
\newtheorem{remark}{Remark}

\newtheorem{theorem}{Theorem}
\newtheorem{corollary}{Corollary}
\newtheorem{lemma}{Lemma}
\newtheorem{proposition}{Proposition}
\theoremstyle{definition}
\newtheorem{definition}{Definition}

\usepackage{algorithm}
\usepackage[noend]{algpseudocode}

\usepackage[table]{xcolor}    

\makeatletter
\let\save@mathaccent\mathaccent
\newcommand*\if@single[3]{%
  \setbox0\hbox{${\mathaccent"0362{#1}}^H$}%
  \setbox2\hbox{${\mathaccent"0362{\kern0pt#1}}^H$}%
  \ifdim\ht0=\ht2 #3\else #2\fi
  }
\newcommand*\rel@kern[1]{\kern#1\dimexpr\macc@kerna}
\newcommand*\wideaccent[2]{\@ifnextchar^{{\wide@accent{#1}{#2}{0}}}{\wide@accent{#1}{#2}{1}}}
\newcommand*\wide@accent[3]{\if@single{#2}{\wide@accent@{#1}{#2}{#3}{1}}{\wide@accent@{#1}{#2}{#3}{2}}}
\newcommand*\wide@accent@[4]{%
  \begingroup
  \def\mathaccent##1##2{%
    \let\mathaccent\save@mathaccent
    \if#42 \let\macc@nucleus\first@char \fi
    \setbox\z@\hbox{$\macc@style{\macc@nucleus}_{}$}%
    \setbox\tw@\hbox{$\macc@style{\macc@nucleus}{}_{}$}%
    \dimen@\wd\tw@
    \advance\dimen@-\wd\z@
    \divide\dimen@ 3
    \@tempdima\wd\tw@
    \advance\@tempdima-\scriptspace
    \divide\@tempdima 10
    \advance\dimen@-\@tempdima
    \ifdim\dimen@>\z@ \dimen@0pt\fi
    \rel@kern{0.6}\kern-\dimen@
    \if#41
      #1{\rel@kern{-0.6}\kern\dimen@\macc@nucleus\rel@kern{0.4}\kern\dimen@}%
      \advance\dimen@0.4\dimexpr\macc@kerna
      \let\final@kern#3%
      \ifdim\dimen@<\z@ \let\final@kern1\fi
      \if\final@kern1 \kern-\dimen@\fi
    \else
      #1{\rel@kern{-0.6}\kern\dimen@#2}%
    \fi
  }%
  \macc@depth\@ne
  \let\math@bgroup\@empty \let\math@egroup\macc@set@skewchar
  \mathsurround\z@ \frozen@everymath{\mathgroup\macc@group\relax}%
  \macc@set@skewchar\relax
  \let\mathaccentV\macc@nested@a
  \if#41
    \macc@nested@a\relax111{#2}%
  \else
    \def\gobble@till@marker##1\endmarker{}%
    \futurelet\first@char\gobble@till@marker#2\endmarker
    \ifcat\noexpand\first@char A\else
      \def\first@char{}%
    \fi
    \macc@nested@a\relax111{\first@char}%
  \fi
  \endgroup
}
\makeatother

\newcommand\widebar{\wideaccent\overline}

\newcommand\widehatoverline[1]{\overline{\widetilde{#1}}}
\newcommand\widebartilde{\wideaccent\widehatoverline}

\usepackage{nomencl}
\makenomenclature
\usepackage{framed}
\usepackage{multicol} 
\renewcommand*\nompreamble{\begin{multicols}{2}}
\renewcommand*\nompostamble{\end{multicols}}

\begin{document}

\begin{frontmatter}

\title{A High-Order Variational Multiscale Approach to Turbulence for Compact Nodal Schemes} 

\author[mymainaddress]{Farshad Navah\corref{mycorrespondingauthor}}
\cortext[mycorrespondingauthor]{Corresponding author}
\ead{farshad.navah@mail.mcgill.ca}

\author[mysecondaryaddress]{Marta de la Llave Plata}
\author[mysecondaryaddress]{Vincent Couaillier}

\address[mymainaddress]{Mechanical Engineering Department, McGill University, Montr\'eal, Canada, H3A 0C3}
\address[mysecondaryaddress]{ONERA - The French Aerospace Lab, FR-92322 Ch\^atillon, France}
\begin{abstract}
This article presents a formulation that extends the variational multiscale modelling for compressible large-eddy simulation to a vast family of compact nodal numerical methods represented by the high-order flux reconstruction scheme. The theoretical aspects of the proposed formulation are laid down via rigorous mathematical derivations which clearly expose the underlying assumptions and approximations and provide sufficient details for accurate reproduction of the methodology. The final form is assessed on a Taylor-Green vortex benchmark with Reynolds number of $5000$ and compared to filtered direct numerical simulation data. 
These numerical experiments exhibit the important role of sufficient de-aliasing, appropriate amount of upwinding from Roe's numerical flux and large/small scale partition, in achieving better agreement with reference data, especially on coarse grids, when compared to the baseline implicit large-eddy simulation.
\end{abstract}

\end{frontmatter}

%


\section{Introduction}
\label{sec:intro}

The advent of high-performance computing and new architectures such as GPU clusters have opened the door to increasingly faithful turbulent flow simulations of industrial interest,  which for decades remained only approachable by affordable Reynolds-averaged Navier-Stokes (RANS) models with well-known deficiencies in capturing statistical unsteadiness present in complex configurations  such as detached flows. While direct numerical simulation (DNS) of Navier-Stokes (NS) equations for practical engineering scenarios remains out of reach for foreseeable future, large-eddy simulation (LES) has a proven potential to provide engineers with cost-effective insight in tackling the challenge of more efficient or even revolutionary design paradigms.

LES via wide-spread second-order (of accuracy) flow solvers suffers from the overly dissipative nature of their underlying spatial discretization. Comparatively, high-order discretization methods  constitute an elegant improvement \cite{Vermeire2015}. Lower discretization error, larger spectral bandwidth and inherent dissipation/dispersion signature of these methods \cite{Vermeire2016,Mengaldo2018} mimicking to some extent the natural subgrid-scale (SGS) dissipation, make high-order methods to be a powerful framework for implicit LES (ILES), which refers to exploiting numerical inaccuracies as an intrinsic turbulence model.  Although ILES has an interesting potential as it eliminates the need for the application of explicit models to all resolved scales, as in a classical LES, there is still no substantial proof for the capability of ILES on coarse grids and large Reynolds numbers, relevant of industrial applications \cite{Flad2017}. A reason for this seems to stem from the limitation of high-order ILES in representing the SGS effects on coarser grids (or for higher Reynolds numbers) \cite{Flad2017,Moura2017,Manzanero2018} where a larger amount of dissipation needs to be applied to a wider range of small resolved scales.

A remedy is provided by variational multiscale (VMS) methods for LES, first introduced by Hughes \textit{et al.} \cite{Hughes2000} for incompressible flows via a two-level scale separation, and later extended by Collis \cite{Collis2001} to a high-order discretization of compressible LES  equations with a three-level scale separation. In contrast to classical LES approaches based on spatial filtering (smoothing), the basic idea of VMS consists of the \textit{a priori} separation of scales present in turbulent flow via variational projection. 
 Furthermore, VMS enables a larger flexibility in modelling the SGS terms by explicitly controlling simultaneously the amount and the spectral range of dissipation, in contrast to the ILES and classical LES approaches. Some other advantages of the VMS-LES formulation are \cite{Collis2001}: a solid mathematical foundation, validity for complex geometries and satisfactory performance using a simple Smagorinsky SGS model even for wall-bounded flows. Finally, exploiting the properties of variational projection, the VMS formulation avoids the errors due to the commutation of filtering and derivation operators which affect the classical LES formalism.  With regards to low-order VMS formulations such as the one in \cite{Koobus2004}, high-order polynomial-based (compact) VMS methods offer the advantage of avoiding costly agglomeration-based scale separation which degrades the parallel efficiency of the numerical algorithm.

The compact high-order VMS-LES formulations are mostly based on modal discontinuous Galerkin (DG)  \cite{Collis2001, Abba2015, Chapelier2016a} and on nodal spectral element \cite{Wasberg2009,Beck2016} methods. Our aim is to present a unifying formulation, allowing the extension of the VMS-LES method to a vast family of compact nodal schemes, represented by the recent flux reconstruction (FR) method, first introduced by Huynh \cite{Huynh2007, Huynh_2009a} on one-dimensional (1D) and tensor-product (TP) elements. This family of schemes includes prominent members such as the nodal DG \cite{Hesthaven-Warburton_2008}, the spectral difference (SD) \cite{Liu-et-al_2006b} and the spectral volume (SV) \cite{Wang2002} methods. An extension of FR to simplices via lifting operations was first proposed by Wang and Gao \cite{Wang-Gao_2009} and called correction procedure via flux reconstruction (CPR/FR) \cite{Haga-et-al_2011}. Through a linear stability analysis, Castonguay \textit{et al.}  \cite{Castonguay-et-al_2012} discovered a continuous range of schemes, determined by a single parameter and labelled energy stable flux reconstruction (ESFR). Finally, connections between ESFR and DG schemes are studied \cite{Allaneau2011,DeGrazia2014}, revealing the equivalence of the former to a version of the latter in which the highest-order modes of the residual are filtered \cite{Mengaldo2016}.

We present the details of the VMS approach to LES by adopting the methodology of Lesieur \textit{et al.} \cite[7.1]{lesieur_metais_comte_2005} in terms of assumptions and approximations made to derive the LES equations for compressible flows.  Nevertheless, in contrast to the latter authors who elect a framework based on spatial filtering, we conduct the derivation within a variational framework, similar to the one proposed by Collis \cite{Collis2001}, with a two-level partitioning of the spatial scales into resolved and unresolved. This first step clearly exhibits the mathematical foundation of the VMS-LES approach and the modelling effort is made explicit for each term by comparing the exact governing equations of the resolved scales versus their modelled counterparts. The model equations are then semi-discretized in time by the explicit third-order Runge-Kutta (RK)  scheme of Shu and Osher \cite{Shu1988}, and in space by the high-order FR/CPR scheme. A second scale separation is then applied to the discrete equations in order to isolate the large and the small resolved scales, thus allowing us to apply the SGS model to the latter only. The derivation of the nodal VMS-LES-FR/CPR formulation is presented with rigorous attention to mathematical justifications such that the validity of the final form is traceable and the reproduction of the results is facilitated.

The proposed nodal VMS-LES-FR/CPR formulation is assessed on a Taylor-Green vortex (TGV) problem at a Reynolds number of 5000 and the effects of a number of parameters involved in the formulation are studied separately. 
These parameters include:  de-aliasing, large/small resolved scales partitioning, upwinding dissipation from Roe's numerical flux, polynomial discretization degree, grid resolution and FR/CPR scheme. The beneficial effect of the VMS approach in improving the agreement with the filtered DNS data on coarse grids is thus made explicit.

The article is structured as follows: starting by an introduction in Section \ref{sec:intro}, we lay down the theoretical aspects of the VMS-LES-FR/CPR formulation in Section \ref{sec:methodology} by presenting the governing equations of compressible flows (Section \ref{sec:methodology_goveqs}), the details of VMS-LES modelling  with a two-level partitioning (Section \ref{sec:methodology_vms2lev}) as well as  a three-level partitioning tailored to the FR/CPR scheme (Section \ref{sec:methodology_VMS3CPR}). In Section \ref{sec:results}, the proposed methodology is assessed on a benchmark problem and its results are discussed. The article is concluded in Section \ref{sec:conclus}.

\section{Methodology}
\label{sec:methodology}
\subsection{Governing equations of compressible flow}
\label{sec:methodology_goveqs}
The compressible (extension of the) Navier-Stokes (CNS) model constitutes a system of five coupled equations with the following concise expression for the $k^{th}$ equation\footnote{Repeated indices in the same term are summed upon following Einstein's convention.}:
\begin{equation}
\label{eq:CNS}
\text{CNS}_k\left(\bm{Q}\right) = 0,
\end{equation} 
where the partial differential operator has the following generic conservative form:
\begin{equation}
\label{eq:CNSop}
\text{CNS}_k\left(\bm{Q}\right) := \frac{\partial Q_k}{\partial t} + \frac{\partial F_{ik}}{\partial x_i},
\end{equation} 
where 
\begin{equation}
\bm{Q} :=  \left[\rho,\, \rho u_1,\, \rho u_2,\, \rho u_3,\, \rho E\right] ^\intercal,
\label{eq:Q}
\end{equation}  is the vector of conservative state variables and $x_i$ is the $i^{th} \in [1,2,3]$ space coordinate. The primitive variables are $\rho$, the density; $u_i$, the $i^{th}$ velocity vector component and $E$, the total energy per unit mass defined via 
\begin{equation}
\rho E := \rho\, C_v T +\frac{1}{2}\rho \left( u_1^2+u_2^2+u_3^2\right),
\label{eq:RhoE}
\end{equation}
where the temperature, $T$, is related to the pressure, $P$, and the density by the ideal gas law: $P=\rho \,R \,T$. The gas constant and the heat capacity at constant volume are respectively denoted by $R$ and $C_v$.

The flux vector component $F_{ik} := F^\text{inv}_{ik}-F^\text{vis}_{ik}$, has inviscid and viscous contributions, respectively defined as
\begin{equation}
\bm{F}^{\text{inv}}_i \left( \bm{Q}\right):= \left[ \rho u_i,\, \rho u_i \,u_1 + P \,\delta_{i1},\,    \rho u_i \,u_2 + P \,\delta_{i2},\,  \rho u_i \,u_3 + P \,\delta_{i3} ,\, (\rho E + P)\,u_i \right] ^\intercal,
\label{eq:F_inv}       
\end{equation}
and
\begin{equation}
\bm{F}^{\text{vis}}_i \left( \bm{Q},\bm{\nabla}  \bm{Q}\right) := \left[ 0,\,\tau_{i1},\, \tau_{i2},\, \tau_{i3},\,  \tau_{ij}  u_j + \lambda \frac{\partial T}{\partial x_i}  \right] ^\intercal,
\label{eq:F_vis}       
\end{equation}
where $\delta_{ij}$ is the Kronecker delta and $\tau_{ij}:=2\,\mu \,A_{ij}$ are the components of the viscous stress tensor, with the deviator of the deformation tensor expressed by
\begin{equation}
A_{ij}(\bm{u}) := \frac{1}{2} \left( \frac{\partial u_j}{\partial x_i} + \frac{\partial u_i}{\partial x_j} \right) -\frac{1}{3} \frac{\partial u_l}{\partial x_l} \delta_{ij},
\label{eq:defor_tesnor}       
\end{equation}
and the dynamic viscosity, $\mu$, prescribed by the Sutherland's law,
\begin{equation}
\mu(T) = \frac{C_1 \,T^{3/2}}{T+T_S},
\label{eq:Suth}       
\end{equation}
where $C_1$ is a coefficient and $T_S$ is the Sutherland's temperature. Finally, the conductivity is related to viscosity via $\lambda =  \frac{C_p}{Pr}\,\mu$ with $C_p$ denoting the heat capacity at constant pressure and $Pr$ standing for the Prandtl number.

\begin{remark}
The dyadic product of the gradient and the solution vectors, $\bm{\nabla}  \bm{Q}$ in $\bm{F}^{\mathrm{vis}}_i \left( \bm{Q},\bm{\nabla}  \bm{Q}\right)$, is meant to remind that the viscous terms depend on the spatial derivatives of the solution and hence the latter need special treatment in discontinuous frameworks as discussed in Remark \ref{remrk:numflx}.
\end{remark}

\subsection{Variational multiscale formulation with a two-level partitioning}
\label{sec:methodology_vms2lev}
The fundamental idea at the basis of the VMS methods for LES  \cite{Hughes2000} consists of the \textit{a priori} separation of the spatial  scales of the flow via projection onto a partitioned functional space. This thus exploits the intrinsic mechanisms of variational formulations such as the finite element method, in which the projection is  arising from the discretization itself. To illustrate this idea, let us define the inner product between $a$ and $b$, two spatiotemporal quantities:
\begin{equation*}
\left<a,b\right> := \int_{\bm{\Omega}} a \, b \,{d\Omega},
\end{equation*}
where $\bm{\Omega}$ denotes a connected domain in $\in {\rm I\!R}^3$.

The variational form of the CNS equations is obtained by projecting them onto a proper functional space, $\mathcal{V}$, via
\begin{equation}
\left<\phi_k, \text{CNS}_k\left(\bm{Q}\right) \right>=0,
\label{eq:weakCNS}
\end{equation}
where $\phi_k \in  \mathcal{V}$ is the test function of the $k^{th}$ equation. The sought weak solution\footnote{Strictly speaking, the weak solution of Eq. \eqref{eq:weakCNS} might be different from the classical solution of Eq. \eqref{eq:CNS}. Nevertheless, we use the same notation for both.} belongs as well to the same functional space: $Q_k \in  \mathcal{V}$.  Assuming a hierarchical basis, we then operate the following spatial decomposition into resolved and unresolved scales, respectively denoted by $\widebartilde{{~}}$ and $\widehat{~}$ accents:
\begin{gather}
\label{eq:scale_sep1}
\mathcal{V} =  \widebartilde{{\mathcal{V}}} \oplus  \widehat{\mathcal{V}}, \quad\quad \phi_k   =  \widebartilde{{\phi}}_k +  \widehat{\phi}_k,  \quad\quad Q_k   =  \widebartilde{{Q}}_k +  \widehat{Q}_k.
\end{gather}

The idea is to use the projection to derive the exact resolved and unresolved scale equations \cite{Ramakrishnan2005}, which respectively read
\begin{equation}
\left<\widebartilde{{\phi}}_k, \text{CNS}_k\left(\bm{Q}\right) \right> = \left<\widebartilde{{\phi}}_k, \mathcal{S}_k (\widebartilde{{\bm{Q}}}) \right> +  \left<\widebartilde{{\phi}}_k, \mathcal{C}_k (\widebartilde{{\bm{Q}}} ,\widehat{\bm{Q}}) \right> +  \left<\widebartilde{{\phi}}_k, \mathcal{R}_k (\widehat{\bm{Q}}) \right>,
\label{eq:large}
\end{equation}
and
\begin{equation}
\left<\widehat{\phi}_k, \text{CNS}_k\left(\bm{Q}\right) \right> = \left<\widehat{\phi}_k, \mathcal{S}_k (\widehat{\bm{Q}}) \right> +  \left<\widehat{\phi}_k, \mathcal{C}_k (\widehat{\bm{Q}},\widebartilde{{\bm{Q}}}) \right> +  \left<\widehat{\phi}_k, \mathcal{R}_k (\widebartilde{{\bm{Q}}}) \right>,
\end{equation}
where $\mathcal{S}$ and $\mathcal{R}$ designate terms  containing solely resolved or unresolved scales whereas $\mathcal{C}$ corresponds to cross terms containing the conjugate interactions between resolved and unresolved components.

In the context of LES, one is interested in the numerical solution of Eq. \eqref{eq:large} only.  This necessitates the modelling of  the SGS terms, i.e. $\mathcal{C}$ and $\mathcal{R}$, in the resolved scale equation to yield
\begin{equation}
\left<\widebartilde{{\phi}}_k, \text{CNS}_k\left(\bm{Q}\right) \right> \approx \left<\widebartilde{{\phi}}_k, \mathcal{S}_k (\widebartilde{{\bm{Q}}}) \right> +  \left<\widebartilde{{\phi}}_k, \mathcal{M}_k (\widebartilde{{\bm{Q}}} ) \right>,
\label{eq:sgs_model}
\end{equation}
where $\mathcal{M}$ represents the model terms such that:
\begin{equation*}
  \left<\widebartilde{{\phi}}_k, \mathcal{M}_k (\widebartilde{{\bm{Q}}} ) \right> \approx  \left<\widebartilde{{\phi}}_k, \mathcal{C}_k (\widebartilde{{\bm{Q}}} ,\widehat{\bm{Q}}) \right> +  \left<\widebartilde{{\phi}}_k, \mathcal{R}_k (\widehat{\bm{Q}}) \right>.
\end{equation*}

A straight-forward approach consists of simply neglecting the SGS terms in \eqref{eq:large}. As exhibited by Collis \cite{Collis2001}, this in fact results in a basic modelling with the assumption of $\mathcal{M}_k (\widebartilde{{\bm{Q}}} ) \approx 0$. Variations of this approach appear in the literature under the designations of \textit{Implicit LES (ILES)}, \textit{model-free LES} as well as \textit{under-resolved DNS}. Due to the absence of modelling effort, this approach is computationally very appealing and, conjointly with high-order discretization schemes, delivers satisfactory results on sufficiently fine h/p resolutions  at moderate Reynolds numbers \cite{vermeire-et-al_2016}. This success often stems from suitable dissipation/dispersion footprint of these schemes in the spectral domain, mimicking the naturally occurring dissipation/dispersion due to SGS terms. However, as noted as well in \cite{Flad2017}, there is still no substantial evidence for the reliability of this approach on coarse resolutions typical of industrial LES applications. 

On the other hand, the adoption of classical SGS models such as Smagorinsky to represent $\mathcal{M}_k (\widebartilde{{\bm{Q}}})$ that is then applied to  all the resolved scales of a high-order two-level VMS formulation, although providing satisfactory results even for wall-bounded flows, often causes an over-dissipation of resolved eddies, thus degrading the potential of the VMS simulation.

 An intermediate approach is obtained by performing a three-level partitioning of the exact solution space by separating the resolved scales into large resolved and small resolved scales and applying the model to the latter set of scales only. We pursue first by presenting the modelling for the resolved scales and then by introducing the second scale separation step.

\subsubsection{Large-eddy simulation modelling for compressible flows}
\label{sec:LES-mod}
In this section we present the detailed derivation of the VMS-LES modelling approach for the set of resolved CNS equations. We proceed\footnote{$(\perp)$ and $(\approx)$ notations respectively refer to cancelation due to orthogonality and to negligibility assumption.} equation by equation and term by term and adopt the modelling approach of Lesieur et al. \cite[7.1]{lesieur_metais_comte_2005} with the major difference of considering variational projection rather than spatial filtering (smoothing) for scale separation. We furthermore assume basis orthonormality\footnote{Orthonormality is defined as: $\forall \, \{\phi,\varphi\}$ bases of $\mathcal{V}$, $\left<\phi,\varphi \right>=1$ if $\phi =\varphi$ and $\left<\phi,\varphi \right>=0$ otherwise.}, although this condition would not strictly apply to all FR/CPR basis functions.

\begin{itemize}
\item Equation of continuity ($k=0$):
\begin{align}
\left<\widebartilde{{\phi}}_k, \text{CNS}_k\left(\bm{Q}\right) \right> &= \left<\widebartilde{{\phi}}_k, \partial_t({{\rho}}) \right> +  \left<\widebartilde{{\phi}}_k, \partial_i({{\rho u}}_i) \right> \approx \left<\widebartilde{{\phi}}_k, \partial_t(\widebartilde{{\rho}} )\right> +   \left<\widebartilde{{\phi}}_k, \partial_i(\widebartilde{{\rho u}}_i )\right>,
\label{eq:LESderiv_cont}
\end{align}
since
\begin{equation*}
\left<\widebartilde{{\phi}}_k, \partial_t({{\rho}}) \right> = \left<\widebartilde{{\phi}}_k, \partial_t(\widebartilde{{\rho}} + \widehat{{\rho}}) \right> = \left<\widebartilde{{\phi}}_k, \partial_t(\widebartilde{{\rho}} )\right>+ \cancelto{0\,(\perp)}{\left<\widebartilde{{\phi}}_k,  \partial_t(\widehat{{\rho}}) \right>},
\end{equation*}
and
\begin{equation*}
\left<\widebartilde{{\phi}}_k, \partial_i({{\rho u}}_i) \right> = \left<\widebartilde{{\phi}}_k, \partial_i(\widebartilde{{\rho u}}_i + \widehat{{\rho u}}_i) \right> \approx \left<\widebartilde{{\phi}}_k, \partial_i(\widebartilde{{\rho u}}_i )\right>+ \cancelto{0\,(\approx)}{\left<\widebartilde{{\phi}}_k,  \partial_i(\widehat{{\rho u}}_i) \right>}.
\end{equation*}
\item Equations of momentum ($k\in [1,2,3]$):
\begin{align}
\left<\widebartilde{{\phi}}_k, \text{CNS}_k\left(\bm{Q}\right) \right> &= \left<\widebartilde{{\phi}}_k, \partial_t({{\rho u_k}}) \right> +  \left<\widebartilde{{\phi}}_k, \partial_i({{\rho u}}_i {u}_k + P \,\delta_{ik}) \right> -  \left<\widebartilde{{\phi}}_k, \partial_i(\tau_{ik}) \right>, \nonumber\\
\begin{split}
&\approx \left<\widebartilde{{\phi}}_k, \partial_t(\widebartilde{{\rho u}}_k )\right> + \left<\widebartilde{{\phi}}_k, \partial_i( \widebartilde{{\rho u}}_i \widebartilde{{u}}'_k + \widebartilde{{P}}''\,\delta_{ik}) \right> -\left<\widebartilde{{\phi}}_k, \partial_i(\widebartilde{{\tau}}'_{ik}) \right> 
- \left<\widebartilde{{\phi}}_k, \partial_i(\tau^\text{SGS}_{ik}) \right>,
\label{eq:LESderiv_mom}
\end{split}
\end{align}
considering
\begin{equation*}
\left<\widebartilde{{\phi}}_k, \partial_t({{\rho u_k}}) \right> = \left<\widebartilde{{\phi}}_k, \partial_t(\widebartilde{{\rho u}}_k + \widehat{{\rho u}}_k) \right> = \left<\widebartilde{{\phi}}_k, \partial_t(\widebartilde{{\rho u}}_k )\right>+ \cancelto{0\,(\perp)}{\left<\widebartilde{{\phi}}_k,  \partial_t(\widehat{{\rho u}}_k) \right>}, 
\end{equation*}
\begin{align}
\left<\widebartilde{{\phi}}_k, \partial_i({{\rho u}}_i u_k) \right> &= \left<\widebartilde{{\phi}}_k, \partial_i( (\widebartilde{{\rho u}}_i +\widehat{\rho u}_i)  (\widebartilde{{u}}'_k+\widehat{u}'_k)  )\right> = \left<\widebartilde{{\phi}}_k, \partial_i( \widebartilde{{\rho u}}_i \widebartilde{{u}}'_k - \mathcal{T}^\text{SGS}_{ik}  )\right> \nonumber\\
&=\left<\widebartilde{{\phi}}_k, \partial_i( \widebartilde{{\rho u}}_i \widebartilde{{u}}'_k - \tau^\text{SGS}_{ik}  )\right> + \left<\widebartilde{{\phi}}_k, \partial_i(-\frac{1}{3}\mathcal{T}^\text{SGS}_{ll} \,\delta_{ik}  )\right>,
\label{eq:LESderiv_mom_nonlin}
\end{align}
where analogously to Favre averaging, we define\footnote{The prime symbol is meant to recall that for example $\widebartilde{{u}}'_k$ does not necessarily belong to the space of $\widebartilde{{\mathcal{V}}}$, and similarly for $\widehat{u}'_k$.} $\widebartilde{{u}}'_k := \widebartilde{{\rho u}}_k / \widebartilde{{\rho}}$ and $\widehat{u}'_k:= u_k-\widebartilde{{u}}'_k$; the subgrid stress tensor with components $\mathcal{T}^\text{SGS}_{ij} := -{\rho u}_i {u}_j + \widebartilde{{\rho u}}_i \widebartilde{{u}}'_j$, is separated into an isotropic and a deviatoric contribution, respectively designated by: $\frac{1}{3}\mathcal{T}^\text{SGS}_{ll} \,\delta_{ij}$ and $\tau^\text{SGS}_{ij}:=\mathcal{T}^\text{SGS}_{ij}-\frac{1}{3}\mathcal{T}^\text{SGS}_{ll}\,\delta_{ij}$; the pressure term is approximated as 
\begin{equation*}
\left<\widebartilde{{\phi}}_k, \partial_i(P \,\delta_{ik}) \right> \approx \left<\widebartilde{{\phi}}_k, \partial_i(\widebartilde{{P}}''\,\delta_{ik}) \right>,
\end{equation*}
where following \cite{lesieur_metais_comte_2005}, we define a macro-pressure, $\widebartilde{{P}}'' := \widebartilde{{P}}' -\frac{1}{3}\mathcal{T}^\text{SGS}_{ll}  $, by incorporating the isotropic portion of the SGS stress tensor from the last term of \eqref{eq:LESderiv_mom_nonlin} into $\widebartilde{{P}}':= R \,\widebartilde{{\rho}}\, \widebartilde{{ T}}' $ with
\[\widebartilde{{ T}}' := \left( {\widebartilde{{\rho}} \,C_v} \right)^{-1} \left( \widebartilde{{\rho E}} - \frac{1}{2}\left(\widebartilde{{\rho u_i}} \widebartilde{{u}}_i - \mathcal{T}^\text{SGS}_{ll} \right) \right),\]
which can be further simplified by defining a macro-temperature, $ \widebartilde{{ T}}'' :=  \widebartilde{{ T}}' -\left(2\,\widebartilde{{ \rho}}\,C_v  \right)^{-1} \mathcal{T}^\text{SGS}_{ll} $, to absorb the trace of the SGS stress tensor; we relate the macro-pressure to the macro-temperature by assuming an extended ideal gas law: $\widebartilde{{P}}'' \approx R \, \widebartilde{{\rho}} \, \widebartilde{{T}}''$,
and finally, we assume the following approximation for the projection of the stress tensor:
\begin{equation*}
\left<\widebartilde{{\phi}}_k, \partial_i(\tau_{ik}) \right>  \approx \left<\widebartilde{{\phi}}_k, \partial_i(\widebartilde{{\tau}}'_{ik}) \right>,
\end{equation*}
where $\widebartilde{{\tau}}'_{ik}:=2\,\widebartilde{{\mu}}' \,\widebartilde{{A}}'_{ik}$ with $\widebartilde{{\mu}}' := \mu(\widebartilde{{T}}'')$ computed from \eqref{eq:Suth}, and $\widebartilde{{A}}'_{ik} := A_{ik}(\widebartilde{{\bm{u}}}')$ evaluated from \eqref{eq:defor_tesnor}.

\item Equation of energy ($k=4$):
\begin{align}
\left<\widebartilde{{\phi}}_k, \text{CNS}_k\left(\bm{Q}\right) \right> &= \left<\widebartilde{{\phi}}_k, \partial_t({{\rho E}}) \right>+ \left<\widebartilde{{\phi}}_k, \partial_i((\rho E + P)\,u_i ) \right> -  \left<\widebartilde{{\phi}}_k, \partial_i(\tau_{ij}  u_j + \lambda \,{\partial_i(T)} ) \right>, \nonumber\\
\begin{split}
&\approx \left<\widebartilde{{\phi}}_k, \partial_t(\widebartilde{{\rho E}} )\right> +\left<\widebartilde{{\phi}}_k, \partial_i( (\widebartilde{{\rho E}}+\widebartilde{{P}}'')\widebartilde{{u_i}}') \right> -\left<\widebartilde{{\phi}}_k, \partial_i(\widebartilde{{\tau}}'_{ij} \widebartilde{{u_j}}' + \widebartilde{{\lambda}}' \,{\partial_i(\widebartilde{{T}}'')} ) \right> -\left<\widebartilde{{\phi}}_k, \partial_i(\mathcal{H}^\text{SGS}_i  ) \right>,
\label{eq:LESderiv_ener}
\end{split}
\end{align}
since
\begin{align*}
\left<\widebartilde{{\phi}}_k, \partial_t({{\rho E}}) \right> &= \left<\widebartilde{{\phi}}_k, \partial_t(\widebartilde{{\rho E}} + \widehat{{\rho E}}) \right> = \left<\widebartilde{{\phi}}_k, \partial_t(\widebartilde{{\rho E}} )\right>+ \cancelto{0\,(\perp)}{\left<\widebartilde{{\phi}}_k,  \partial_t(\widehat{{\rho E}}) \right>}, 
\end{align*}
\begin{align*}
 \left<\widebartilde{{\phi}}_k, \partial_i((\rho E+P)\,u_i ) \right> =  \left<\widebartilde{{\phi}}_k, \partial_i((\rho E+P )\,(\widebartilde{{u}}'_i + \widehat{ u}'_i) ) \right> = \left<\widebartilde{{\phi}}_k, \partial_i( (\widebartilde{{\rho E}}+\widebartilde{{P}}'')\widebartilde{{u}}'_i -\mathcal{H}^\text{SGS}_i ) \right>, 
\end{align*}
where $\mathcal{H}^\text{SGS}_i:= -(\rho E+P)\,u_i +(\widebartilde{{\rho E}}+\widebartilde{{P}}'')\widebartilde{{u}}'_i $ is the $i^{th}$ component of the subgrid heat flux vector; 
and finally
\begin{align*}
\left<\widebartilde{{\phi}}_k, \partial_i(\tau_{ij}  u_j + \lambda \,{\partial_i(T)} ) \right> \approx \left<\widebartilde{{\phi}}_k, \partial_i(\widebartilde{{\tau}}'_{ij}\widebartilde{{u}}'_j  + \widebartilde{{\lambda}}' \,{\partial_i(\widebartilde{{T}}'')} ) \right>,
\end{align*}
where $\widebartilde{{\lambda}}' :=\frac{C_p}{Pr}\,\widebartilde{{\mu}}' $.

\end{itemize}

Through these developments, we have derived the set of model equations governing the resolved scales, i.e., Eqs. \eqref{eq:LESderiv_cont}, \eqref{eq:LESderiv_mom} and \eqref{eq:LESderiv_ener} and thus by comparison to the VMS-LES formulation of Eqs. \eqref{eq:large} and \eqref{eq:sgs_model}, we can identify the resolved terms, the remaining SGS terms   to be modelled and the model terms to respectively be 
\begin{align*}
\mathcal{S}_k (\widebartilde{{\bm{Q}}}) &:= \text{CNS}_k (\widebartilde{{\bm{Q}}}) = \frac{\partial \widebartilde{{Q}}_k}{\partial t} + \frac{\partial {F}^{\text{inv}}_{ik} ( \widebartilde{{\bm{Q}}})}{\partial x_i} - \frac{\partial {F}^{\text{vis}}_{ik} (\widebartilde{{\bm{Q}}}, \bm{\nabla} \widebartilde{{\bm{Q}}})}{\partial x_i},\\
\mathcal{C}_k (\widebartilde{{\bm{Q}}} ,\widehat{\bm{Q}}) +\mathcal{R}_k (\widehat{\bm{Q}}) &:= - \frac{\partial {F}^{\text{SGS}}_{ik}}{\partial x_i},\\
\mathcal{M}_k (\widebartilde{{\bm{Q}}}) &:= - \frac{\partial {F}^{\text{mod}}_{ik} (\widebartilde{{\bm{Q}}}, \bm{\nabla} \widebartilde{{\bm{Q}}})}{\partial x_i},
\end{align*}
where $\text{CNS}_k (\widebartilde{{\bm{Q}}})$ designates the system of compressible NS equations applied to the resolved conservative variables along with the primitive variables, $\widebartilde{{\rho}}$ and $\widebartilde{{u}}_i'$, macro-quantities, $\widebartilde{{P}}''$ and $\widebartilde{{T}}''$, as well as $\widebartilde{{\mu}}'$ and $\widebartilde{{\lambda}}'$. 
The subgrid flux,
\begin{equation*}
{F}^{\text{SGS}}_{i}  := \left[ 0,\,\tau^\text{SGS}_{i1},\, \tau^\text{SGS}_{i2},\, \tau^\text{SGS}_{i3},\,  \mathcal{H}^\text{SGS}_i  \right] ^\intercal,
\label{eq:SGS_flux}
\end{equation*}
is approximated by the model flux,
\begin{align}
{F}^{\text{mod}}_{i} (\widebartilde{{\bm{Q}}}, \bm{\nabla} \widebartilde{{\bm{Q}}}) := &\left[ 0,\,\tau^\text{mod}_{i1},\, \tau^\text{mod}_{i2},\, \tau^\text{mod}_{i3},\,  \mathcal{H}^\text{mod}_i  \right] ^\intercal
\approx \,{F}^{\text{SGS}}_{i},
 \label{eq:model}
\end{align}
with the model stress tensor components defined via the Boussinesq assumption as
\begin{equation}
\label{eq:SGS_tau}
\tau^\text{mod}_{ik} := 2 \,\widebartilde{{\rho}}\, \nu_t(\widebartilde{{\bm{u}}}')\, A_{ik}(\widebartilde{{\bm{u}}}'),
\end{equation}
and the model heat flux components expressed by
\begin{equation}
\label{eq:SGS_H}
\mathcal{H}^\text{mod}_i  := \frac{C_p}{Pr_t}\, \widebartilde{{\rho}}\, \nu_t(\widebartilde{{\bm{u}}}') \frac{\partial \, \widebartilde{{T}}''}{\partial x_i},
\end{equation}
where $Pr_t$ is the turbulent Prandtl number and $\nu_t$ is the turbulent (eddy) viscosity.

\subsubsection{Eddy viscosity model}
\label{sec:eddyviscmod}
The system of compressible LES equations is closed by providing a proper model for $\nu_t$. We adopt the classical Smagorinsky-Lilly model from \cite{Pope2000}: 
\begin{equation}
\nu_t(\widebartilde{{\bm{u}}}') = (C_S\,\Delta)^2 \,|S(\widebartilde{{\bm{u}}}')|,
\label{eq:smag}
\end{equation}
which relates the SGS viscosity to the cutoff length scale via $\Delta$, the local filter width in units of length, and to the contraction norm,
$|S(\widebartilde{{\bm{u}}}')| := \sqrt{2\,S_{ij}(\widebartilde{{\bm{u}}}') S_{ij}(\widebartilde{{\bm{u}}}')}$, of the isotropic part of the deformation tensor, $S_{ij}(\widebartilde{{\bm{u}}}') := \frac{1}{2} \left( \frac{\partial \widebartilde{{{u}}}'_j}{\partial x_i} + \frac{\partial \widebartilde{{{u}}}'_i}{\partial x_j} \right)$. In Section \ref{sec:SGS3level}, we will discuss in more details our choice of values for the Smagorinsky constant, $C_S$, and for $\Delta$ when introducing the separation in resolved scales. 

\subsubsection{Final VMS-LES form with two-level partitioning}
The VMS-LES formalism with a two-level partitioning reads
\begin{equation}
\left<\widebartilde{{\phi}}_k, \text{LES}_k (\widebartilde{{\bm{Q}}})\right>=0,
\label{eq:VMS-LES}
\end{equation}
where
\begin{equation}
\text{LES}_k (\widebartilde{{\bm{Q}}}) := \frac{\partial \widebartilde{{Q}}_k}{\partial t} + \frac{\partial {F}^{\text{inv}}_{ik} ( \widebartilde{{\bm{Q}}})}{\partial x_i} - \frac{\partial {F}^{\text{vis}}_{ik} (\widebartilde{{\bm{Q}}}, \bm{\nabla} \widebartilde{{\bm{Q}}})}{\partial x_i} - \frac{\partial {F}^{\text{mod}}_{ik} (\widebartilde{{\bm{Q}}}, \bm{\nabla} \widebartilde{{\bm{Q}}})}{\partial x_i}.
\label{eq:LES}
\end{equation}

\subsection{VMS formulation with a three-level partitioning via the FR/CPR scheme}
\label{sec:methodology_VMS3CPR}
\subsubsection{High-order FR/CPR scheme}
\label{sec:methodology_CPR}
To illustrate the idea of the FR/CPR scheme, let us start from the VMS-LES formulation \eqref{eq:VMS-LES}-\eqref{eq:LES} derived in the previous section for the $k^{th}$ governing equation and cluster all the fluxes together into\footnote{The prime symbol is again meant to recall that $\widebartilde{{F}}'_{ik} \in \widebartilde{{\mathcal{V}}}$ is not necessarily true. The inviscid flux is actually often non-polynomial for compressible flows.} $\widebartilde{{F}}'_{ik} \equiv F_{ik}(\widebartilde{{\bm{Q}}})$  to obtain:
\begin{equation*}
\int_{\bm{\Omega}}  \widebartilde{{\phi}} \left(\frac{\partial \widebartilde{{Q}}}{\partial t} + \frac{\partial \widebartilde{{F}}'_{i}}{\partial x_i} \right) {d\Omega} = 0.
\end{equation*}
where the $k$ indices are omitted for simplicity.
Integrating the divergence term by parts once and employing the divergence theorem yield the variational formulation in the Green's form,
\begin{equation}
\int_{\bm{\Omega}}  \widebartilde{{\phi}} \,\frac{\partial \widebartilde{{Q}}}{\partial t} {d\Omega} - \int_{\bm{\Omega}} \frac{\partial \widebartilde{{\phi}}}{\partial x_i} \, \widebartilde{{F}}'_{i} \,{d\Omega} + \int_{\bm{\Gamma}}  \widebartilde{{\phi}} \, \widebartilde{{F}}'^n \,{{d\Gamma}}= 0,
\label{eq:greenform}
\end{equation}
where $\bm{\Gamma}$ is the boundary of the spatial domain and $\widebartilde{{F}}'^n\equiv \widebartilde{{F}}'_{i}\,n_i$, with $n_i$ denoting the $i^{th}$ component of the local unit outward normal vector, $\bm{n}$. 

Let us note that this formulation is the one at the basis of the classical continuous and discontinuous finite element discretization methods. In both cases, the domain is approximated and partitioned into a tessellation, $\bm{\Omega}^{hp}$, of $N_{el}$  elements denoted by $\bm{\Omega}_{e_i}$, i.e., $\bm{\Omega}^{hp} := \cup_{e_i=1}^{N_{el}} \bm{\Omega}_{e_i} \approx \bm{\Omega}$, with the difference that in the discontinuous case, the test function and the solution are defined on a compact elemental support such that $\forall \,\bm{x}\in \bm{\Omega}_{e_i}$, $\{\widebartilde{{\phi}} (\bm{x}),\widebartilde{{Q}}(t,\bm{x})\} \in \widebartilde{{\mathcal{V}}}(\bm{\Omega}_{e_i})$. For the rest of this work, we furthermore assume that $\forall \,\bm{x}\in \bm{\Omega}_{e_i}$, $\widebartilde{{Q}}(t,\bm{x})  \in {{\rm I\!P}}^{\widebartilde{{p}}}(\bm{\Omega}_{e_i}) \subset \widebartilde{{\mathcal{V}}}(\bm{\Omega}_{e_i})$ where ${\rm I\!P}^p(\bm{\Omega}_{e_i})$ is the space of polynomials of degree $p$ or less with compact support on $\bm{\Omega}_{e_i}$ and $\widebartilde{{p}}$ is the cutoff degree of the resolved scales. In order to couple the internal solution of each element to external information from  neighbouring elements or  boundary conditions, we take advantage of numerical fluxes, denoted\footnote{ $^-$ and $^+$ exponents respectively refer to internal and external quantities at the element boundaries.} $\widebartilde{{\mathcal{F}}}'^n \equiv \widebartilde{{\mathcal{F}}}' (\widebartilde{{\bm{Q}}}_{}^-,\widebartilde{{\bm{Q}}}_{}^+, \bm{n})\approx \widebartilde{{F}}'^{n}$,  at the basis of finite volume schemes, to replace the discontinuous flux in the surface integral of \eqref{eq:greenform}:
\begin{equation*}
\sum_{e_i}\left(\int_{\bm{\Omega}_{e_i}}  \widebartilde{{\phi}}\, \frac{\partial \widebartilde{{Q}}}{\partial t} {d\Omega} - \int_{\bm{\Omega}_{e_i}} \frac{\partial \widebartilde{{\phi}}}{\partial x_i}  \widebartilde{{F}}'_{i} \,{d\Omega} + \int_{\bm{\Gamma}_{e_i}}  \widebartilde{{\phi}} \,\widebartilde{{\mathcal{F}}}'^n  \,{{d\Gamma}} \right)= 0.
\end{equation*}

By applying the integration by parts and the divergence theorem once again, the variational formulation in the divergence form is obtained that for each element reads
\begin{equation}
\int_{\bm{\Omega}_{e_i}}  \widebartilde{{\phi}}\, \frac{\partial \widebartilde{{Q}}}{\partial t} {d\Omega} + \int_{\bm{\Omega}_{e_i}} \widebartilde{{\phi}}\,  \frac{\partial \widebartilde{{F}}'_{i} }{\partial x_i}  \,{d\Omega} + \int_{\bm{\Gamma}_{e_i}}  \widebartilde{{\phi}} \,(\widebartilde{{\mathcal{F}}}'^n-\widebartilde{{F}}'^n)  \,{{d\Gamma}}= 0.
\label{eq:divform}
\end{equation}
One can further project the surface integrand onto the elemental volume and thus recover a correction field, $ \widebartilde{{\mathscr{C}}} \in {\rm I\!P}^{\widebartilde{{p}}}(\bm{\Omega}_{e_i})$, via the following lifting operation:
\begin{equation}
 \int_{\bm{\Omega}_{e_i}}  \widebartilde{{\phi}} \, \widebartilde{{\mathscr{C}}}  \,{d\Omega} = \int_{\bm{\Gamma}_{e_i}}  \widebartilde{{\phi}}\, (\widebartilde{{\mathcal{F}}}'^n-\widebartilde{{F}}'^n)  \,{{d\Gamma}},
\label{eq:lifting1}
\end{equation}
which substituted to Eq. \eqref{eq:divform} yields
\begin{equation*}
\int_{\bm{\Omega}_{e_i}}  \widebartilde{{\phi}}\, \left(\frac{\partial \widebartilde{{Q}}}{\partial t} + \,  \frac{\partial \widebartilde{{F}}'_{i} }{\partial x_i}  +   \,\widebartilde{{\mathscr{C}}} \right) \,{d\Omega}= 0,
\end{equation*}
that with a proper choice of test function can be expressed as a differential scheme in nodal form,
\begin{equation}
\frac{\partial \widebartilde{{Q}}}{\partial t} + \,  \frac{\partial \widebartilde{{F}}'_{i} }{\partial x_i}  +   \,\widebartilde{{\mathscr{C}}} = 0.
\label{eq:CPR1}
\end{equation}

\begin{remark}
The lifting operator is a variational equivalent to the derivatives of the analytical correction functions in the original tensor-product\footnote{Designations \textit{tensor product} and \textit{Kronecker product} are employed equivalently in this study.} FR formulation \cite{Huynh2007}. To illustrate this link more clearly, let us
segment the elemental boundary into $N_f$ non-overlapping and connected facettes, noted $\bm{\Gamma}_f$ such that $\bm{\Gamma}_{e_i} = \cup_{f=1}^{N_{f}} \bm{\Gamma}_{f}$,  and thus rewrite the resultant correction field as
\begin{equation*}
 \widebartilde{{\mathscr{C}}} = \sum_f  \widebartilde{{\mathscr{C}}}_f,
\end{equation*}
where each facette's correction field is
\begin{equation*}
\widebartilde{{\mathscr{C}}}_f:= \mathscr{L}_f \left(\widebartilde{{\mathcal{F}}}'^n{|_{\bm{\Gamma}_f}}-\widebartilde{{F}}'^n{|_{\bm{\Gamma}_f}}\right),
\end{equation*}
with $\mathscr{L}_f$ designating the lifting operator  \eqref{eq:lifting1} for the facette $\bm{\Gamma}_f$. We can hence rewrite the formulation \eqref{eq:CPR1} as follows:
\begin{equation}
\frac{\partial \widebartilde{{Q}}}{\partial t} + \,  \frac{\partial \widebartilde{{F}}'_{i} }{\partial x_i}  +   \,\sum_f  \mathscr{L}_f \left(\widebartilde{{\mathcal{F}}}'^n{|_{\bm{\Gamma}_f}}-\widebartilde{{F}}'^n{|_{\bm{\Gamma}_f}}\right) = 0.
\label{eq:CPR2}
\end{equation}
In the original FR, instead of computing a correction field added to the flux divergence, one corrects the flux itself by the addition of a flux correction per facette that reads
\begin{equation*}
F^{\mathscr{C}_f}_i:= g_f(\bm{x}) \, \left(\widebartilde{{\mathcal{F}}}'^n{|_{\bm{\Gamma}_f}}-\widebartilde{{F}}'^n{|_{\bm{\Gamma}_f}}\right)_i,
\end{equation*}
where the index $i$ in the right-hand side refers to the $i^{th}$ space coordinate's contribution to the term $(\widebartilde{{\mathcal{F}}}'^n{|_{\bm{\Gamma}_f}}-\widebartilde{{F}}'^n{|_{\bm{\Gamma}_f}})$ evaluated at discrete points along the facette, called flux points, and   $g_f(\bm{x})$ is an analytically defined polynomial correction function of degree ${\widebartilde{{p}}+1}$ associated to the face  $\bm{\Gamma}_f$  (see \cite{Huynh2007,Castonguay-et-al_2012} and \ref{app:Fr_corrfuncs} for definitions).
The corrected flux is thus defined as
\begin{equation*}
F^C_{i} := \widebartilde{{F}}'_{i} + \sum_f F^{\mathscr{C}_f}_i.
\end{equation*}
This results in the FR scheme to be expressed as
\begin{equation*}
\frac{\partial \widebartilde{{Q}}}{\partial t} + \,  \frac{\partial F^C_{i} }{\partial x_i}   = 0,
\end{equation*}
or in an expanded form as
\begin{equation}
\frac{\partial \widebartilde{{Q}}}{\partial t} + \,  \frac{\partial \widebartilde{{F}}'_{i} }{\partial x_i}  +   \,\sum_f \left(\partial_i \left(g_f(\bm{x}) \right)\, \left(\widebartilde{{\mathcal{F}}}'^n{|_{\bm{\Gamma}_f}}-\widebartilde{{F}}'^n{|_{\bm{\Gamma}_f}}\right)_i \right)=0.
\label{eq:CPR3}
\end{equation}
Finally, by comparing formulations \eqref{eq:CPR2} and \eqref{eq:CPR3}, the equivalence is reduced to
\begin{equation*}
\partial_i \left(g_f(\bm{x}) \right)\, \left(\widebartilde{{\mathcal{F}}}'^n{|_{\bm{\Gamma}_f}}-\widebartilde{{F}}'^n{|_{\bm{\Gamma}_f}}\right)_i = \mathscr{L}_f \left(\widebartilde{{\mathcal{F}}}'^n{|_{\bm{\Gamma}_f}}-\widebartilde{{F}}'^n{|_{\bm{\Gamma}_f}}\right),
\end{equation*}
which assumes that the quadrature points used to compute the surface integral in the lifting-based formulation are the same as the flux points of the original FR.  The equivalence is expressed in a concise form as
\begin{equation*}
\frac{\partial F^{\mathscr{C}_f}_i}{\partial x_i} = \widebartilde{{\mathscr{C}}}_f.
\end{equation*}
We note that  it is advantageous to use the formulation based on correction functions since analytical expressions for the latter are readily available in the literature for a variety of schemes,  thus sparing the need to define the corresponding test function for each scheme in the formulation based on lifting operation.

\end{remark}

\begin{remark}[]
\label{remrk:numflx}
The numerical flux $\widebartilde{{\mathcal{F}}}'^n$ in Eqs.  \eqref{eq:lifting1} and \eqref{eq:CPR3} is computed differently for each flux type. For the advective flux, we use the Roe's approximate Riemann solver based on \cite[4.3.3]{blazek-2001} which is shown  \cite{Winters2017}  to produce superior LES results compared to the local Lax-Friedrichs (LLF) interface function and  has the following concise form:
\begin{equation}
\label{eq:alpharoe}
\widebartilde{{\mathcal{F}}}'^{n} = \frac{1}{2}\left( F_{i}(\widebartilde{{\bm{Q}}}_{}^- )\,n_i+ F_{i}(\widebartilde{{\bm{Q}}}_{}^+)\,n_i - \alpha\, \mathcal{D}(\widebartilde{{\bm{Q}}}_{}^- ,\widebartilde{{\bm{Q}}}_{}^+,\bm{n}) \right),
\end{equation}
where $\mathcal{D}(\widebartilde{{\bm{Q}}}_{}^- ,\widebartilde{{\bm{Q}}}_{}^+,\bm{n}) $ is an upwinding dissipation function and we introduce the coefficient $\alpha$ to calibrate the amount of dissipation to lower values for low-Mach flows and consequently designate the formulation \eqref{eq:alpharoe}  by "{$\alpha$-Roe}" in this study.

As for the viscous and model fluxes, we adopt the tensor-product formulation of the second Bassi and Rebay (BR2) scheme detailed in \cite{Wang-et-al_2011a,Navah2017a} which penalizes the spatial gradients of the solution as a function of solution jumps at the interface.

\end{remark}

\begin{remark}[]
\label{remk:3}
The flux divergence is the only term in Eq. \eqref{eq:CPR1} that does not necessarily belong to the solution space, i.e., $\frac{\partial \widebartilde{{F}}'_{i} }{\partial x_i}  \notin {\rm I\!P}^{\widebartilde{{p}}}(\bm{\Omega}_{e_i})$, due to primed and double-primed terms as well as product terms in the flux, presented in Section \ref{sec:LES-mod}, which often result in non-polynomial expressions for compressible flows.
Projection methods have been introduced \cite{Wang-Gao_2009,Gao-Wang_2013}  (see  Section \ref{sec:proj}) to yield $\frac{\partial \widebartilde{{F}}_{i} }{\partial x_i} \in {\rm I\!P}^{\widebartilde{{p}}}(\bm{\Omega}_{e_i})$ where $\frac{\partial \widebartilde{{F}}_{i} }{\partial x_i} \approx \frac{\partial \widebartilde{{F}}'_{i} }{\partial x_i} $  designates the projected flux divergence.
\end{remark}

\subsubsection{Elemental projection operators}
\label{sec:proj}
We have already introduced the lifting operator \eqref{eq:lifting1} which projects the normal flux difference at the boundary  onto the test function and recovers a volume correction field. Here, we generalize and clarify further the details of elemental projection and recovery operations that will be useful for the regularization of the flux divergence mentioned in Remark \ref{remk:3} as well as for the presentation of the second scale separation mechanism in Section \ref{sec:VMS-FR-3level}.

A signal or term $\forall \, \widebartilde{{\mathscr{S}}} \in{\rm I\!P}^{\widebartilde{{p}}}(\bm{\Omega}_{e_i})$ can be decomposed as $ \widebartilde{{\mathscr{S}}}(\bm{x}) :=  \widebartilde{{\mathscr{S}}}_{m}\,\widebartilde{{\phi}}_m(\bm{x})$ where $\widebartilde{{\phi}}_m(\bm{x})$,  with $m \in \left[0,...,\widebartilde{{N}}_\text{DOFs}-1\right]$ and $\widebartilde{{N}}_\text{DOFs}:=(\widebartilde{{p}}+1)^3$, is a basis of ${\rm I\!P}^{\widebartilde{{p}}}(\bm{\Omega}_{e_i})$ and $\widebartilde{{\mathscr{S}}}_{m}$ is its associated weight  coefficient. For example, we consider the following decomposition of the resolved state variable: $\widebartilde{{Q}}(t,\bm{x}) := \widebartilde{{Q}}_{m}(t)\,\widebartilde{{\phi}}_m(\bm{x})$. 

Two useful basis sets of ${\rm I\!P}^{\widebartilde{{p}}}(\bm{\Omega}_{e_i})$ are:
\begin{itemize}
\item Lagrange ($\mathcal{L}$) polynomials,  $\widebartilde{{\phi}}_l^\mathcal{L}$, constitute a nodal basis with interpolation property (See Definition \ref{def:intrp} ) such that $\widebartilde{{\phi}}_l^\mathcal{L} (\bm{x}_n) = \delta_{ln}$ where $\bm{x}_n $ with $n \in \left[0,...,\widebartilde{{N}}_\text{DOFs}-1\right]$ are a set of \textit{solution nodes} as well as $\forall \bm{x} \in \bm{\Omega}_{e_i}, \, \sum_l \widebartilde{{\phi}}_l^\mathcal{L} (\bm{x}) = 1$.  We use the Gauss-Lobatto-Legendre (GLL) set (see Remark \ref{def:3dgll}) along with its associated quadrature for the numerical evaluation of integrals.
\item Normalized Legendre ($L$) polynomials, $\widebartilde{{\phi}}_l^{L} $,  form a hierarchical modal basis with orthonormality  property, $\left<\widebartilde{{\phi}}_l^{L} ,\widebartilde{{\phi}}_m^{L}\right> = \delta_{lm}$.
\end{itemize}

\begin{definition}
\label{def:intrp}
Interpolation of a signal $\mathscr{S}(\bm{x})$ is defined as $I^p(\mathscr{S})\in {\rm I\!P}^{p}:= \mathscr{S}_l \phi^\mathcal{L}_l(\bm{x})$ where $\mathscr{S}_l:= \mathscr{S}(\bm{x}_l)$.
\end{definition}

\begin{remark}
\label{rmrk:nodmod}
Using the Lagrange/Legendre basis to define both the resolved state variable and test function gives rise to a nodal/modal scheme.
\end{remark}

\begin{remark}
\label{def:3dgll}
The three-dimensional nodal set is formed by tensor-products of the 1D GLL set.
\end{remark}

The projection and recovery of an arbitrary signal, $\mathscr{S}(\bm{x})$, onto ${\rm I\!P}^{\widebartilde{{p}}}(\bm{\Omega}_{e_i})$ can be achieved via the following system of $\widebartilde{{N}}_\text{DOFs}$ equations (indexed by $l$), in $\widebartilde{{N}}_\text{DOFs}$ unknowns ($\widebartilde{{\mathscr{S}}}_m^\mathcal{L}$ coefficients):
\begin{equation}
\left<\widebartilde{{\phi}}_l^L ,\widebartilde{{\phi}}_m ^\mathcal{L}\widebartilde{{\mathscr{S}}}_m ^\mathcal{L}\right>\, =\, \left<\widebartilde{{\phi}}_l^L ,\mathscr{S}\right>, 
\label{eq:project_recov}
\end{equation}
noted more concisely as $\widebartilde{{\phi}}_m^\mathcal{L} \widebartilde{{\mathscr{S}}}_m^\mathcal{L}= \widebartilde{{\mathcal{P}}}(\mathscr{S})$ where $\widebartilde{{\mathcal{P}}}$ designates the projection and recovery operator, the details of which are discussed in \ref{app:De-alaising}.

\begin{remark}
\label{rmrk:proj-alias}
Note that the recovered signal will suffer from \textit{aliasing errors} if the original signal in the right-hand side of \eqref{eq:project_recov} is not properly sampled, for example whenever it is approximated by an interpolation  $I^p (\mathscr{S}) \approx \mathscr{S}$ with insufficiently high $p$ (see Remark \ref{rmrk:proj-interp}), or when the integrals of \eqref{eq:project_recov} are numerically computed with quadratures of insufficient precision. In \ref{app:De-alaising}, we present the technique used in this study to reduce aliasing errors.
\end{remark}

\begin{remark}
\label{rmrk:proj-interp}
Note that: $ \widebartilde{{\mathcal{P}}} \left( I^{\widebartilde{{p}}} (\mathscr{S}) \right) = I^{\widebartilde{{p}}} (\mathscr{S}) $, and hence proper de-aliasing often requires a sampling via $I^p (\mathscr{S}) $ with $p>{\widebartilde{{p}}}$.
\end{remark}

Employing the concepts introduced in this section, we present now the regularization of the flux divergence mentioned in Remark \ref{remk:3}. Two methods are considered:
\begin{itemize}
\item Lagrange polynomial (LP): $
\partial_i \widebartilde{{F}}_{ik}:= {\partial_i I^{\widebartilde{{p}}}(\widebartilde{{F}}'_{ik})}$ which is an economical substitute to  ${\partial_i \widebartilde{{\mathcal{P}}}(\widebartilde{{F}}'_{ik})}$ which should ideally serve to represent ${\partial_i  \widebartilde{{F}}'_{ik}}$. 
We use this method to compute the viscous and model flux divergences, i.e., respectively $\partial_i \widebartilde{{F}}'^{\text{vis}}_{ik}$ and $\partial_i \widebartilde{{F}}'^{\text{mod}}_{ik}$.

\item Chain rule (CR):
$
{\partial_i \widebartilde{{F}}_{ik}}:=I^{\widebartilde{{p}}}\left(\frac{\partial \widebartilde{{F}}'_{ik}}{ \partial \widebartilde{{Q}}_r} \frac{ \partial \widebartilde{{Q}}_r}{ \partial x_i}   \right)$ which is an economical substitute to $\widebartilde{{\mathcal{P}}}\left(\frac{\partial \widebartilde{{F}}'_{ik}}{ \partial \widebartilde{{Q}}_r} \frac{ \partial \widebartilde{{Q}}_r}{ \partial x_i}   \right)$ which should ideally serve to represent ${\partial_i  \widebartilde{{F}}'_{ik}}. $
CR is shown \cite{Wang-Gao_2009} to be more accurate than LP but to be non-conservative. A fix for the conservation is provided in \cite{Gao-Wang_2013} which is not applied here since it is not crucial for the  considered flows.
This method is used to compute the inviscid flux divergence, $\partial_i \widebartilde{{F}}'^{\text{inv}}_{ik}$.
\end{itemize}

\subsubsection{VMS-LES-FR/CPR with two-level partitioning}
\label{sec:VMS-FR-2level}
Using the regularized flux divergences, the final form of the VMS-LES with a two-level partitioning, Eq. \eqref{eq:LES}, discretized via the FR/CPR scheme, Eq. \eqref{eq:CPR1}, is:
\begin{align}
 \frac{\partial \widebartilde{{Q}}_k}{\partial t} + \frac{\partial  \widebartilde{{F}}_{ik}^{\text{inv}}}{\partial x_i} - \frac{\partial \widebartilde{{F}}_{ik}^{\text{vis}}}{\partial x_i} - \frac{\partial \widebartilde{{F}}_{ik}^{\text{mod}}}{\partial x_i}
+ \widebartilde{{\mathscr{C}}}_k^{\text{inv}} - \widebartilde{{\mathscr{C}}}_k^{\text{vis}} - \widebartilde{{\mathscr{C}}}_k^{\text{mod}} =0,
\label{eq:LES2}
\end{align}
where all the terms belong to the space ${\rm I\!P}^{\widebartilde{{p}}}(\bm{\Omega}_{e_i})$.
\begin{remark}
 Although the VMS framework has the flexibility to allow for a different test function per governing equation, as considered in the derivations of Section \ref{sec:LES-mod}, we nevertheless choose the same test function (and hence the same correction function) for all of the equations and consequently drop the index $k$ for the test function.
\end{remark}
\subsubsection{VMS-LES-FR/CPR with three-level partitioning}
\label{sec:VMS-FR-3level}
The discrete formulation for the VMS-LES-FR/CPR is presented in Section \ref{sec:VMS-FR-2level} for a two-level partitioning of all scales present in the turbulent flow into  resolved and unresolved (SGS) scales (see  \eqref{eq:scale_sep1}). Following the approach of Collis \cite{Collis2001}, we go one step further by applying a second scale separation to isolate the large resolved and small resolved scales, respectively denoted by $\bar{{~}}$ and $\widetilde{~}$ accents:
\begin{gather}
\label{eq:scale_sep2}
\mathcal{V} =   \underbrace{\widebar{{\mathcal{V}}} \oplus {\widetilde{\mathcal{V}}}}_{ \widebartilde{{\mathcal{V}}} } \,  \oplus \, \widehat{\mathcal{V}}, \quad\quad  \phi =   \underbrace{\widebar{{\phi}} +  {\widetilde{\phi}}}_{ \widebartilde{{\phi}}}   \,+\,  \widehat{\phi},  \quad\quad  Q_k =   \underbrace{ \widebar{{Q}}_k + {\widetilde{Q}}_k}_{ \widebartilde{{Q}}_k }   \,+\,  \widehat{Q}_k.
\end{gather}
This separation provides an additional degree of flexibility in turbulence modelling since the SGS model flux can thus be applied to a select range of scales. 

Similarly to the first separation, the second separation mechanism relies as well on variational projection with the difference that we exploit the fact that  the resolved scales are in the space ${\rm I\!P}^{\widebartilde{{p}}}(\bm{\Omega}_{e_i})$, to define large and small scales by projecting onto the filtered modal basis. We further clarify this idea through the following definitions and demonstrations:
\begin{definition}
The mapping operator between 1D Lagrange and normalized Legendre polynomials (see \ref{app:bases} for definitions), respectively noted as  $ \widebartilde{{\phi}}_a^\mathcal{L\text{1D}}$ and $\widebartilde{{\phi}}_b^{L\text{1D}}$, is defined as $\widebartilde{{C}}_{ba}^\text{1D} := \left<\widebartilde{{\phi}}_b^{L\text{1D}}, \widebartilde{{\phi}}_a^\mathcal{L\text{1D}} \right>_\text{1D}$ such that $ \widebartilde{{\phi}}_a^\mathcal{L\text{1D}} = \widebartilde{{\phi}}_b^{L\text{1D}} \, \widebartilde{{C}}_{ba}^\text{1D}$ for $\{a,b\}\in \left[0,..., \widebartilde{{p}}\right]$. 
\end{definition}

\begin{lemma}
\label{lem:C1d}
The matrix $\widebartilde{{C}}_{ba}^\text{1D}$ is invertible.
\begin{proof}
The polynomial sets $ \widebartilde{{\phi}}_a^\mathcal{L\text{1D}}$ and $\widebartilde{{\phi}}_b^{L\text{1D}}$ form  two separate complete bases of ${\rm I\!P}^{\widebartilde{{p}}}(\Delta x)$ and hence any polynomial of either set is linearly independent with regards to all the rest of the same set. Since each column of $\widebartilde{{C}}_{ba}^\text{1D}$ contains the coefficients defining $ \widebartilde{{\phi}}_a^\mathcal{L\text{1D}}$, it is therefore linearly independent of the other columns. We hence have: $ \widebartilde{{\phi}}_a^\mathcal{L\text{1D}} \, \widebartilde{{C}}_{ab}^{-\text{1D}} = \widebartilde{{\phi}}_b^{L\text{1D}}$.
\end{proof}
\end{lemma}

\begin{definition}
\label{def:1dlowpassfil}
The 1D low-pass modal filter operator is defined as
\begin{equation*}
   \widebar{\mathrm{F}}^\text{1D}_{ab} := 
\begin{cases}
    \delta_{ab},& \text{if } a < \widebar{m}  \text{~or~} b < \widebar{m} \\
    0,              & \text{otherwise}
\end{cases}
\end{equation*}
where $\delta_{ab}$ is the Kronecker delta and  $\widebar{m} \in[0,...,\widebartilde{{p}}+1]$ is a fixed value for the large-scale cutoff mode and indices are $\{a,b\}\in \left[0,...,\widebartilde{{p}}\right]$. This results in the largest degree in the output to be defined as\footnote{The value of  $\widebar{p}=-1$ refers to the absence of large modes in the output.} $\widebar{p}:=\widebar{m}-1$.
\end{definition}

\begin{remark}
We employ affine tensor-product elements in this study. For general elements, one would rather define the 3D modal polynomials directly via an orthonormalization process such as the Gram-Schmidt and define related 3D operators accordingly. This difference aside, the derivations will be closely similar to the ones presented here. The details of the treatment of generalized elements is presented in \ref{app:mappings}.
\end{remark}

\begin{definition}
The Lagrange and Legendre polynomials on 3D TP elements are defined via Kronecker products of their 1D  counterparts as respectively  $\widebartilde{{\phi}}_l ^\mathcal{L}(\bm{x}) :=  \widebartilde{{\phi}}_a^\mathcal{L\text{1D}} (x) \, \widebartilde{{\phi}}_b^\mathcal{L\text{1D}} (y) \,  \widebartilde{{\phi}}_c ^\mathcal{L\text{1D}}(z)$ and $ \widebartilde{{\phi}}_l^{L} (\bm{x}) :=  \widebartilde{{\phi}}_a^{L\text{1D}} (x) \, \widebartilde{{\phi}}_b^{L\text{1D}} (y) \, \widebartilde{{\phi}}_c^{L\text{1D}} (z)$  with $l:=a(\widebartilde{{p}}+1)^2+b(\widebartilde{{p}}+1)+c$ and $\{a,b,c\}\in \left[0,..., \widebartilde{{p}}\right]$ .
\end{definition}

\begin{lemma}
\label{lem:3dmmdiag}
The 3D modal mass matrix, $\widebartilde{{M}}_{lm}^L:= \left< \widebartilde{{\phi}}_l^{L} , \widebartilde{{\phi}}_m^{L}\right> $, is diagonal.
\begin{proof}
$\left< \widebartilde{{\phi}}_l^{L} , \widebartilde{{\phi}}_m^{L}\right> = \delta_{lm}$ due to the orthonormality of the Legendre polynomials. 
\end{proof}
\end{lemma}

\begin{definition}
\label{def:3dC}
On 3D TP elements, the nodal-modal mapping and the low-pass filter operators are respectively defined via  Kronecker products of their 1D  counterparts as $\widebartilde{{C}}_{gh} := \widebartilde{{C}}_{ad}^\text{1D}\,\widebartilde{{C}}_{be}^\text{1D}\,\widebartilde{{C}}_{cf}^\text{1D}$  and $\widebar{\mathrm{F}}_{gh} := \widebar{\mathrm{F}}^\text{1D}_{ad} \,\widebar{\mathrm{F}}^\text{1D}_{be} \, \widebar{\mathrm{F}}^\text{1D}_{cf}$ with $g:=a(\widebartilde{{p}}+1)^2+b(\widebartilde{{p}}+1)+c$, $h:=d(\widebartilde{{p}}+1)^2+e(\widebartilde{{p}}+1)+f$  and $\{a,b,c,d,e,f\}\in \left[0,...,\widebartilde{{p}}\right]$ such that we have $\widebartilde{{\phi}}_{h}^\mathcal{L} = \widebartilde{{\phi}}_{g}^{L} \,\widebartilde{{C}}_{gh} $.
\end{definition}

\begin{lemma}
\label{lem:3dFdiag}
The 3D low-pass filter matrix is diagonal.
\begin{proof}
$\widebar{\mathrm{F}}_{gh} \equiv \widebar{\mathrm{F}}^\text{1D}_{ad} \,\widebar{\mathrm{F}}^\text{1D}_{be} \, \widebar{\mathrm{F}}^\text{1D}_{cf} \neq 0$ only if $g=h$, i.e. $a=d$, $b=e$ and $c=f$,  and $a<\widebar{m}$, $b<\widebar{m}$ and $c<\widebar{m}$ via Definition \ref{def:1dlowpassfil}.
\end{proof}
\end{lemma}

\begin{definition}
\label{def:high-passfilter}
The 3D high-pass filter operator is defined as
\begin{equation}
\label{eq:high-passfilter}
   \widetilde{\mathrm{F}}_{gh} :=   \delta_{gh} -  \widebar{\mathrm{F}}_{gh}.
\end{equation}
\end{definition}

Using the filter operator \eqref{eq:high-passfilter}, we can now isolate the 3D large and small resolved modes via the following definition:
\begin{definition}
\label{def:resolved}
The large and small resolved modes are respectively defined to be
\begin{gather*}
\widebar{\phi}_g^{L}  := \widebar{\mathrm{F}}_{gh} \, \widebartilde{{\phi}}_h^{L} \quad \text{and} \quad
\widetilde{\phi}_g^{L} := \widetilde{\mathrm{F}}_{gh} \, \widebartilde{{\phi}}_h^{L} .
\end{gather*}
\end{definition}
This definition is adopted from \cite{Wasberg2009} and translates to labelling as \textit{small resolved scales} all conjugate terms featuring at least one small 1D mode. As such, the large resolved scales contain  terms with strictly large 1D modes. We can now decompose $\forall \, \widebartilde{{\mathscr{S}}} \in {\rm I\!P}^{\widebartilde{{p}}}(\bm{\Omega}_{e_i})$ into its large and small modal contributions  as $\widebartilde{{\mathscr{S}}} = \widebar{{\mathscr{\mathscr{S}}}}+{\widetilde{\mathscr{S}}}$ where  $ \widebar{\mathscr{S}} =  \widebartilde{{\phi}}_l^{L} \,\widebar{\mathscr{S}}_l^L$ with $ \widebar{\mathscr{S}}^L_l =\, \left< \widebar{{\phi}}^{L}_l , \widebartilde{{\mathscr{S}}}\right> $ and   $ \widetilde{\mathscr{S}} = \widebartilde{{\phi}}_l^{L} \, \widetilde{\mathscr{S}}_l^L$ with $ \widetilde{\mathscr{S}}_l^L =\, \left< \widetilde{{\phi}}^{L}_l , \widebartilde{{\mathscr{S}}}\right> $.

\begin{lemma}
\label{lem:3dCinv}
The 3D basis mapping matrix is invertible.
\begin{proof}
Since $\widebartilde{{C}}_{gh}$ is defined as a Kronecker product, we can use the latter's property to obtain $\widebartilde{{C}}_{gh}^{-1} = \widebartilde{{C}}_{ad}^{-\text{1D}}\,\widebartilde{{C}}_{be}^{-\text{1D}}\,\widebartilde{{C}}_{cf}^{-\text{1D}}$ and the proof is completed since the 1D matrix is invertible as proved in Lemma \ref{lem:C1d}.
\end{proof}
\end{lemma}

\begin{lemma}
\label{lem:3dnodMinv}
The 3D nodal mass matrix, $\widebartilde{{M}}_{lm}^\mathcal{L}:= \left< \widebartilde{{\phi}}_l^\mathcal{L} , \widebartilde{{\phi}}_m^\mathcal{L} \right> $, is invertible.
\begin{proof}
$\widebartilde{{M}}_{lm}^{\mathcal{L}} = \widebartilde{{C}}_{lh}^\intercal \widebartilde{{M}}_{hg}^{{L}} \widebartilde{{C}}_{gm}$ via Definition \ref{def:3dC} and hence $\widebartilde{{M}}_{lm} ^{-\mathcal{L}}= \widebartilde{{C}}_{lh}^{-1} \delta_{hg} {\widebartilde{{C}}_{gm}^{-\intercal}} $ due to Lemmas \ref{lem:3dmmdiag} and \ref{lem:3dCinv}.
\end{proof}
\end{lemma}

\begin{definition}
\label{def:filtlag}
We define\footnote{The prime notation in ${ \widetilde{\phi}}^{\prime\mathcal{L}}_l $ is meant to signify that the filtered Lagrange polynomial is different than the small resolved scale Lagrange polynomial.} the filtered Lagrange polynomials as
\begin{equation}
{ \widetilde{\phi}}^{\prime\mathcal{L}}_l :=   \widebartilde{{\phi}}_m^\mathcal{L} \,\widebartilde{{C}}_{mg}^{-1} \,\widetilde{F}_{gh} \, \widebartilde{{C}}_{hl},
 \label{eq:filtrd_Lag}
 \end{equation}
 using  Lemma \ref{lem:3dCinv}.
\end{definition}

Before presenting the main result of this section, let us introduce a modal formulation of the  VMS-LES-FR/CPR equation \eqref{eq:LES2} by first noting the latter concisely as  $ \widebartilde{{\mathscr{S}}}+\widebartilde{{\mathscr{M}}}=0$ where $\widebartilde{{\mathscr{M}}}$ refers to the model terms and $\widebartilde{{\mathscr{S}}}$ to all the rest. We then project the equation onto the modal test space and apply a high-pass filter to the projected model terms to activate them only on the small resolved scales, thus obtaining:
 \begin{equation}
 \left<\widebartilde{{\phi}}_q^L \,,\,\widebartilde{{\mathscr{S}}} \right> + {\widetilde{F}}_{qm}\left<\widebartilde{{\phi}}_m^L \,,\, \widebartilde{{\mathscr{M}}} \right> \, \equiv  \,\left<\widebartilde{{\phi}}_q^L \,,\, \widebartilde{{\mathscr{S}}} \right> + \left<{{\widetilde{\phi}}}^L_q \,,\, \widebartilde{{\mathscr{M}}} \right> \,=\,0. 
 \label{eq:modalCPRVMS}
 \end{equation}
 
\begin{theorem}
\label{thrm:scheme}
The modal VMS-LES-FR/CPR formulation \eqref{eq:modalCPRVMS} where the model terms are projected onto a filtered modal space is equivalent to a nodal formulation where the model terms are interpolated using the filtered Lagrange polynomials. In other words, the system $\left<\widebartilde{{\phi}}_q^L \,,\, \widebartilde{{\mathscr{S}}} \right> + \left<{{\widetilde{\phi}}}^L_q \,,\, \widebartilde{{\mathscr{M}}} \right> \,=\,0$ is equivalent to the system $\widebartilde{{{\mathscr{S}}}}_q^\mathcal{L} \,+\, { \widetilde{\phi}}^{\prime\mathcal{L}}_h(\bm{x}_q)\, \widebartilde{{{\mathscr{M}}}}_h^\mathcal{L} = 0$.
\begin{proof}
Let us recall that since $ \{\widebartilde{{{\mathscr{S}}}},\widebartilde{{{\mathscr{M}}}}\} \in {\rm I\!P}^{\widebartilde{{p}}}(\bm{\Omega}_{e_i})$, hence $\widebartilde{{{\mathscr{S}}}} = I^{\widebartilde{{p}}} (\widebartilde{{\mathscr{S}}}) \equiv\widebartilde{{\phi}}_l^\mathcal{L} \,  \widebartilde{{\mathscr{S}}}_l^\mathcal{L}$ without any approximation and similarly for $\widebartilde{{{\mathscr{M}}}}$. We thus have
\begin{align}
\label{eq:theorem1}
\nonumber
&\left<\widebartilde{{\phi}}_q ^L\,,\, \widebartilde{{\mathscr{S}}} \right> + \left<{{\widetilde{\phi}}}_q ^L \,,\, \widebartilde{{\mathscr{M}}} \right>
\,  =  \, \left<\widebartilde{{\phi}}_q ^L \,,\, \widebartilde{{\phi}}_l^\mathcal{L} \right>   \widebartilde{{\mathscr{S}}}_l^\mathcal{L}\,+\, {\widetilde{F}}_{qm}\left<\widebartilde{{\phi}}_m^L \,,\, \widebartilde{{\phi}}_l^\mathcal{L} \right>  \widebartilde{{\mathscr{M}}}_l^\mathcal{L} \\
\,=& \left<\widebartilde{{\phi}}_q^L \,,\, \widebartilde{{\phi}}_l^\mathcal{L}  \right>  \widebartilde{{\mathscr{S}}}_l^\mathcal{L} \,+\, {\widetilde{F}}_{qm}\left<\widebartilde{{\phi}}_m^L \,,\, \widebartilde{{\phi}}_l^{L}\right>  \widebartilde{{C}}_{lh} \,\widebartilde{{\mathscr{M}}}_h^\mathcal{L}  = 0,
\end{align}
via Definitions \ref{def:resolved} and \ref{def:3dC}. We then multiply Eq. \eqref{eq:theorem1} by $\widebartilde{{C}}^\intercal_{oq}$, which is proved to be non-singular in Lemma \ref{lem:3dCinv}, to obtain
\begin{align*}
&\widebartilde{{C}}^\intercal_{oq}\left<\widebartilde{{\phi}}_q^L \,,\, \widebartilde{{\phi}}_l^\mathcal{L}  \right>  \widebartilde{{\mathscr{S}}}_l^\mathcal{L} \, \,+\,
\widebartilde{{C}}^\intercal_{oq}\,  \widetilde{F}_{qm}\left<\widebartilde{{\phi}}_m^L \,,\, \widebartilde{{\phi}}_l^{L} \right>\widebartilde{{C}}_{lh} \,  \widebartilde{{\mathscr{M}}}_h^\mathcal{L} \\
= \,&\widebartilde{{C}}^\intercal_{oq}\left<\widebartilde{{\phi}}_q^L \,,\, \widebartilde{{\phi}}_l^\mathcal{L} \right> \widebartilde{{\mathscr{S}}}_l^\mathcal{L} \, \,+\,
\widebartilde{{C}}^\intercal_{oq}\left<\widebartilde{{\phi}}_q^L \,,\, \widebartilde{{\phi}}_l^{L}  \right>  \widetilde{F}_{lg} \,\widebartilde{{C}}_{gh} \,  \widebartilde{{\mathscr{M}}}_h^\mathcal{L} \\
= \,&\left<\widebartilde{{\phi}}_o^\mathcal{L} \,,\, \widebartilde{{\phi}}_l^\mathcal{L}  \right> \,  \widebartilde{{\mathscr{S}}}_l^\mathcal{L} \, \,+\,
\left<\widebartilde{{\phi}}_o^\mathcal{L} \,,\, \widebartilde{{\phi}}_l^\mathcal{L} \right> \,\widebartilde{{C}}_{lm}^{-1}  \widetilde{F}_{mg} \,\widebartilde{{C}}_{gh} \,  \widebartilde{{\mathscr{M}}}_h^\mathcal{L} = 0,
\end{align*}
where the modal mass matrix and the filter are switched by exploiting the fact that they are diagonal (see Lemmas \ref{lem:3dmmdiag} and \ref{lem:3dFdiag} and Definition \ref{def:high-passfilter}). We now multiply by $\widebartilde{{M}}_{qo}^{-\mathcal{L}}$, the inverse of the nodal mass matrix, viz.
\begin{align*}
&\widebartilde{{M}}_{qo}^{-\mathcal{L}} \widebartilde{{M}}_{ol}^{\mathcal{L}} \, \widebartilde{{\mathscr{S}}}_l ^\mathcal{L}\,  + \widebartilde{{M}}_{qo}^{-\mathcal{L}} \widebartilde{{M}}_{ol}^{\mathcal{L}} \,   \widebartilde{{C}}_{lm}^{-1} \widetilde{F}_{mg}\,\widebartilde{{C}}_{gh} \,  \widebartilde{{\mathscr{M}}}_h^\mathcal{L} \, \\
=\,&\widebartilde{{\phi}}_l^{\mathcal{L}}(\bm{x}_q)\,  \widebartilde{{\mathscr{S}}}_l ^\mathcal{L}\,  + \widebartilde{{\phi}}_l^{\mathcal{L}}(\bm{x}_q)\, \widebartilde{{C}}_{lm}^{-1} \widetilde{F}_{mg}\,\widebartilde{{C}}_{gh} \,  \widebartilde{{\mathscr{M}}}_h^\mathcal{L} \, \equiv\, \widebartilde{{\mathscr{S}}}_q^\mathcal{L} + { \widetilde{\phi}}^{\prime\mathcal{L}}_h(\bm{x}_q)\, \widebartilde{{\mathscr{M}}}_h^\mathcal{L} = 0,
 \end{align*}
by employing Lemma \ref{lem:3dnodMinv}, Definition \ref{def:filtlag} and the property $\widebartilde{{\phi}}_l^\mathcal{L}(\bm{x}_q)= \delta_{ql}$.
\end{proof}
\end{theorem}

Starting from Eq. \eqref{eq:LES2}, the result of Theorem \ref{thrm:scheme} allows us to express the nodal VMS-LES-FR/CPR formulation with three-level partitioning as
\begin{align}
\left( \frac{\partial \widebartilde{{Q}}_k}{\partial t} + \frac{\partial  \widebartilde{{F}}_{ik}^{\text{inv}}}{\partial x_i}  - \frac{\partial \widebartilde{{F}}_{ik}^{\text{vis}}}{\partial x_i} +  \widebartilde{{\mathscr{C}}}_k^{\text{inv}}  -  \widebartilde{{\mathscr{C}}}_k^{\text{vis}} \right)_q^\mathcal{L} -  { \widetilde{\phi}}_h^{\prime\mathcal{L}}(\bm{x}_q)\,  \left( \frac{\partial \widebartilde{{F}}_{ik}^{\text{mod}}}{\partial x_i} 
+  \widebartilde{{\mathscr{C}}}_k^{\text{mod}}\right)^\mathcal{L}_h =0,
\label{eq:LES3}
\end{align}
where the exponent $\mathcal{L}$ is reminded to refer to Lagrange coefficients whereas $q$ and $h$ indices are those of solution nodes.
Hence in comparison to Eq. \eqref{eq:LES2}, the additional aspect of this formulation   consists  simply of interpolating the SGS terms using the filtered Lagrange polynomials.

\begin{remark}
Theorem \ref{thrm:scheme} implies that the VMS-LES-FR/CPR approach is an analog of the filtered version of the VMS method proposed by Vreman \cite{Vreman2003} in which the VMS-LES equations are obtained by applying an explicit high-pass filter directly to the model term that appears in the equations. In our formulation, this filter is given by the high-pass modal filter operator of Definition \ref{def:high-passfilter}.
\end{remark}

\subsubsection{SGS modelling  with three-level partitioning}
\label{sec:SGS3level}
Now that the resolved scales are separated into large and small, we can go back to the SGS closure terms, i.e., Eqs. \eqref{eq:SGS_tau} and \eqref{eq:SGS_H}, and elect a combination of scales  to evaluate them. There are three classical approaches available in the literature \cite{Hughes2001a,Ramakrishnan2005,Sagaut2005} referring to the scales used in these equations to evaluate the dynamic eddy viscosity, $\widebartilde{{\rho}} \nu_t(\widebartilde{{\bm{\rho u}}}/\widebartilde{{\rho}})$ with $\bm{\rho u}:=[\rho u_1,\rho u_2, \rho u_3]^\intercal $, and the deviator of the deformation tensor, $A_{ik}(\widebartilde{{\bm{\rho u}}}/\widebartilde{{\rho}})$:
\begin{itemize}
\item
$\text{small-small:}\; {\widetilde{\rho}}\nu_t({\widetilde{\bm{\rho u}}}/{\widetilde{\rho}})$; $A_{ik}({\widetilde{\bm{\rho u}}}/{\widetilde{\rho}}),$
\item 
$\text{large-small:}\; \widebar{{\rho}} \nu_t(\widebar{{\bm{\rho u}}}/\widebar{{\rho}})$; $A_{ik}({\widetilde{\bm{\rho u}}}/{\widetilde{\rho}}),$
\item 
$
\text{large-large:} \; \widebar{{\rho}} \nu_t(\widebar{{\bm{\rho u}}}/\widebar{{\rho}})$; $A_{ik}(\widebar{{\bm{\rho u}}}/\widebar{{\rho}}).$
\end{itemize}
More recently, an all-all closure is proposed \cite{Chapelier2016a}  in the context of a high-order modal DG method:
\begin{itemize}
\item
$\text{all-all:}\; \widebartilde{{\rho}} \nu_t(\widebartilde{{\bm{\rho u}}}/\widebartilde{{\rho}})$; $A_{ik}(\widebartilde{{\bm{\rho u}}}/\widebartilde{{\rho}}),$
\end{itemize}
 which has the advantage of being computationally less expensive than the previously mentioned approaches since it does not need additional projections to separate the large and small state variables. It is furthermore shown in \cite{Chapelier2016a}, to produce very similar results to the small-small approach for a Taylor-Green vortex problem while being slightly less dissipative. We hence adopt the all-all approach. 
 
 Another aspect of the SGS modelling left untreated in Section  \ref{sec:eddyviscmod} is the values of Smagorinsky model parameters $\Delta$ and $C_s$ in  Eq. \eqref{eq:smag} that depend on the discretization context. We define the filter width based on a local estimation of the effective h/p grid resolution for element $e_i$ as $\Delta \equiv \Delta^{h/p}_{e_i} := ({{\Omega}_{e_i}/\widebartilde{{N}}_\text{DOFs})^{1/3}} $ where ${\Omega}_{e_i}$ is the volume of the element. Regarding the Smagorinsky coefficient, the classical value of $C_s=0.1$, recommended for freely evolving turbulence, is deemed convenient for the h/p resolutions of this study. 
 The reader is referred to \cite{Beck2016} for a discussion on the effects of $C_s$ variations.

\section{Results and discussion}
\label{sec:results}

\subsection{Taylor-Green vortex}

\begin{figure}[!hbt]
\centering
\includegraphics[trim = 1mm 1mm 1mm 1mm, clip,width=0.5\linewidth]{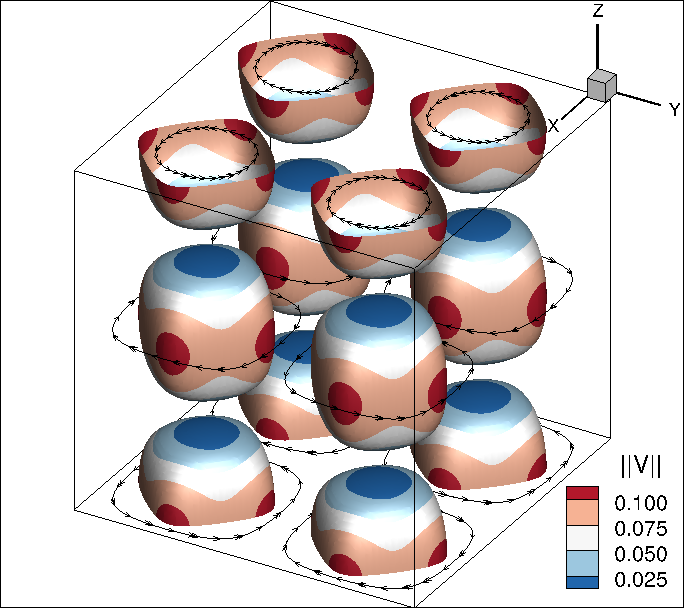}
\caption{Initial TGV fields;  Iso-surfaces of  Q-criterion intensity (Q=0.002), coloured by velocity magnitude. }
\label{fig:VMS_Qcriterioninit}
\end{figure}

The proposed methodology is assessed on the Taylor-Green vortex problem with an initial Reynolds number of 5000 (Fig \ref{fig:VMS_Qcriterioninit}).
The domain is defined as $\bm{\Omega}:=[-\pi L, \pi L]^3$ with triply-periodic boundary conditions and the initial fields of primitive variables are:
\begin{align}
u_1 (\bm{x}, 0) &= U_0 \sin (x_1/L) \cos (x_2/L) \cos (x_3/L),\\
u_2 (\bm{x}, 0) &= -U_0 \cos (x_1/L) \sin (x_2/L) \cos (x_3/L),\\
u_3(\bm{x}, 0) &= 0,\\
P (\bm{x}, 0) &= P_0 + \frac{\rho_0 U_0^2}{16} (\cos(2\,x_1/L)+\cos(2\,x_2/L)) (\cos(2\,x_3/L)+2) .
\end{align}
This case is originally devised as an incompressible one, prescribing $\rho(\bm{x},t) = \rho_0$. We however adopt the version proposed in \cite{Wokrshop3d-tgv} for compressible solvers with $\rho(\bm{x},0) = P(\bm{x},0)/(R \,T(\bm{x},0))$ where $T(\bm{x},0) = T_0:=P_0/(R\, \rho_0)$. We choose to work with dimensionless values of $L=1$, $\rho_0=1$, $U_0=1$, ${Ma}_0 := U_0/(\sqrt{\gamma\,P_0/\rho_0}) =0.1$ and $Re_0 := L\,\rho_0\,U_0/\mu_0 = 5000$ completing the definition of the initial TGV fields, noting as well that the characteristic advective time, $t_c:=L/U_0$ serves to non-dimensionalize the time variable.

We furthermore use the following dimensionless values for the parameters of the LES equations: $\gamma=1.4$, $R=1$, $C_p : = \gamma R/(\gamma-1)$, $C_v : = R/(\gamma-1)$, $Pr=0.71$ and $Pr_t=0.6$. The value of dynamic viscosity is fixed by $\mu:=\mu_0= 1/Re_0$. In fact, we tested the impact of enabling the compressible modelling of $\mu$ in terms of temperature via the Sutherland's law, Eq. \eqref{eq:Suth}, with $C_1=1.461 \times 10^{-6} \,(kg/(m\,s\,\sqrt{K}))$ and $T_S = 110.3 \,(K)$ to yield a dimensional viscosity, $\mu^D(T^D)$, in terms of a dimensional temperature, $T^D:=T/T_0 \,T^r$ where the reference temperature is $T^r= 300\, (K)$. This dimensionless viscosity is thus computed as $\mu := \mu_0 \,\mu^D/\mu^r$ where the reference viscosity is $\mu^r:= \mu^D(T^r)$. Nevertheless, enabling the Sutherland's model did not generate any significant difference for the TGV case considered here.

\subsection{Nomenclature of simulations}

The nomenclature employed to characterize each simulation discussed in this section has the structure shown by the following example: 
\begin{center}
\textbf{P}6\_\textbf{R}84\_\textbf{VMS}2\_\textbf{DP}8\_$\bm{\alpha}$0.05, 
\end{center}
where the components mean:
\begin{itemize}
\item [\textbf{P}6] provides ${\widebartilde{{p}}}$, the degree of the resolved space. For this example, the solution is $\widebartilde{{Q}} \in {\rm I\!P}^{6}(\bm{\Omega}_{e_i})$. Since we are using TP elements, this also means that the number of degrees of freedom of each element is $\widebartilde{{N}}_\text{DOFs} = (6+1)^3 = 343$.
\item [\textbf{R}84] indicates the h/p grid resolution 
 which is the number of total degrees of freedom in each spatial direction. This for instance entails 84 = 12 (elements) $\times$ 7 (P6 approximation).
\item [\textbf{VMS}2] refers to the number of  small scale modes, $\widetilde{m}:=\widebartilde{{p}}+1 - \widebar{m}$, (see Definition \ref{def:1dlowpassfil}) out of the total number of 1D resolved modes. This example has $\widetilde{m}=2$. Consequently, the cases \textbf{VMS}0 and \textbf{VMS}7 respectively correspond to  ILES and classical Smagorinsky LES.
\item [\textbf{DP}8] refers to $\dot{p}$ the degree of the enriched polynomial space used to perform de-aliasing as explained in \ref{app:De-alaising}. In this example, the enriched residuals are computed in $ {\rm I\!P}^{8}$ and \textbf{DP}6 would mean that no-dealiasing is applied (baseline collocated scheme).
\item [$\bm{\alpha}$0.05] is the value of $\alpha$, the reduction coefficient of the upwind dissipation of the Roe flux; see Eq. \eqref{eq:alpharoe}. $\alpha=0.05$ is prescribed in this example.
\end{itemize}

\subsection{Reference DNS data}
The reference DNS  solution for this case, requires $1280^3$ degrees of freedom per equation and is kindly provided by the authors of \cite{Dairay2017}. The filtered DNS data are obtained by applying two sharp low-pass filters in the Fourier domain with $(1280/16=80)^3$ and $(1280/8=160)^3$ degrees of freedom and respectively labelled as $\text{F}80$-DNS and $\text{F}160$-DNS.

\subsection{Analysis metrics}
In order to assess the simulation results, we will be looking at the following parameters: 
\begin{itemize}
\item The evolution of the volume-averaged kinetic energy, $E_K$, and that of the components of $-D_t(E_K)$, the total kinetic energy dissipation, which are: $ \varepsilon$, the resolved viscous dissipation, $\varepsilon_d$, the bulk viscosity dissipation, $\varepsilon_c$, the pressure dilatation dissipation and finally $\varepsilon_\text{SGS}$, the subgrid-scale dissipation. The derivation and the exact definition of these components is presented in \ref{app:disscomp}.
\item The spectrum of the kinetic energy at $t=14$. The approach employed to evaluate this metric is presented in \ref{app:spec}.
\item Iso-surfaces of Q-criterion with an intensity of $Q=0.01$ at $t=14$.
\item Maximum stable time-step.
\end{itemize}

\subsection{Effect of de-aliasing}

As mentioned in Section \ref{sec:proj},  the collocation projection at the basis of the FR/CPR scheme, although very appealing from a mere computational perspective, causes aliasing errors when the projected terms reside in a richer space than the solution, often translating to considerable degradation of physical fidelity and numerical stability. Aliasing refers to the spurious injection of energy from higher-order unresolved modes, excited due to non-linearities, into lower resolved modes. These errors are especially considerable on coarse high-order grids typical of LES and for predominantly advective (high Reynolds) flows. 

In \ref{app:De-alaising}, we present the polynomial de-aliasing technique adopted in this work which briefly consists of first sampling the residuals computed in an  enriched polynomial space of degree $\dot{p}$, to then obtain their contributions to the solution space modes by $L_2$ projection. The latter relies on a sufficiently accurate integration, tackled here by the collocated quadratures of the enriched space, to properly truncate spurious higher-wavenumber modes and prevent them from polluting the resolved space.

As pointed out by \cite{Beck2016}, for weakly compressible flows such as the ones considered here, $\dot{p} \approx 3/2\,p$ is sufficient to minimize aliasing errors. However this comes with a considerable cost that is estimated to grow at least by $(\dot{p}/p)^3$ when only the fluxes are projected \cite{Beck2016}. The impact of increasing the de-aliasing polynomial on the computational cost is presented in Table \ref{table:dealeff} which shows a faster growth in the actual cost compared to the ideal rate of $(\dot{p}/p)^3$. This is due to two factors: firstly, we de-aliase the whole residual  instead of (inviscid) fluxes only; secondly, there is a cost associated to the projection and recovery operations (\ref{app:De-alaising}) which supplements  the mere increase due to residual computation in  enriched space.

\begin{table}[!hbt]
\centering
     \rowcolors{2}{gray!25}{white}
  \begin{tabular}{lccc}
    \rowcolor{gray!50}
    \textbf{Case} & Total core-hours & Observed growth factor & Ideal growth factor\\
    \textbf{DP}6		&	  \phantom{0}346	&  \phantom{0}1.0		&	1.0	\\
    \textbf{DP}7   	&			  1328 	&  \phantom{0}3.8		&	1.6 \\    
    \textbf{DP}8	 	&			  2128 	&  \phantom{0}6.2		&	2.4\\
    \textbf{DP}9	 	&			  4043 	& 11.7				&	3.4\\ 
    \textbf{DP}10 	&			  7166 	& 21.0				&	4.6\\            
  \end{tabular}  
\caption{Effect of de-aliasing on  the computational cost of the cases \textbf{P}6\_\textbf{R}84\_\textbf{VMS}0\_\textbf{DP}{\color{blue}X}\_$\bm{\alpha}$1.0, computed on 192 cores.}
  \label{table:dealeff}
\end{table}

 \begin{figure}[!hbt]
 \centering
\subfloat[$E_K$]{
\includegraphics[trim = 3mm 5mm 0mm 0mm, clip,width=0.5\linewidth]{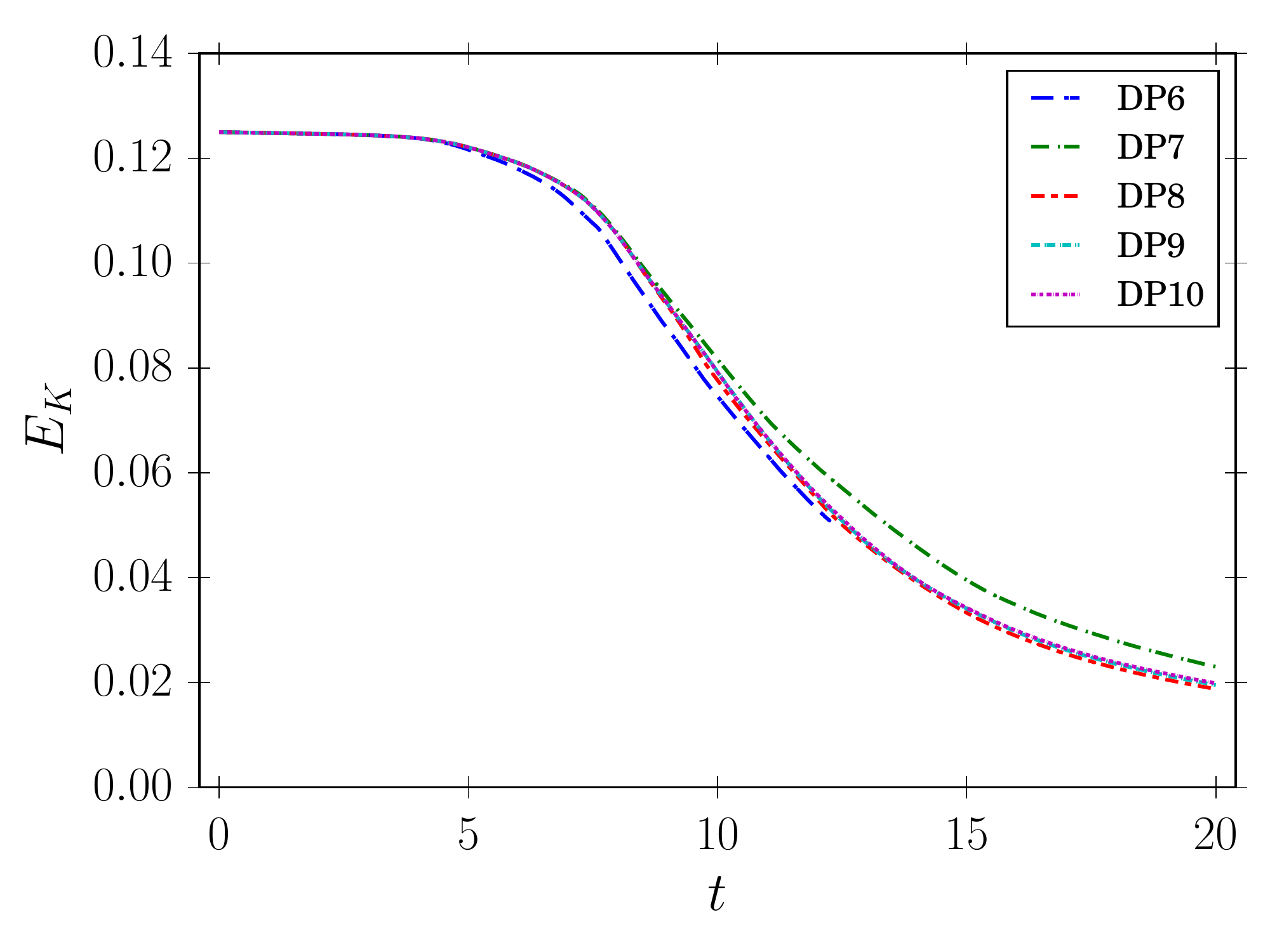}}~
\subfloat[$\varepsilon$]{
\includegraphics[trim = 3mm 5mm 0mm 0mm, clip,width=0.5\linewidth]{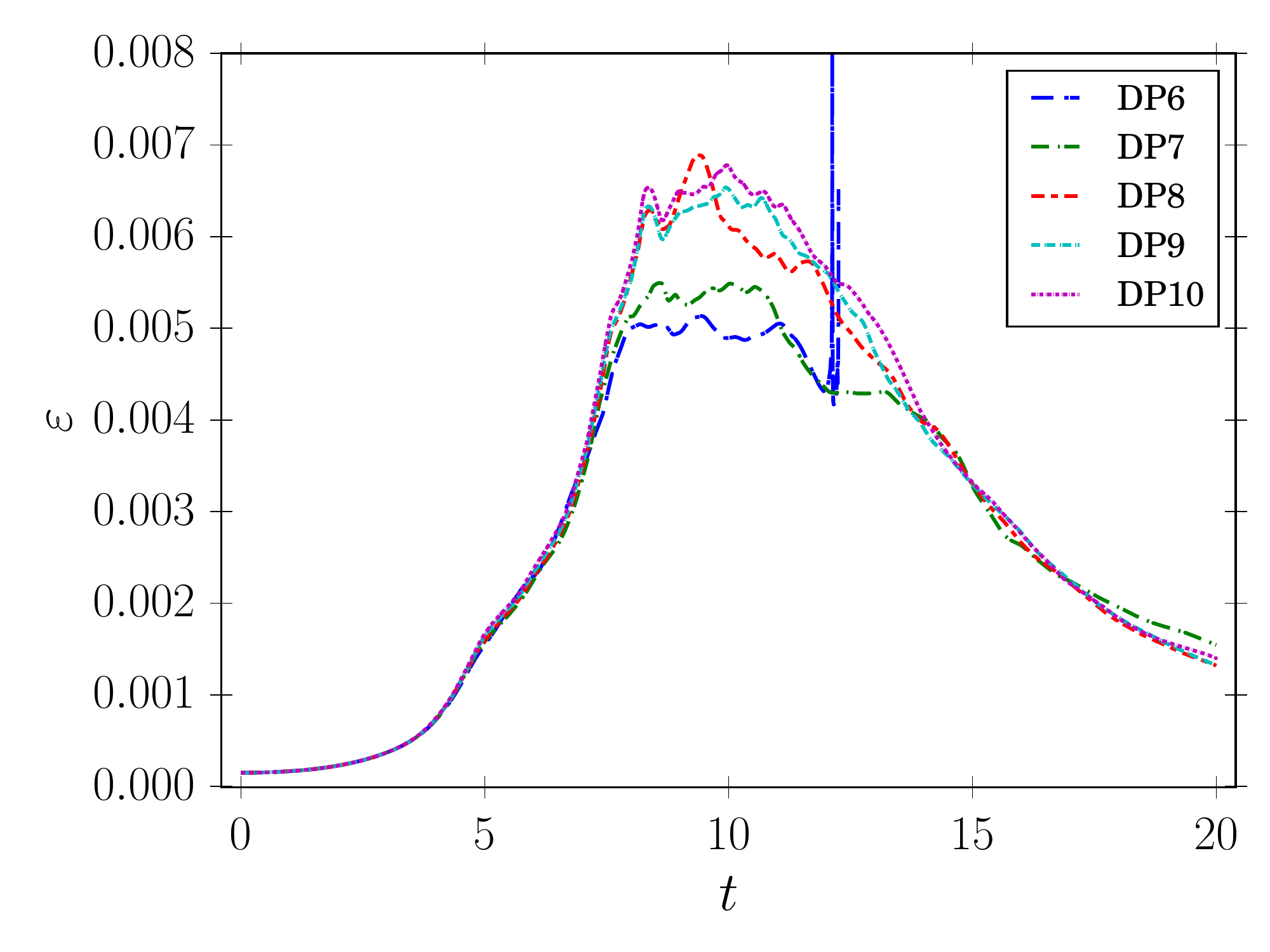}}
\caption{Effect of de-aliasing on  the cases \textbf{P}6\_\textbf{R}84\_\textbf{VMS}0\_\textbf{DP}{\color{blue}X}\_$\bm{\alpha}$1.0.}
\label{fig:de-alias_eps1}
\end{figure}
Figure \ref{fig:de-alias_eps1} shows the effect of different $\dot{p}$ values on the evolution of the kinetic energy and its dissipation for the TGV case.  We first observe that the case \textbf{DP}6, i.e. the baseline collocation projection with no de-aliasing, is unstable. Increasing $\dot{p}$ allows the simulation to be stable, yet $\dot{p}\geq8$ is needed to significantly reduce aliasing errors. Based on these results, we adopt \textbf{DP}8 for all \textbf{P}6 simulations, to achieve an efficient balance between de-aliasing and computational cost. As for \textbf{P}8 simulations, a de-aliasing corresponding to \textbf{DP}12 is  applied.
\subsection{Effect of VMS partitioning}
\begin{figure}[!hbt]
\centering
\subfloat[$E_K$]{
\includegraphics[trim = 1mm 1mm 1mm 1mm, clip,width=0.5\linewidth]{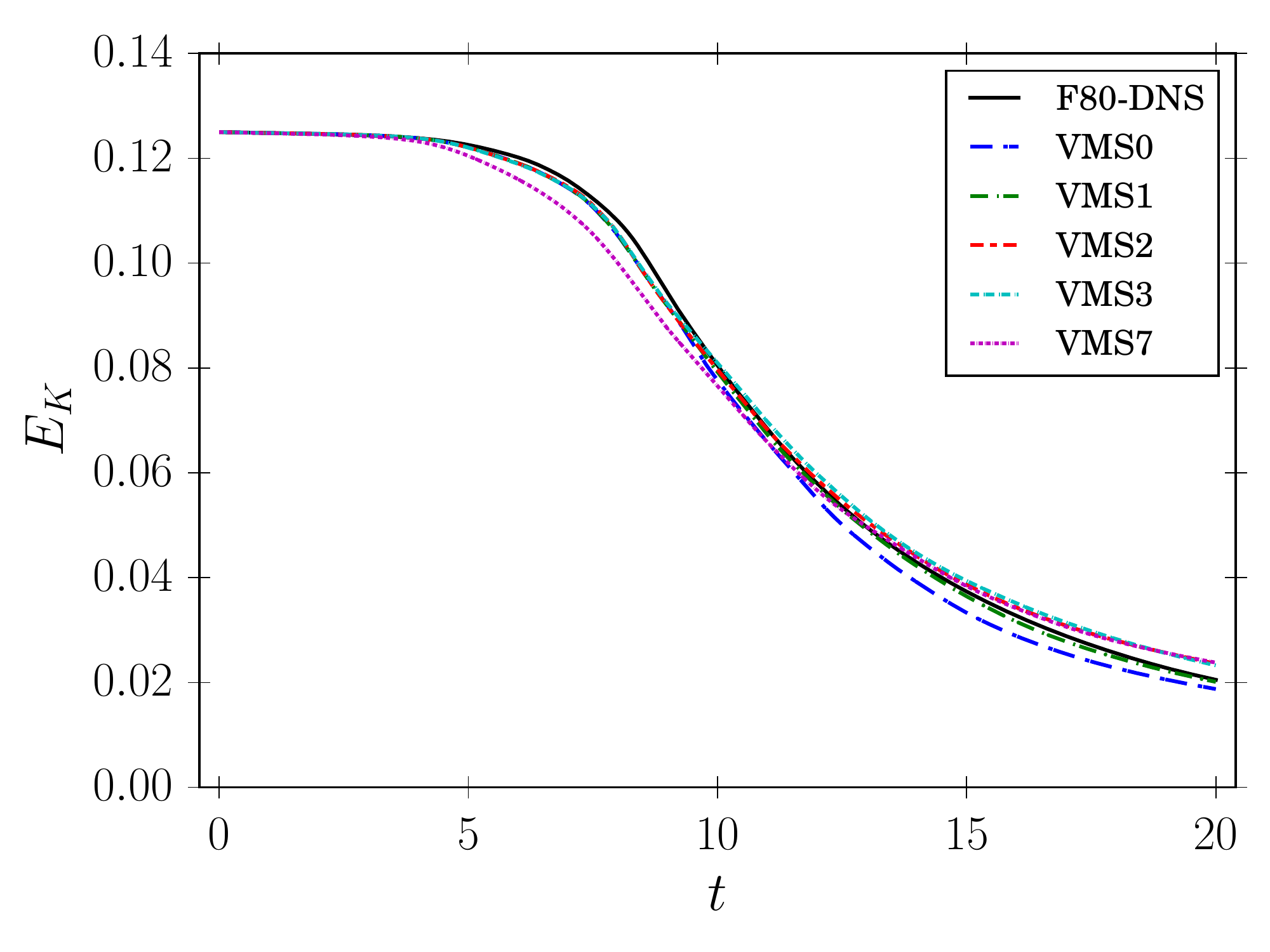}}\\
\subfloat[$\varepsilon$]{
\includegraphics[trim = 1mm 1mm 1mm 1mm, clip,width=0.5\linewidth]{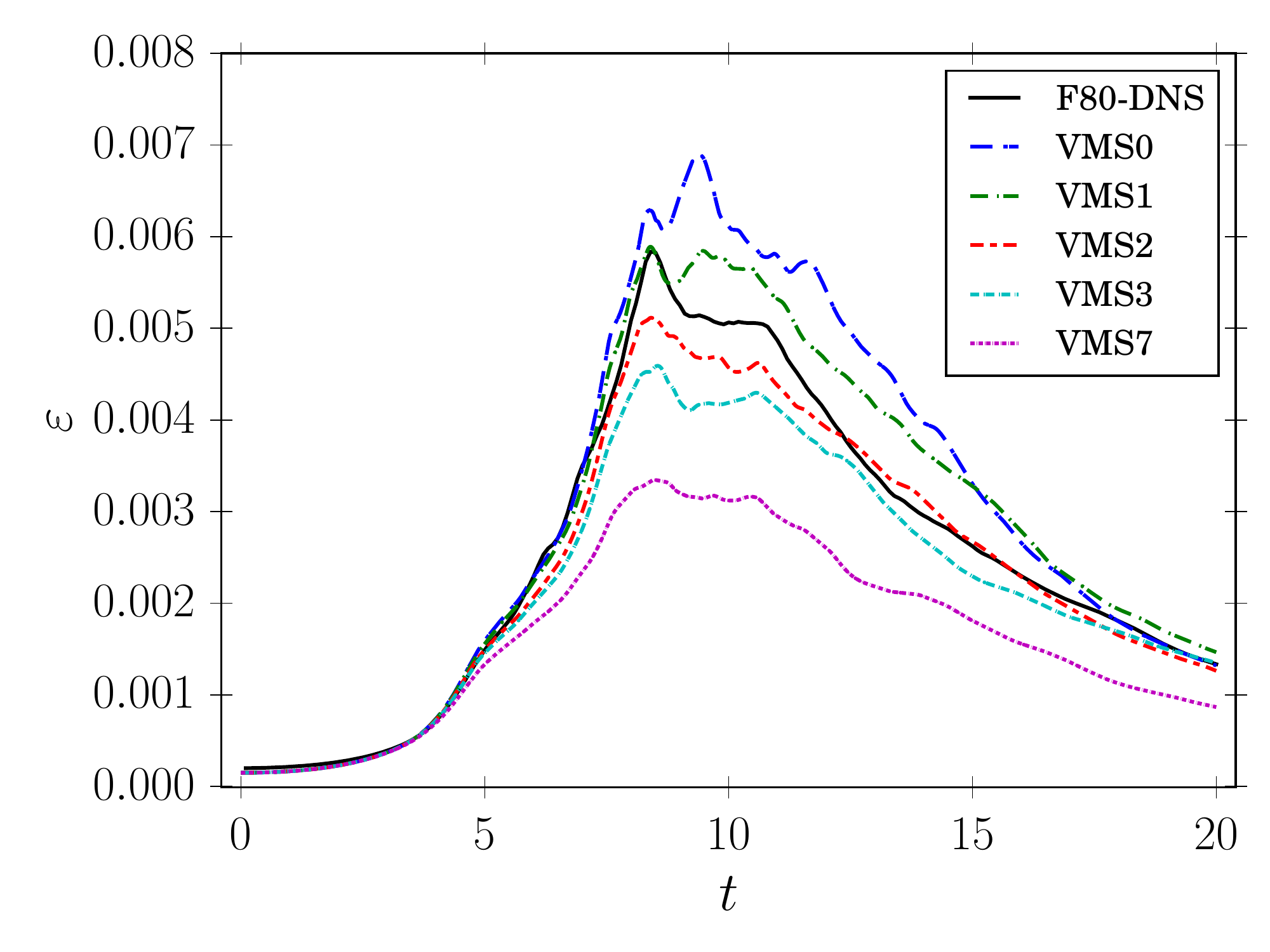}}~
\subfloat[$\varepsilon_\text{SGS}$]{
\includegraphics[trim = 1mm 1mm 1mm 1mm, clip,width=0.5\linewidth]{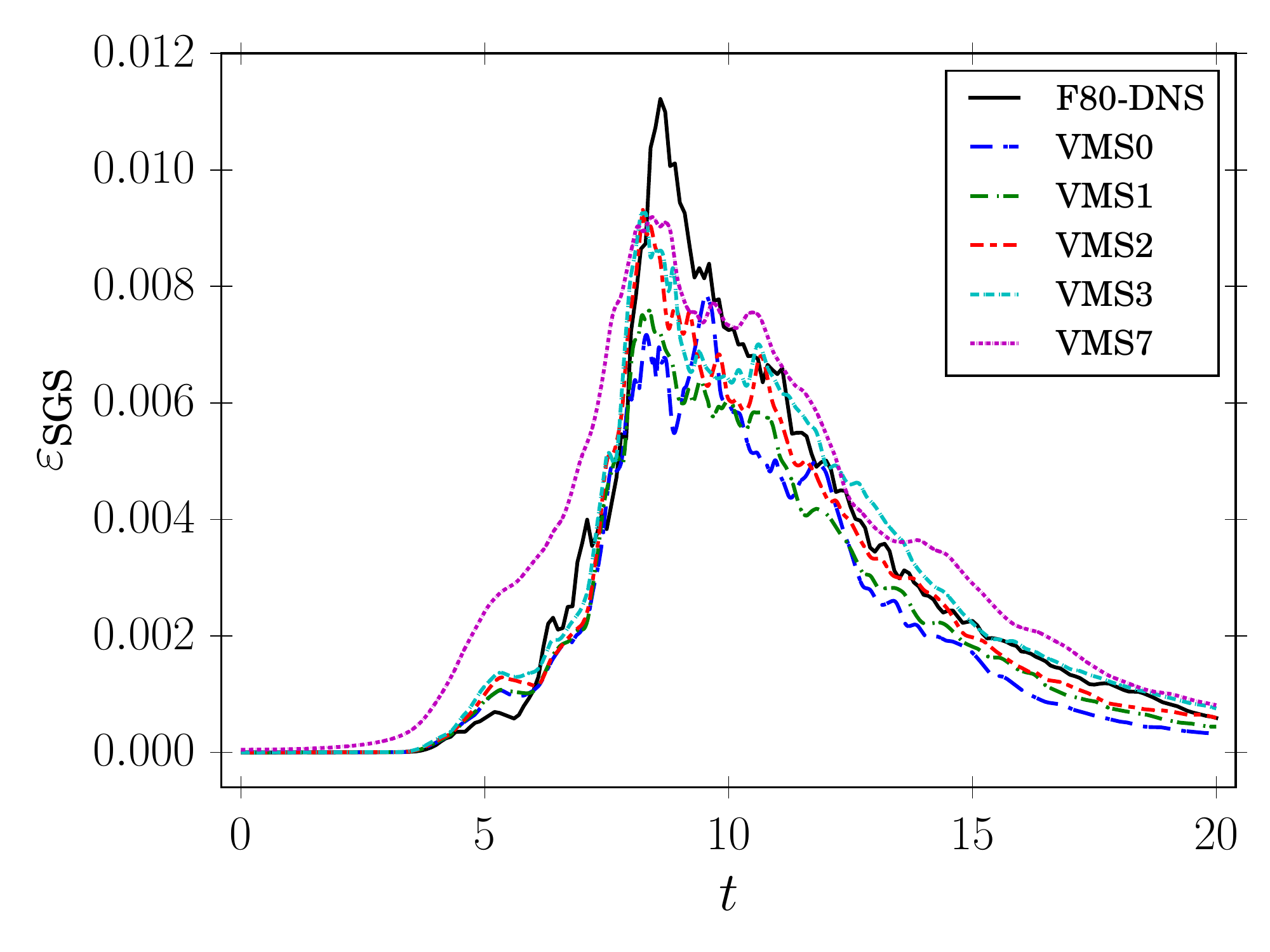}}
\caption{Effect of large/small VMS cutoff, $\widetilde{m}$, on the cases \textbf{P}6\_\textbf{R}84\_\textbf{VMS}{\color{blue}X}\_\textbf{DP}8\_$\bm{\alpha}$1.0; Kinetic energy dissipation components.}
\label{fig:VMS_effect}
\end{figure}

The motivation for the VMS method relies on a twofold observation from Figure \ref{fig:VMS_effect}: on one hand, the implicit LES (\textbf{VMS}0), relying on the inherent dissipation of the scheme, fails to produce sufficient subgrid-scale dissipation, $\varepsilon_\text{SGS}$, which results in the under-dissipation of the small eddies. This in turn causes an over-prediction of the viscous dissipation, $\varepsilon$, especially for $t>8.5$, i.e., past the peak of dissipation when the smallest eddies are formed \cite{Dairay2017}. On the other hand, the classical Smagorinsky LES (\textbf{VMS}7), producing an overly intense  $\varepsilon_\text{SGS}$, eliminates the small eddies precociously and thus disrupts the resolved scale dissipation mechanism, an effect reflected in an under-predicted $\varepsilon$ and an over-predicted kinetic energy past the peak. Via the application of the model dissipation to a select range of smallest resolved scales, the proposed VMS formulation compensates for the lack of dissipation of the no-model LES and thus enables one to  acquire a control on the combined (numerical and model) SGS dissipation, resulting in more faithful simulations.

\begin{figure}[!hbt]
\centering
\subfloat[\textbf{VMS}0 (ILES)]{
\includegraphics[trim = 1mm 1mm 1mm 1mm, clip,width=0.5\linewidth]{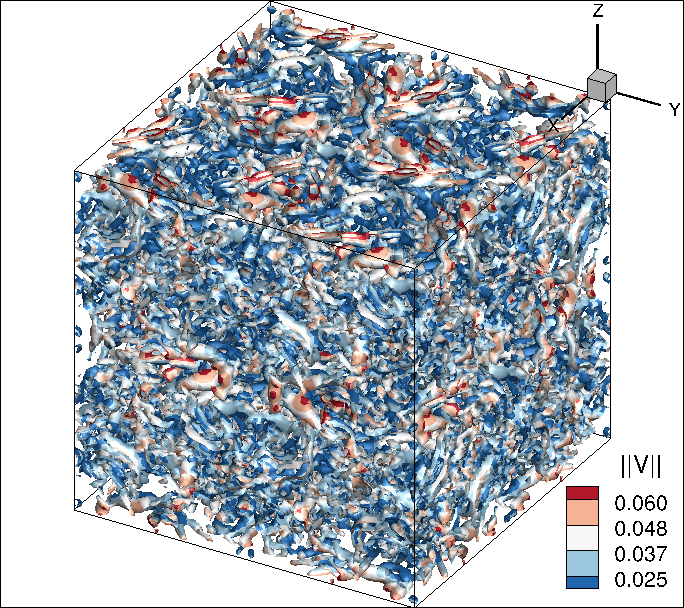}}~
\subfloat[\textbf{VMS}2]{
\includegraphics[trim = 1mm 1mm 1mm 1mm, clip,width=0.5\linewidth]{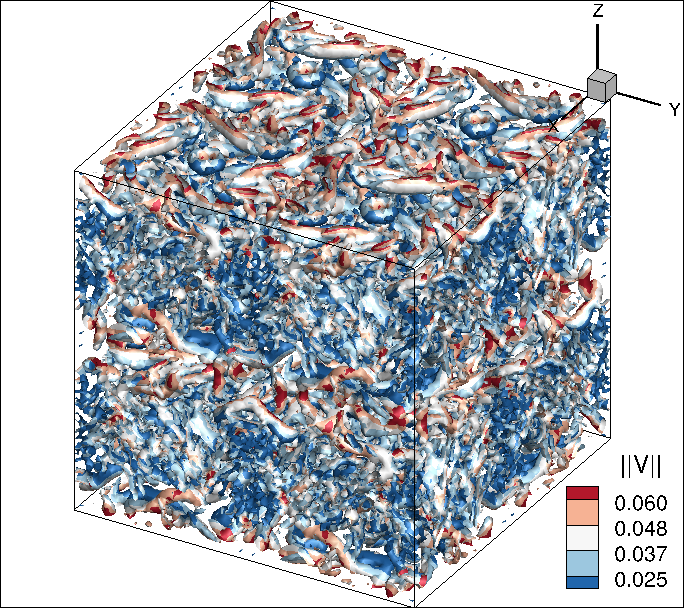}}\\
\subfloat[\textbf{VMS}7 (Smag-LES)]{
\includegraphics[trim = 1mm 1mm 1mm 1mm, clip,width=0.5\linewidth]{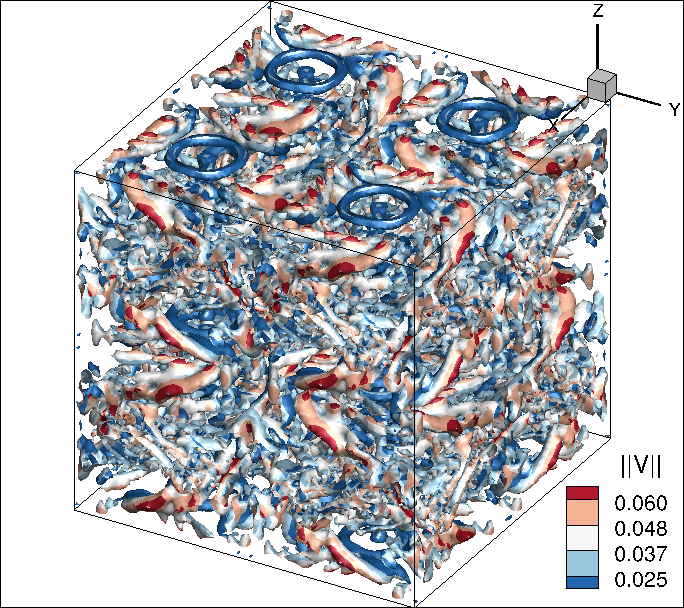}}
\caption{Effect of large/small VMS cutoff, $\widetilde{m}$, on the cases \textbf{P}6\_\textbf{R}84\_\textbf{VMS}{\color{blue}X}\_\textbf{DP}8\_$\bm{\alpha}$1.0; Iso-surfaces of Q-criterion intensity (Q=0.01), coloured by velocity magnitude at $t=14$. }
\label{fig:VMS_Qcriterion}
\end{figure}

\begin{figure}[!hbt]
\centering
\includegraphics[trim = 1mm 1mm 1mm 1mm, clip, width=0.6\linewidth]{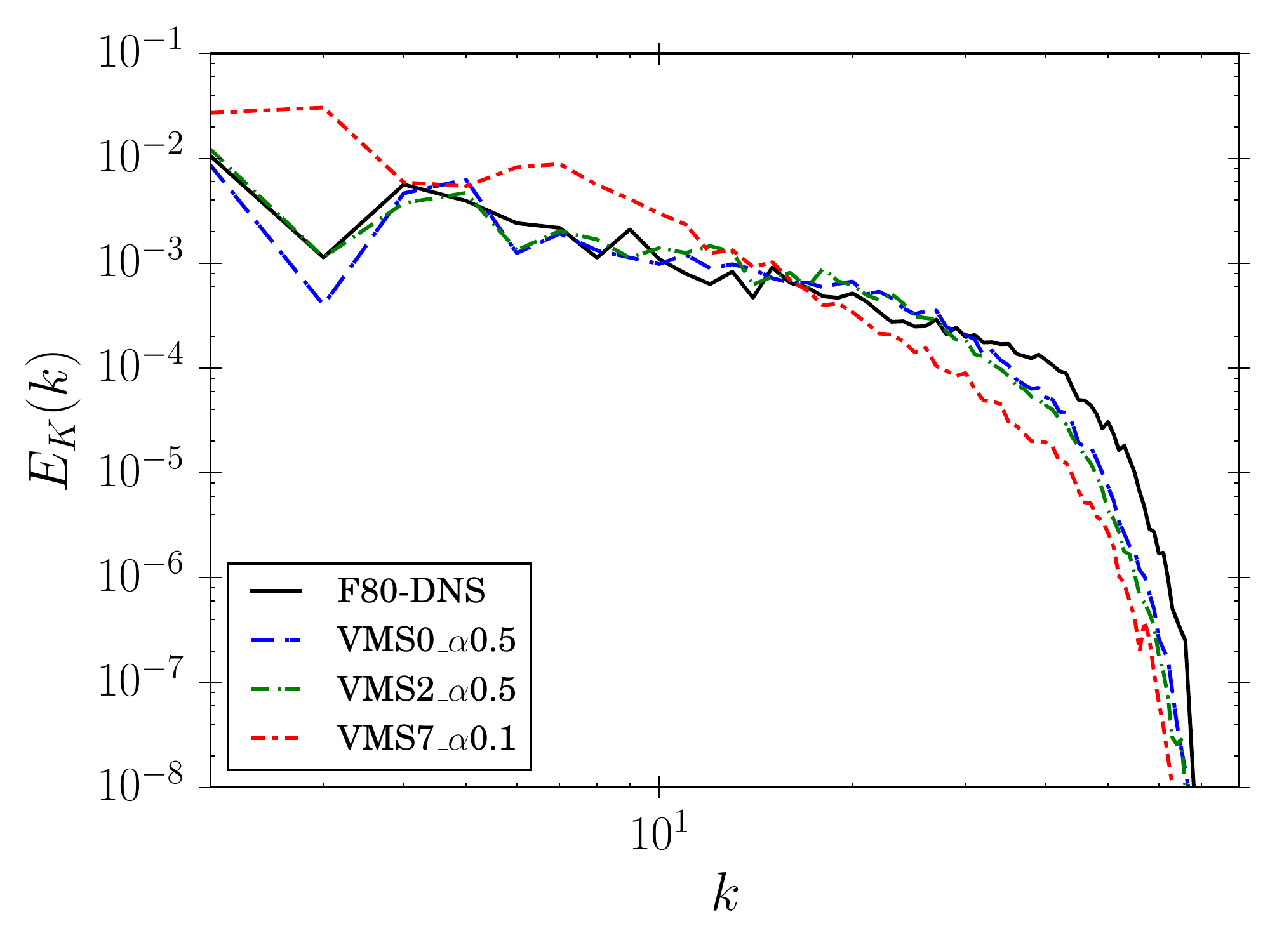}
\caption{Effect of large/small VMS cutoff, $\widetilde{m}$, on the cases \textbf{P}6\_\textbf{R}84\_\textbf{VMS}{\color{blue}X}\_\textbf{DP}8\_$\bm{\alpha}${\color{blue}X}; Kinetic energy spectrum at $t=14$. }
\label{fig:VMS_spectrum}
\end{figure}

The iso-surfaces of Q-criterion at $t =14$ in Figure \ref{fig:VMS_Qcriterion} show the presence of too many and respectively too little small eddies in the \textbf{VMS}0 and \textbf{VMS}7 simulations, responsible for the respectively over- and under-predicted $\varepsilon$ values past the dissipation peak. These effects are also reflected in the energy spectra of Figure \ref{fig:VMS_spectrum} which exhibits the exaggerated accumulation of energy in the largest scales of \textbf{VMS}7 while its smallest scales are over-dissipated. In the same figure, it is notable that the value of $\alpha$, the Roe's upwinding coefficient, is reduced to better emphasize the role of  the VMS large/small cutoff in adjusting the total dissipation. We discuss the impact of $\alpha$ at length in the next section.

\begin{table}[!hbt]
\centering
     \rowcolors{2}{gray!25}{white}
  \begin{tabular}{lc}
    \rowcolor{gray!50}
    \textbf{Case} & Total core-hours \\
     ILES (\textbf{VMS}0)	&	  2128	\\
    \textbf{VMS}1		&	  2531	\\
    \textbf{VMS}2		&	  2656	\\
    \textbf{VMS}3		&	  2617	\\
     Smag-LES (\textbf{VMS}7)	&	  1995	\\
  \end{tabular}
\caption{Effect of large/small VMS cutoff, $\widetilde{m}$, on the computational cost of the cases \textbf{P}6\_\textbf{R}84\_\textbf{VMS}{\color{blue}X}\_\textbf{DP}8\_$\bm{\alpha}$1.0, computed on 192 cores.}
\label{tab:vmscost}
\end{table}

The comparison of computational cost of the VMS method in Table \ref{tab:vmscost} to the ILES and classical LES (Smag) reveals $\approx25\%$ increase due to  interpolation via filtered Lagrange polynomials in the \textbf{VMS}1/2/3 cases which is absent from the ILES/Smag simulations. The Smag case should in general be slightly more expensive than ILES since an explicit SGS model is computed in the former. However, this difference is veiled by fluctuations in the performance of the cluster in terms of latency.

\subsection{Effect of $\alpha$-Roe}
In this section, we study the impact of upwind dissipation from the Roe's Riemann solver on VMS-LES-FR/CPR solutions. The effect of $\beta$, the upwinding parameter of the LLF numerical flux, on the eigen-footprint of a number of FR schemes  is discussed in \cite{Mengaldo2018} and it is shown in  \cite{Plata2017} that lowering the LLF upwinding for low-Mach flows improves the VMS solutions in a modal DG context. We consider the role of $\alpha$, the upwind coefficient of Roe flux of Eq. \eqref{eq:alpharoe}, in improving the simulation of low-Mach flows.

\begin{figure}[!hbt]
\centering
\subfloat[\textbf{VMS}1]{
\includegraphics[trim = 1mm 1mm 1mm 1mm, clip,width=0.5\linewidth]{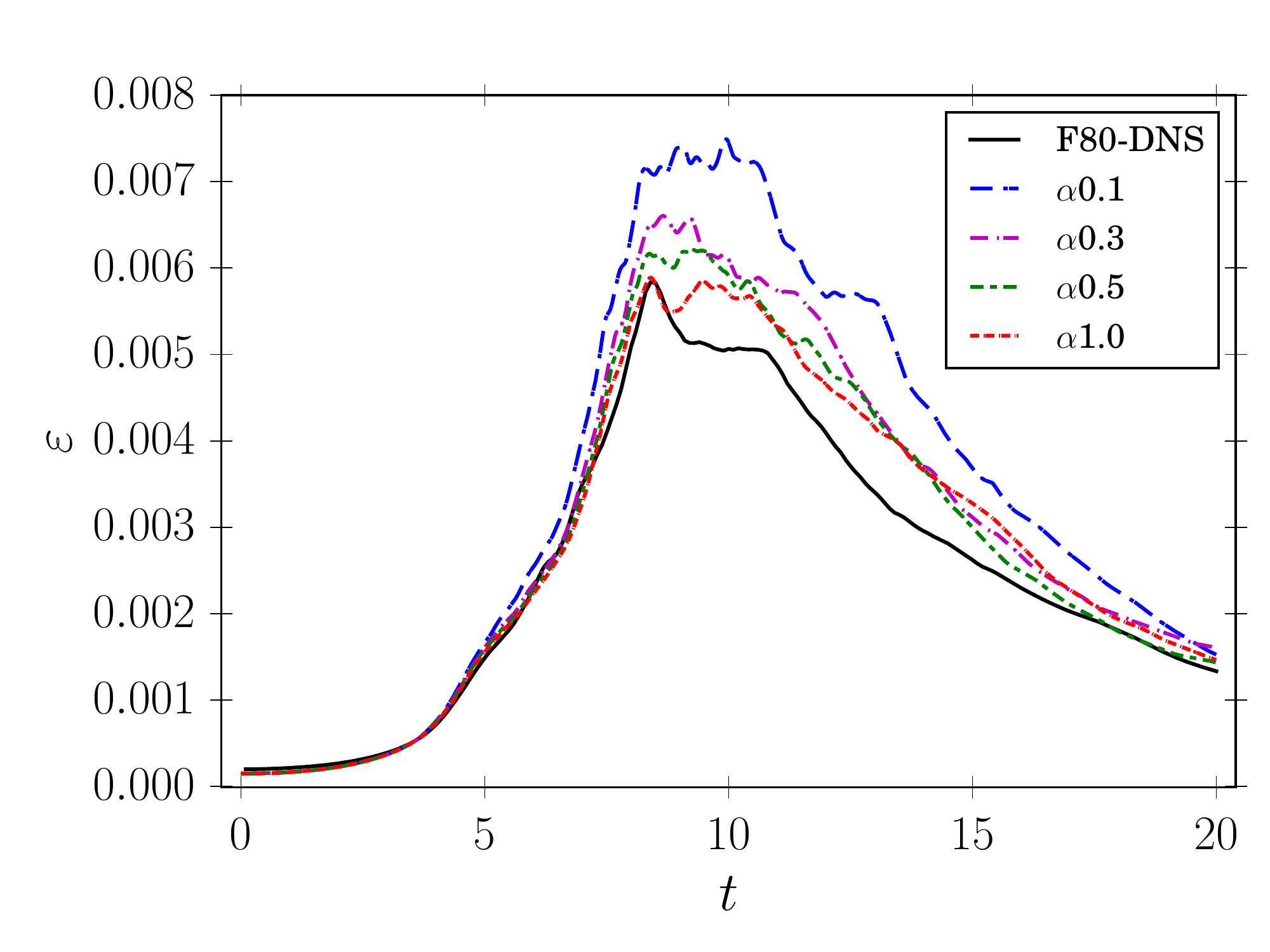}}~
\subfloat[\textbf{VMS}2]{
\includegraphics[trim = 1mm 1mm 1mm 1mm, clip,width=0.5\linewidth]{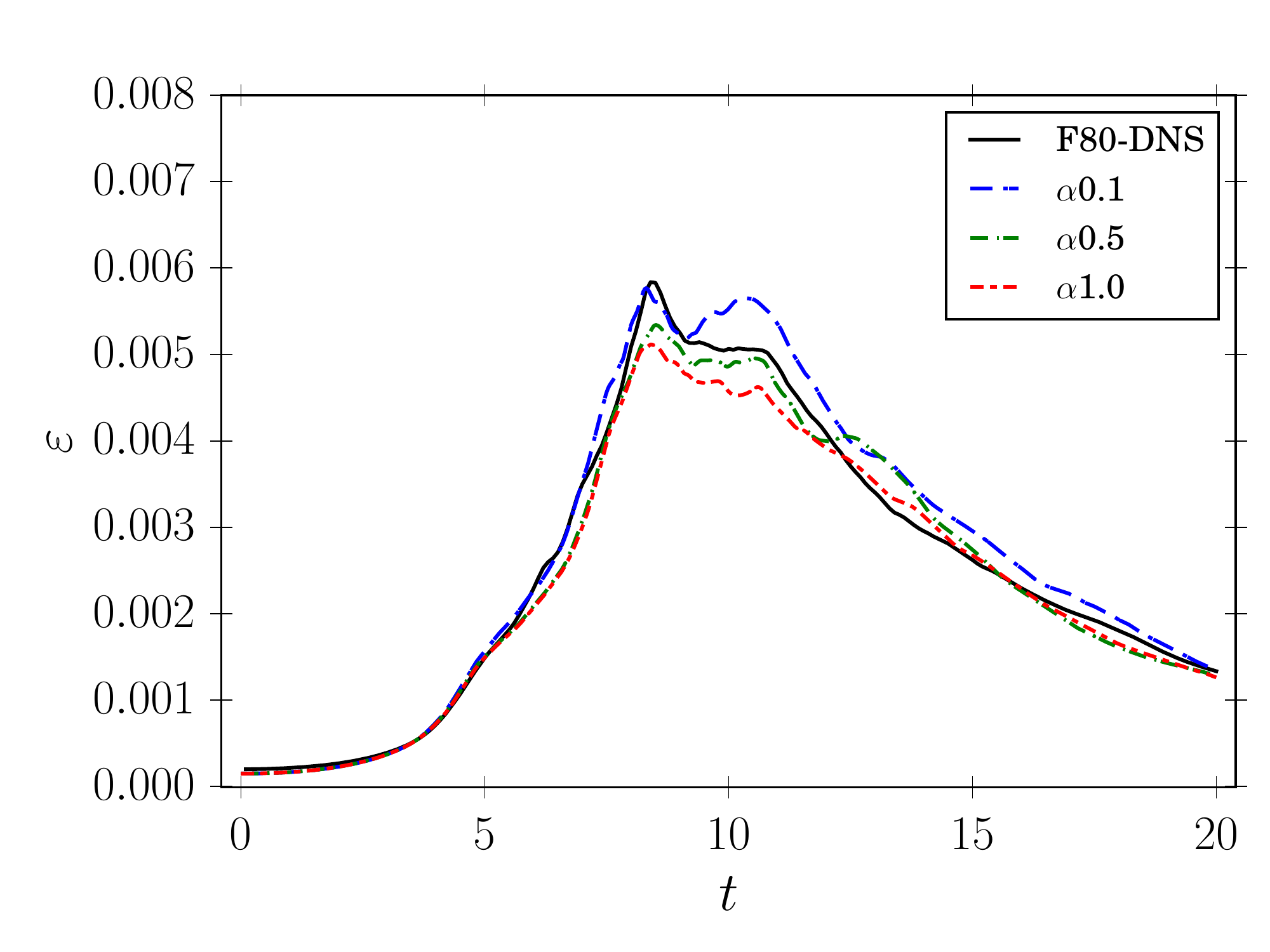}}\\
\subfloat[\textbf{VMS}3]{
\includegraphics[trim = 1mm 1mm 1mm 0mm, clip,width=0.5\linewidth]{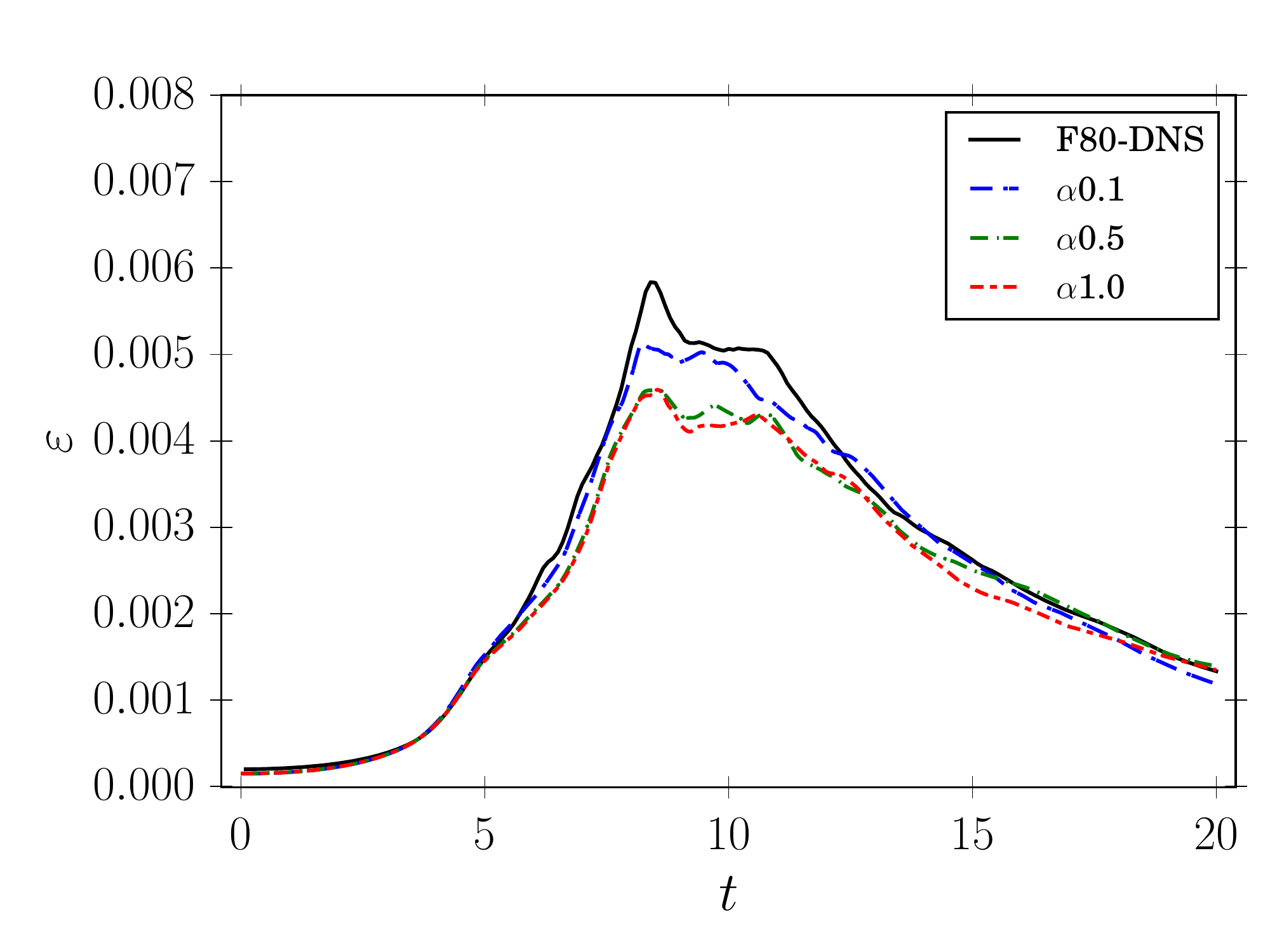}}
\caption{Effect of $\alpha$-Roe on the cases \textbf{P}6\_\textbf{R}84\_\textbf{VMS}{\color{blue}X}\_\textbf{DP}8\_$\bm{\alpha}${\color{blue}X}; Resolved viscous dissipation, $\varepsilon$. }
\label{fig:alphaRoe_eps1}
\end{figure}

Figure \ref{fig:alphaRoe_eps1}, shows the effect of $\alpha$ on resolved viscous dissipation for three different VMS configurations. It can be deduced that lowering $\alpha$ below the default value of $1.0$ increases $\varepsilon$ in general; but whether reducing $\alpha$ is recommendable and the suitable value of $\alpha$ seem to depend on the context in terms of the baseline dissipation produced by the VMS model. For instance, in the \textbf{VMS}1 case, the dissipation produced by $\bm{\alpha}$1.0 is larger than the one from the filtered DNS result. Hence lowering $\alpha$ is counter-productive in this case. Conversely, the \textbf{VMS}3 along with $\bm{\alpha}$1.0 is over-dissipative and diminishing $\alpha$ to $\bm{\alpha}$0.1 is beneficial in that it improves the agreement with the filtered DNS result. The \textbf{VMS}2 configuration is an intermediate one and consistently, the moderate value of $\bm{\alpha}$0.5 produces a satisfactory match although the difference with other considered values is not very significant.

\begin{figure}[!hbt]
\centering
\subfloat[$\varepsilon_\text{SGS}$]{
\includegraphics[trim = 1mm 1mm 1mm 1mm, clip,width=0.5\linewidth]{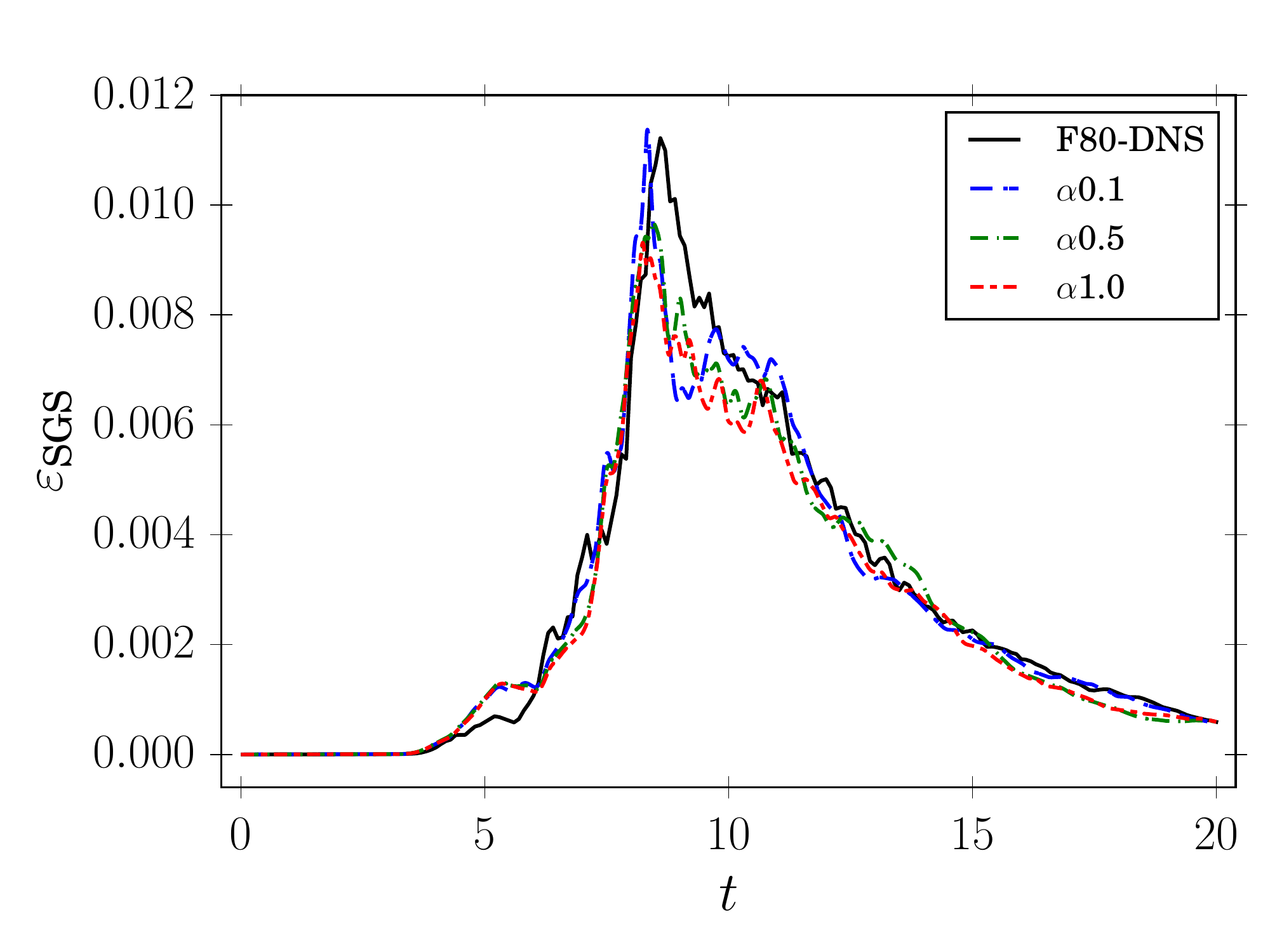}}\\
\subfloat[$\varepsilon_d$]{
\includegraphics[trim = 1mm 1mm 1mm 1mm, clip,width=0.5\linewidth]{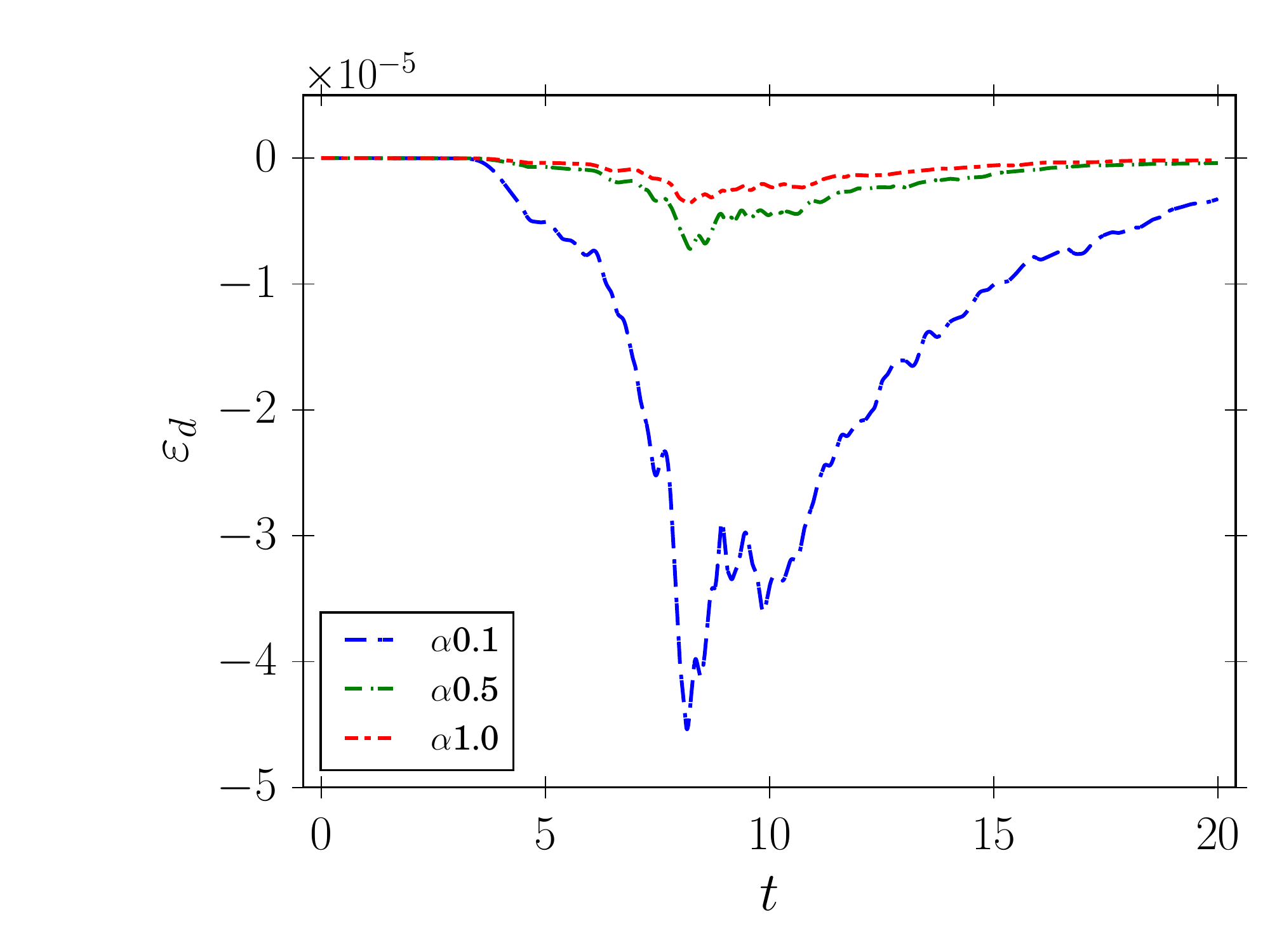}}~
\subfloat[$\varepsilon_c$]{
\includegraphics[trim = 1mm 1mm 1mm 1mm, clip,width=0.5\linewidth]{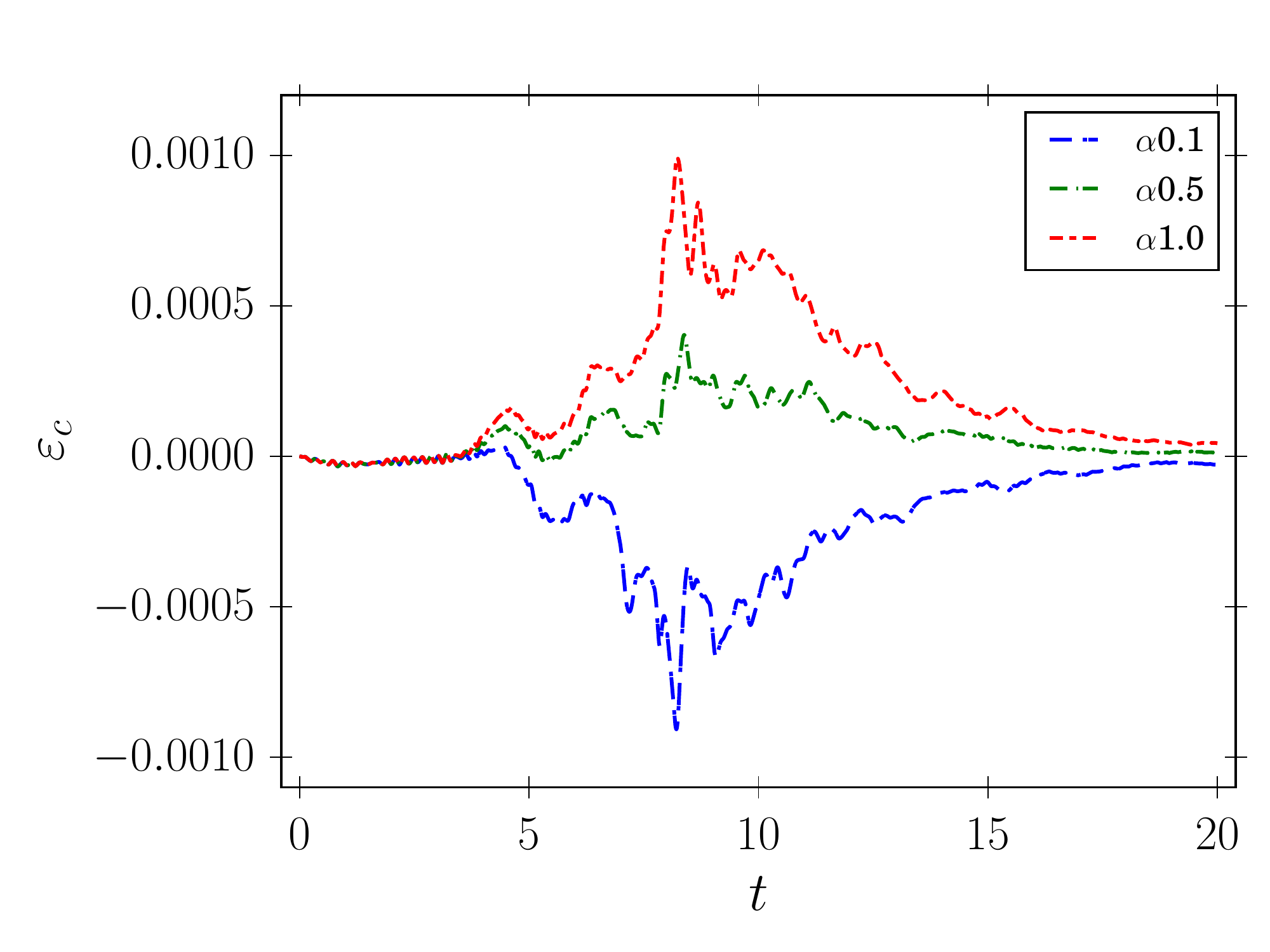}}
\caption{Effect of $\alpha$-Roe on the cases \textbf{P}6\_\textbf{R}84\_\textbf{VMS}2\_\textbf{DP}8\_$\bm{\alpha}${\color{blue}X}; Kinetic energy dissipation components.}
\label{fig:alphaRoe_2}
\end{figure}

It is arduous to identify clear trends with regards to the effect of $\alpha$ on SGS dissipation from Figure \ref{fig:alphaRoe_2}, comparatively to the rather obvious tendencies with regards to the effect of  $\alpha$  on the resolved viscous dissipation extracted from Figure \ref{fig:alphaRoe_eps1}. It hence seems as if $\alpha$ has a more significant impact on the resolved dissipation than on the SGS dissipation.

The effect of $\alpha$ on the bulk viscosity dissipation, $\varepsilon_d$, and on the pressure-dilatation dissipation, $\varepsilon_c$ are rather clearly depicted in Figs. \ref{fig:alphaRoe_2} (b) and \ref{fig:alphaRoe_2} (c) respectively. Both components diminish with  decreasing $\alpha$. The negative values of  $\varepsilon_c$ occurring for very low $\alpha$ act as a source of energy and although this  is strictly a numerical artifact, it interestingly contributes to adjusting the values of resolved viscous dissipation (Figure \ref{fig:alphaRoe_eps1}),  especially around the dissipation peak ($t\approx 8.5$). Negative $\varepsilon_c$ values are as well reported in \cite{Bull2015} for the FR-DG scheme. Finally, we note that further investigation is incumbent in order to acquire a more profound comprehension of the underlying mechanisms triggered by reducing $\alpha$ and its effects on VMS-LES simulations.

We can also verify from Figure \ref{fig:alphaRoe_2} that $|\varepsilon_d| \ll |\varepsilon_c|$ as expected; the former is negligible and the rather small values of the latter reflect the small compressibility of the low-Mach number flow considered here.

\begin{figure}[!hbt]
\centering
\subfloat[\textbf{VMS}0]{
\includegraphics[trim = 1mm 1mm 1mm 1mm, clip,width=0.5\linewidth]{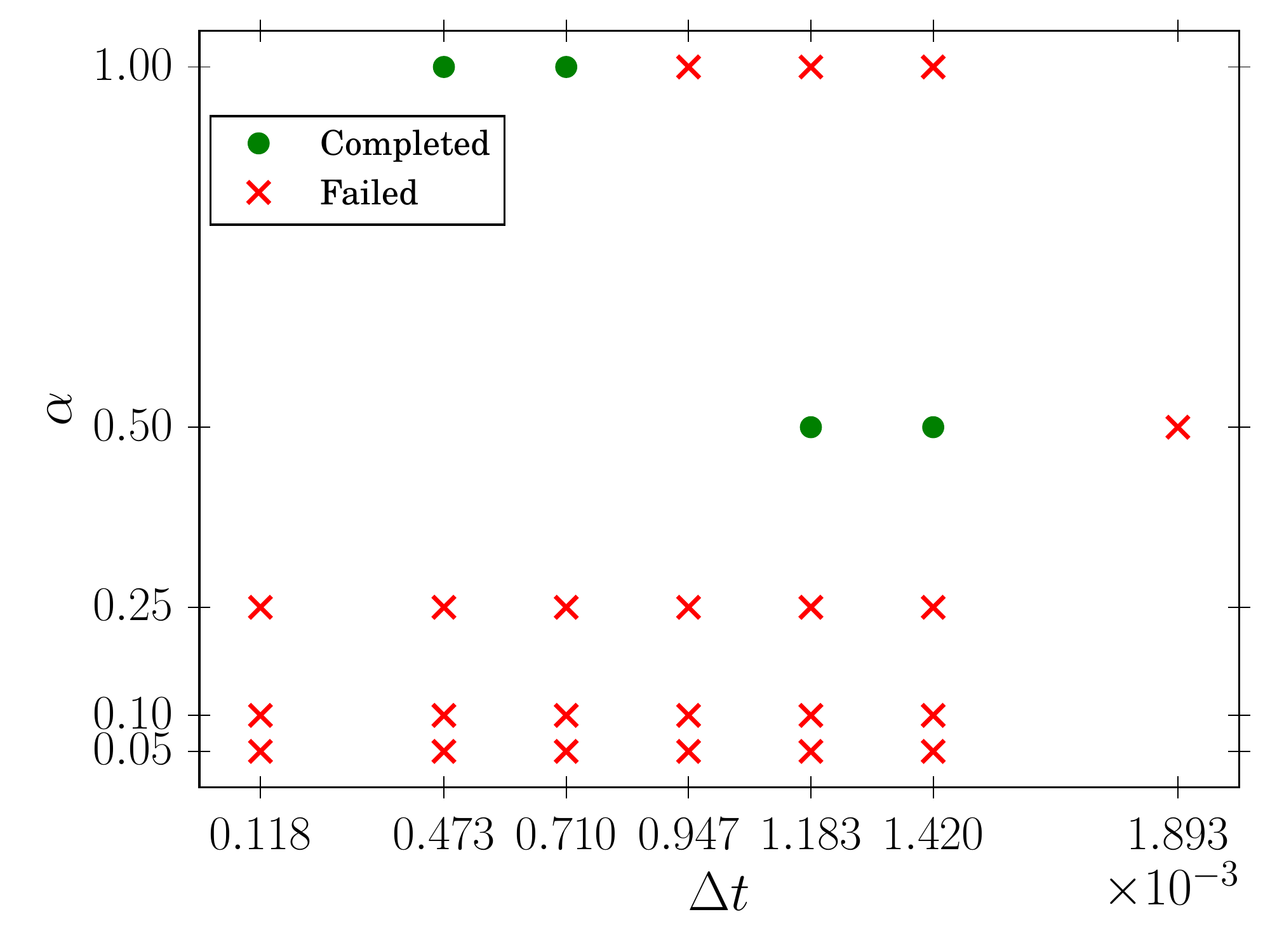}}~
\subfloat[\textbf{VMS}1]{
\includegraphics[trim = 1mm 1mm 1mm 1mm, clip,width=0.5\linewidth]{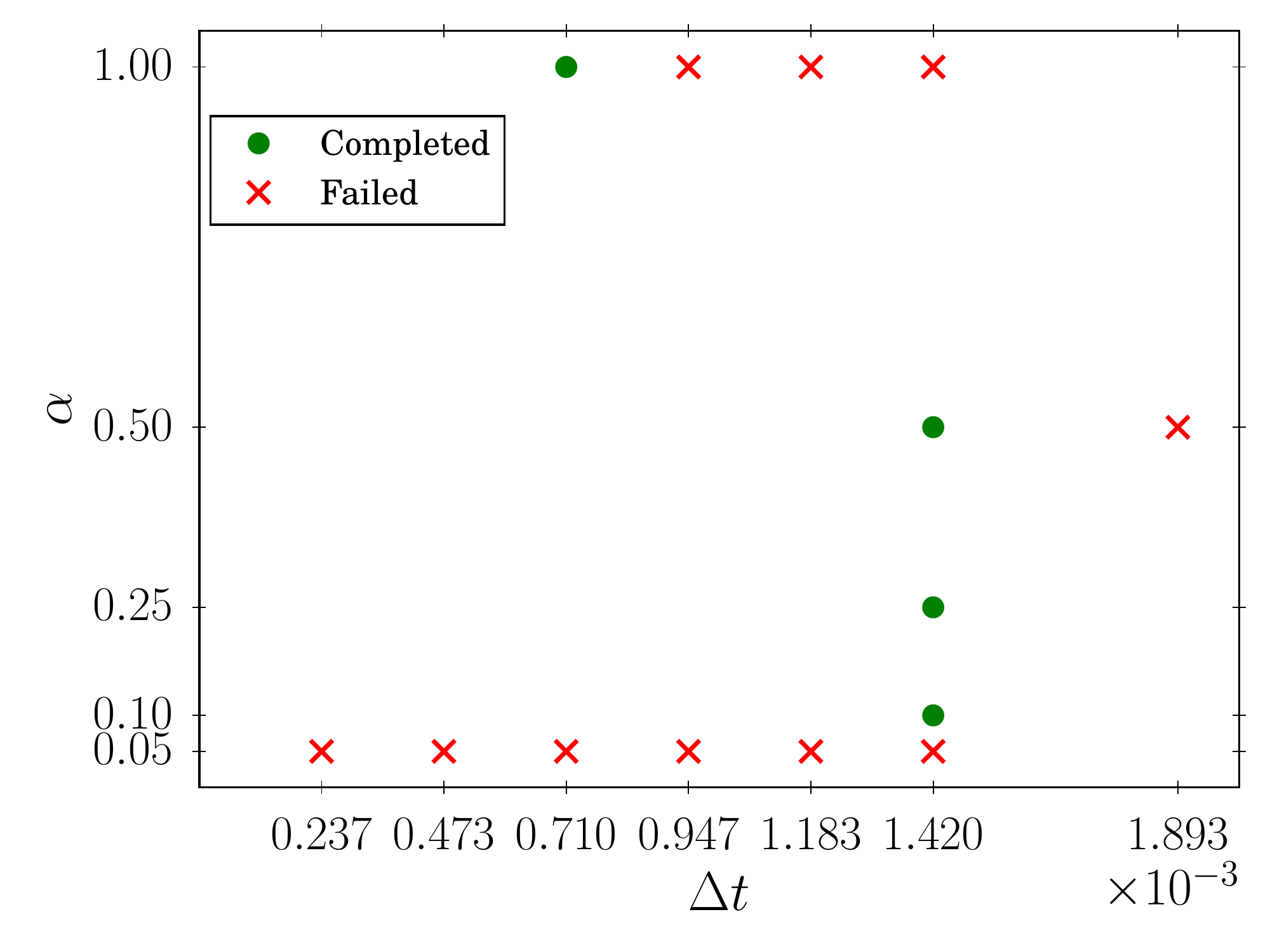}}
\caption{Effect of $\alpha$ on the cases \textbf{P}6\_\textbf{R}84\_\textbf{VMS}{\color{blue}X}\_\textbf{DP}8\_$\bm{\alpha}${\color{blue}X}; The maximum stable time step. }
\label{fig:alpha_dt}
\end{figure}

We have as well studied the effect of the  parameter $\alpha$ on the maximum stable time step. The results in Figure \ref{fig:alpha_dt} show that larger time steps are achieved for moderate values of $\alpha$. For lower values of $\alpha$, the maximum time step can be significantly increased if the model dissipation is applied to a larger number of small scales (e.g. \textbf{VMS}1 instead of \textbf{VMS}0). Indeed, for the case \textbf{P}6\_\textbf{R}84\_\textbf{VMS}2\_\textbf{DP}8\_$\bm{\alpha}$0.05, absent from the figure, the stable time step reached the value of $1.42 \times 10^{-3}$ whereas the case of \textbf{VMS}1 was found to be unstable even with a time step as small as $0.237 \times 10^{-3}$. 

Let us remark that we ensured that the physical time step necessary to capture the physics of the LES problem is larger than the maximum realizable time step dictated by numerical stability constraints. This is achieved by lowering the time step by a factor of 5 and verifying the absence of noticeable differences in the results.

Another parameter of the Roe's approximate Riemann solver is the Harten's entropy correction \cite[4.3.3]{blazek-2001}, meant to prevent the detection of unphysical expansion shocks. We verified that this correction is not activated  for the considered flows.

\subsection{Effect of polynomial discretization degree}

 \begin{figure}[!hbt]
\centering
\subfloat[\textbf{VMS}0]{
\includegraphics[trim = 1mm 1mm 1mm 1mm, clip,width=0.5\linewidth]{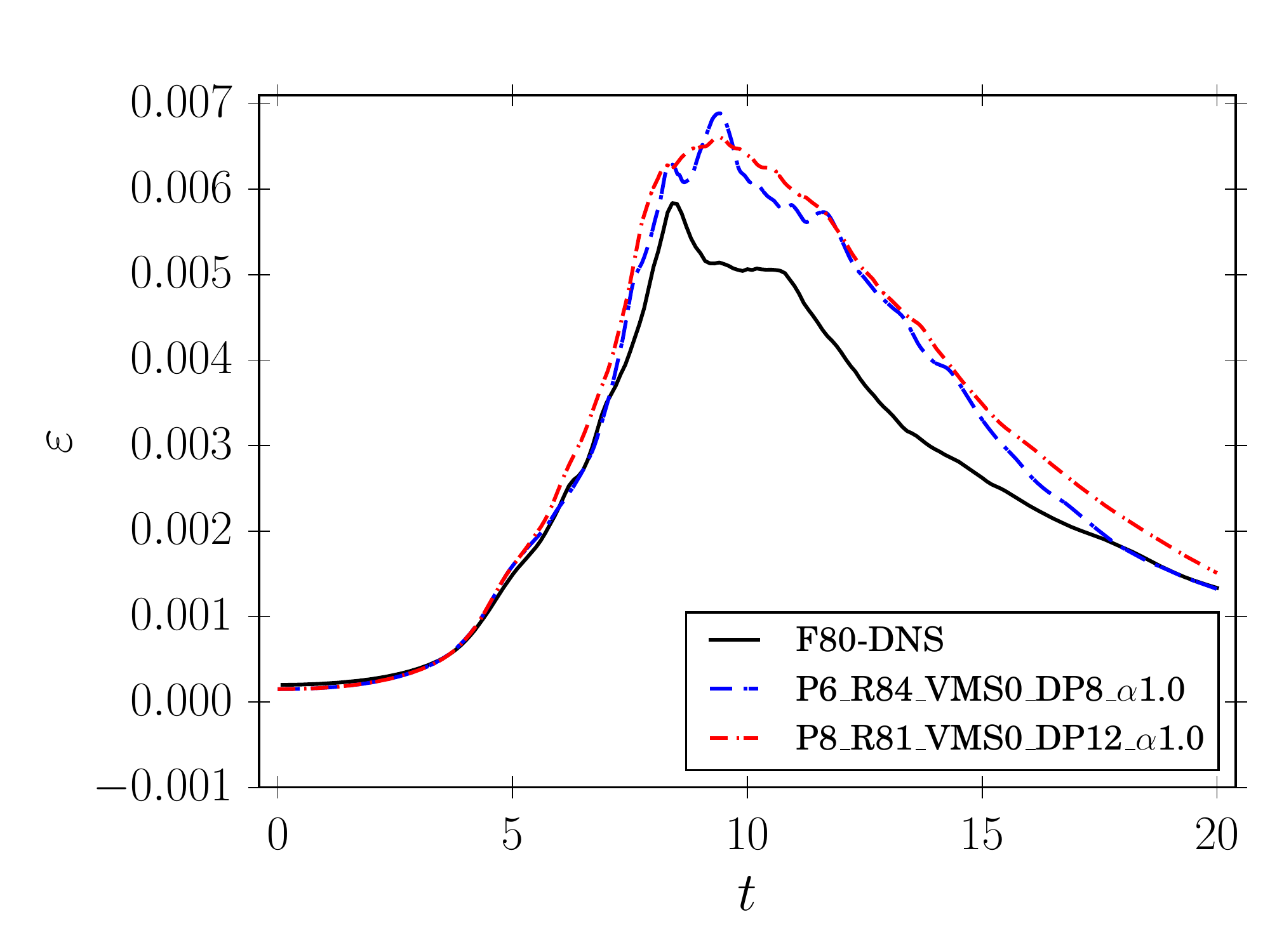}}~
\subfloat[\textbf{VMS}2/3]{
\includegraphics[trim = 1mm 1mm 1mm 1mm, clip,width=0.5\linewidth]{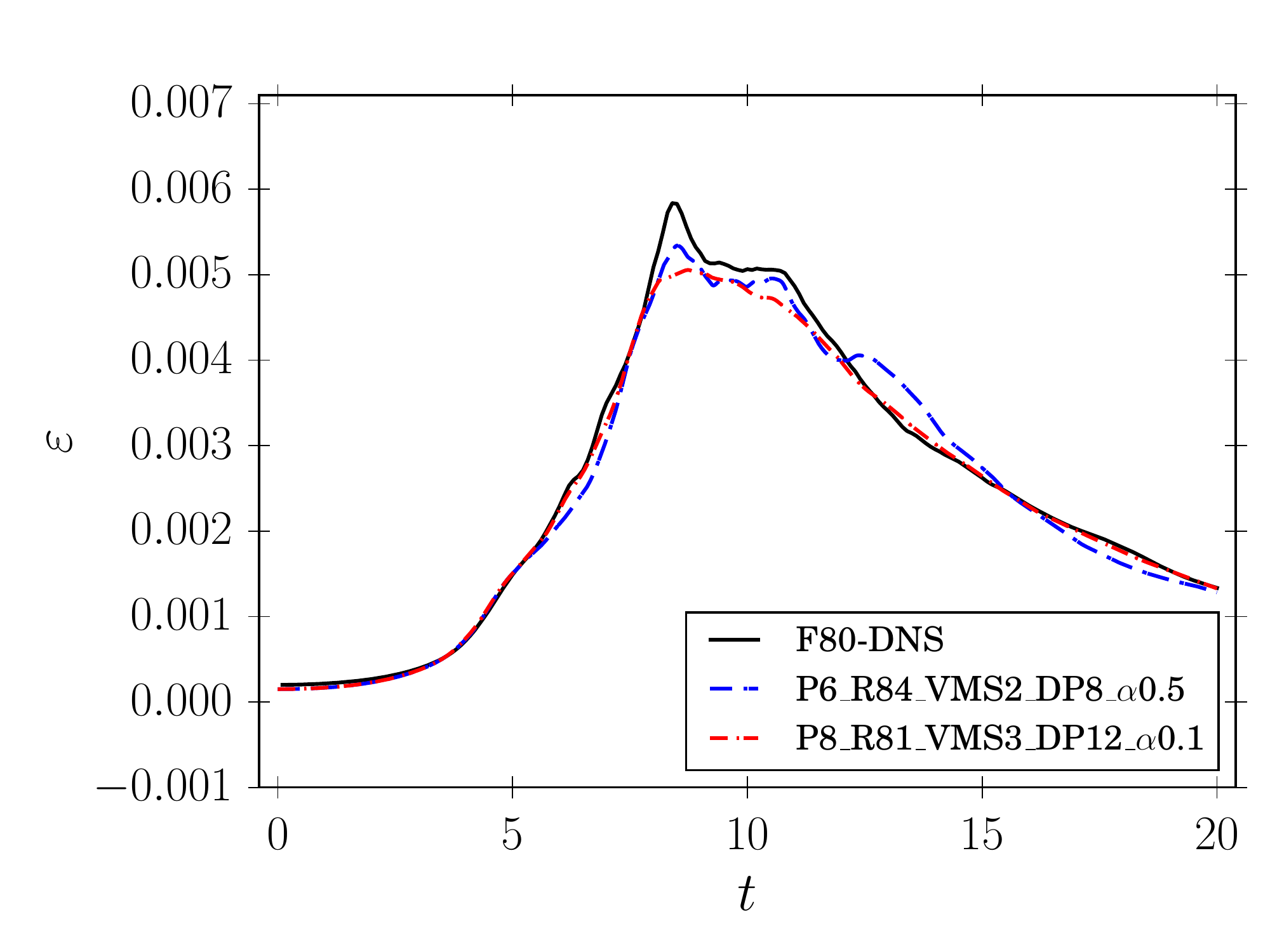}}
\caption{Effect of the polynomial degree of the solution space, $\bar{\tilde{p}}$; Resolved viscous dissipation, $\varepsilon$.}
\label{fig:P_visc_diss}
\end{figure}

 \begin{figure}[!hbt]
\centering
\subfloat[\textbf{VMS}0]{
\includegraphics[trim = 1mm 1mm 1mm 1mm, clip,width=0.5\linewidth]{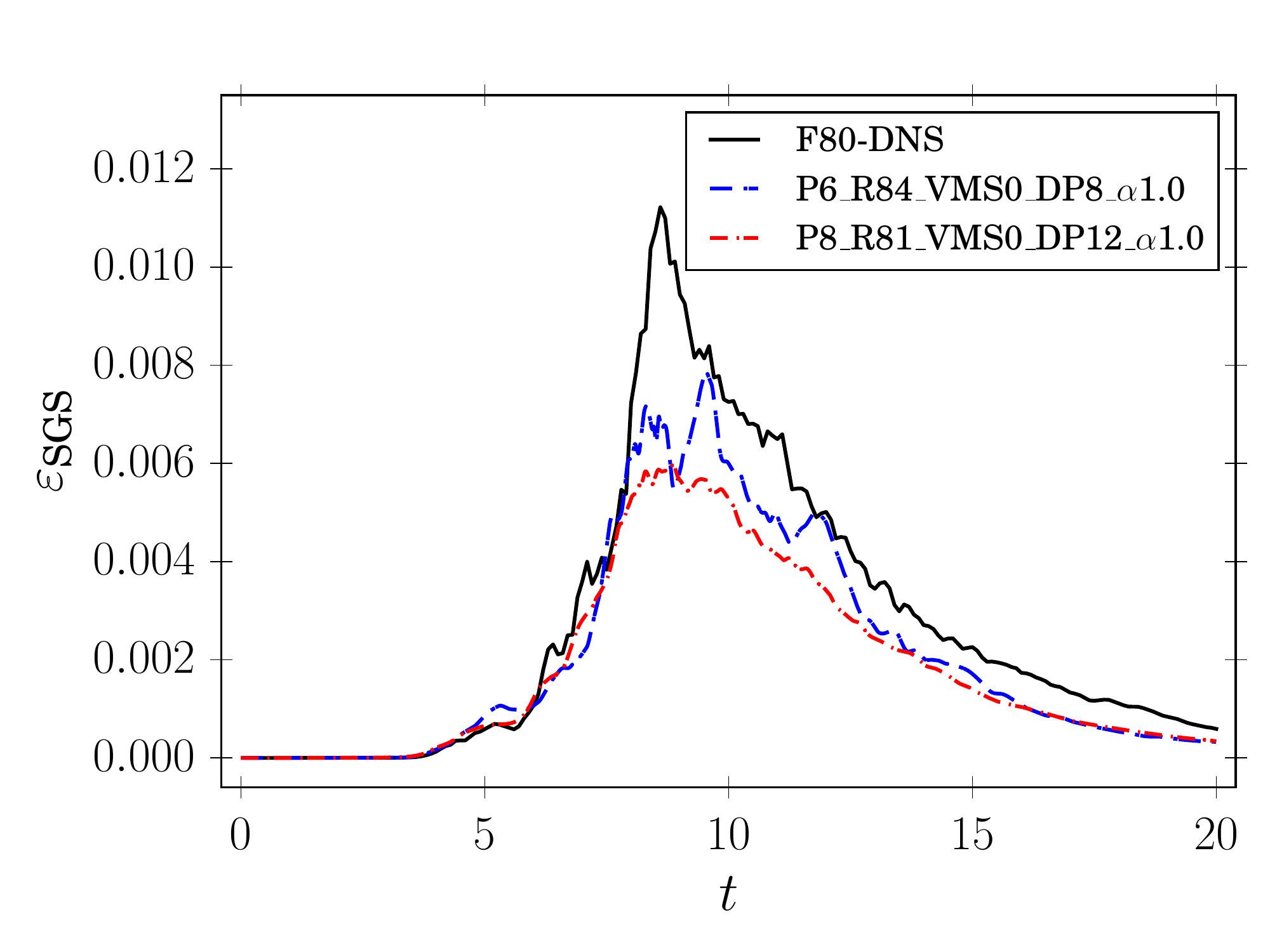}}~
\subfloat[\textbf{VMS}2/3]{
\includegraphics[trim = 1mm 1mm 1mm 1mm, clip,width=0.5\linewidth]{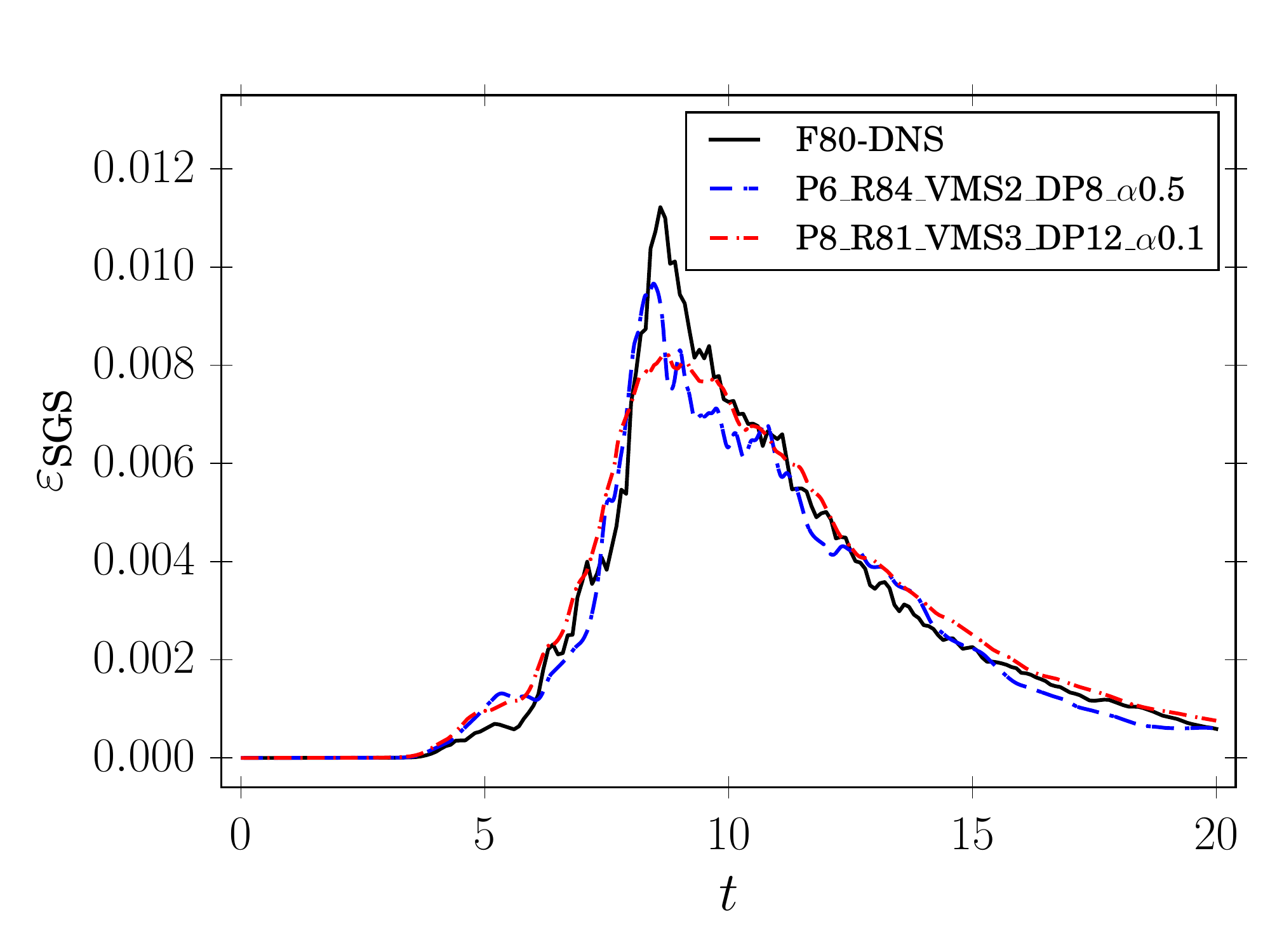}}
\caption{Effect of the polynomial degree of the solution space, $\bar{\tilde{p}}$; Subgrid-scale dissipation, $\varepsilon_{\text{SGS}}$.}
\label{fig:P_SGS_diss}
\end{figure}

 \begin{figure}[!hbt]
\centering
\subfloat[\textbf{VMS}0]{
\includegraphics[trim = 1mm 1mm 1mm 1mm, clip,width=0.5\linewidth]{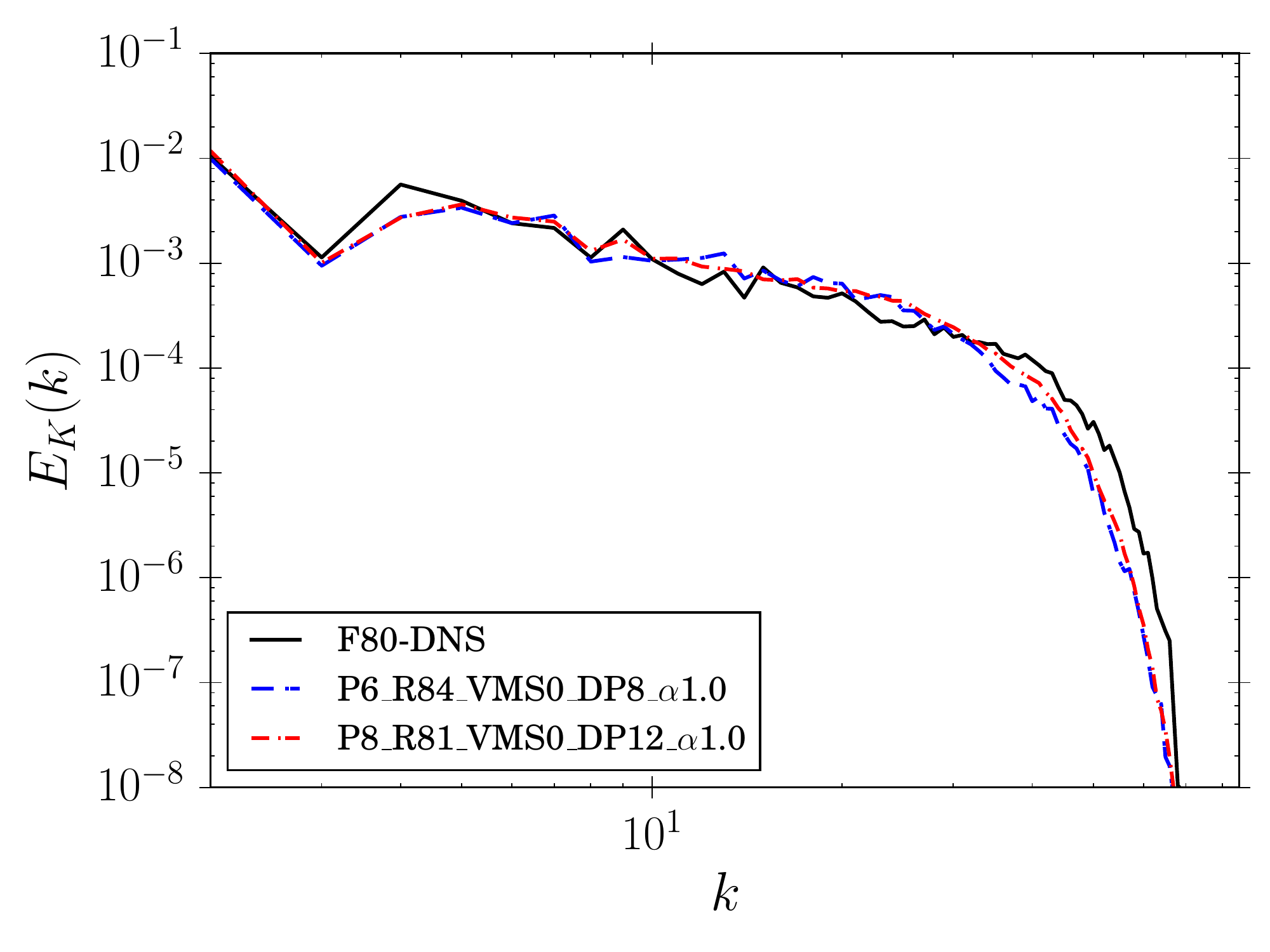}}~
\subfloat[\textbf{VMS}2/3]{
\includegraphics[trim = 1mm 1mm 1mm 1mm, clip,width=0.5\linewidth]{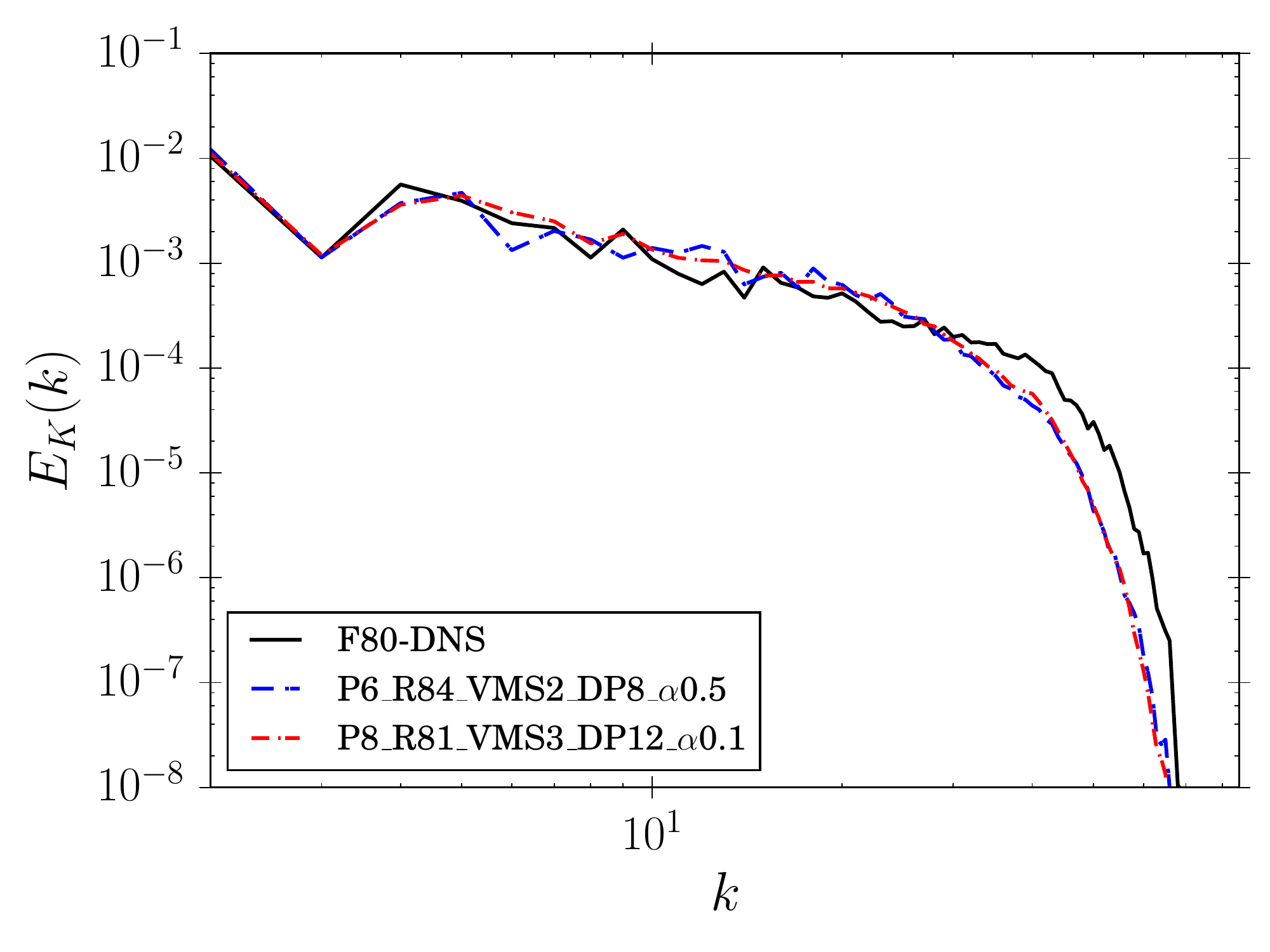}}
\caption{Effect of the polynomial degree of the solution space, $\bar{\tilde{p}}$; Kinetic energy spectrum at $t=14$. }
\label{fig:P_spectrum}
\end{figure}

It is previously shown by Chapelier \textit{et al.} \cite[4.5.]{Chapelier2016a} that to an equal total number of degrees of freedom, using a low-order discretization (\textbf{P}1) results in a significantly lower performance than a high-order scheme. We compare here the capability of two high-order VMS-FR/CPR discretizations, i.e. \textbf{P}6 and \textbf{P}8, in Figures \ref{fig:P_visc_diss}, \ref{fig:P_SGS_diss} and \ref{fig:P_spectrum}. From these figures, it can be assessed that for both the model-free (\textbf{VMS}0) and modelled (\textbf{VMS}2/3) simulations, it is possible to achieve closely comparable results by either discretization with the difference that \textbf{P}8 outputs are slightly smoother. Nevertheless, the \textbf{P}8 simulations are significantly more costly due to the need for higher de-aliasing and also to slightly lower stable time steps.  For example, the case
\textbf{P}8\_\textbf{R}81\_\textbf{VMS}0\_\textbf{DP}12\_$\bm{\alpha}$1.0 required $11,006$ total core-hours  on 243 cores versus $2,128$ core-hours on 192 cores in the case of \textbf{P}6\_\textbf{R}84\_\textbf{VMS}0\_\textbf{DP}{8}\_$\bm{\alpha}1.0$ (the former is $\sim$4 times more expensive).

As a side note, we can remark again the better performance of the \textbf{VMS}2/3 simulations compared to the ILES in terms of dissipation components in Figures \ref{fig:P_visc_diss} and \ref{fig:P_SGS_diss}, whereas, the ILES runs more faithfully reproduce the spectral footprint of small resolved scales  in Figure \ref{fig:P_spectrum} with regards to the filtered DNS result.

Finally, we note that based on our tests,  in the \textbf{P}6  and \textbf{P}8 cases, the best results were obtained for large/small partitions of respectively \textbf{VMS}2 and \textbf{VMS}3  which correspond to respectively $71 \%$  and $67 \%$ large/total modes, which is in agreement with the recommendation provided in \cite{Holmen2004,Chapelier2016a}.

\subsection{Effect of grid resolution}
In this section we focus on the effect of increasing the grid resolution, for a fixed polynomial degree, on the VMS simulations. To this end, we compare high-resolution \textbf{R}161 results to lower resolutions of \textbf{R}81/84.
 
\begin{figure}[!hbt]
\centering
\subfloat[$\varepsilon$]{
\includegraphics[trim = 1mm 1mm 1mm 1mm, clip,width=0.5\linewidth]{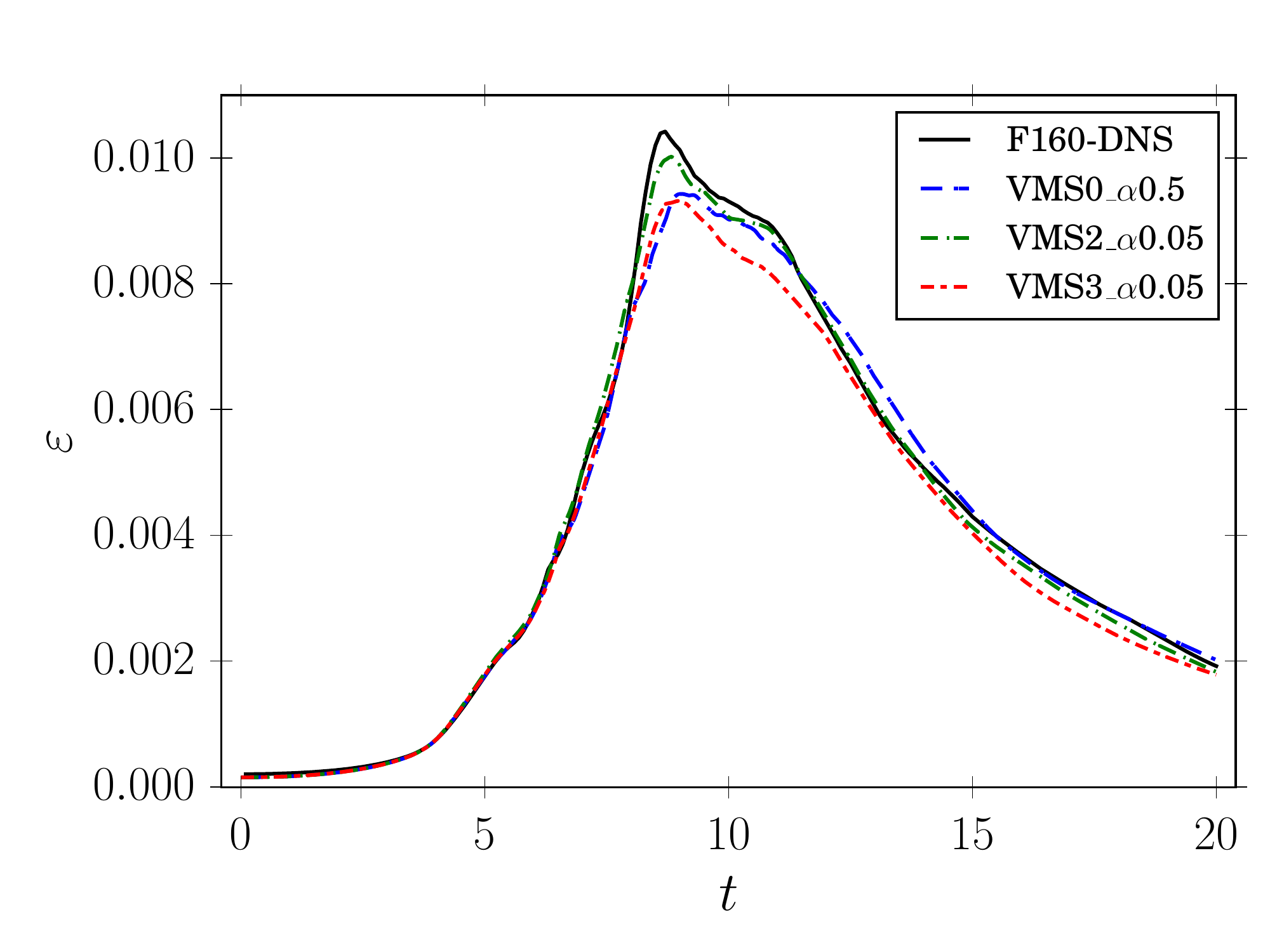}}~
\subfloat[$\varepsilon_\text{SGS}$]{
\includegraphics[trim = 1mm 1mm 1mm 1mm, clip,width=0.5\linewidth]{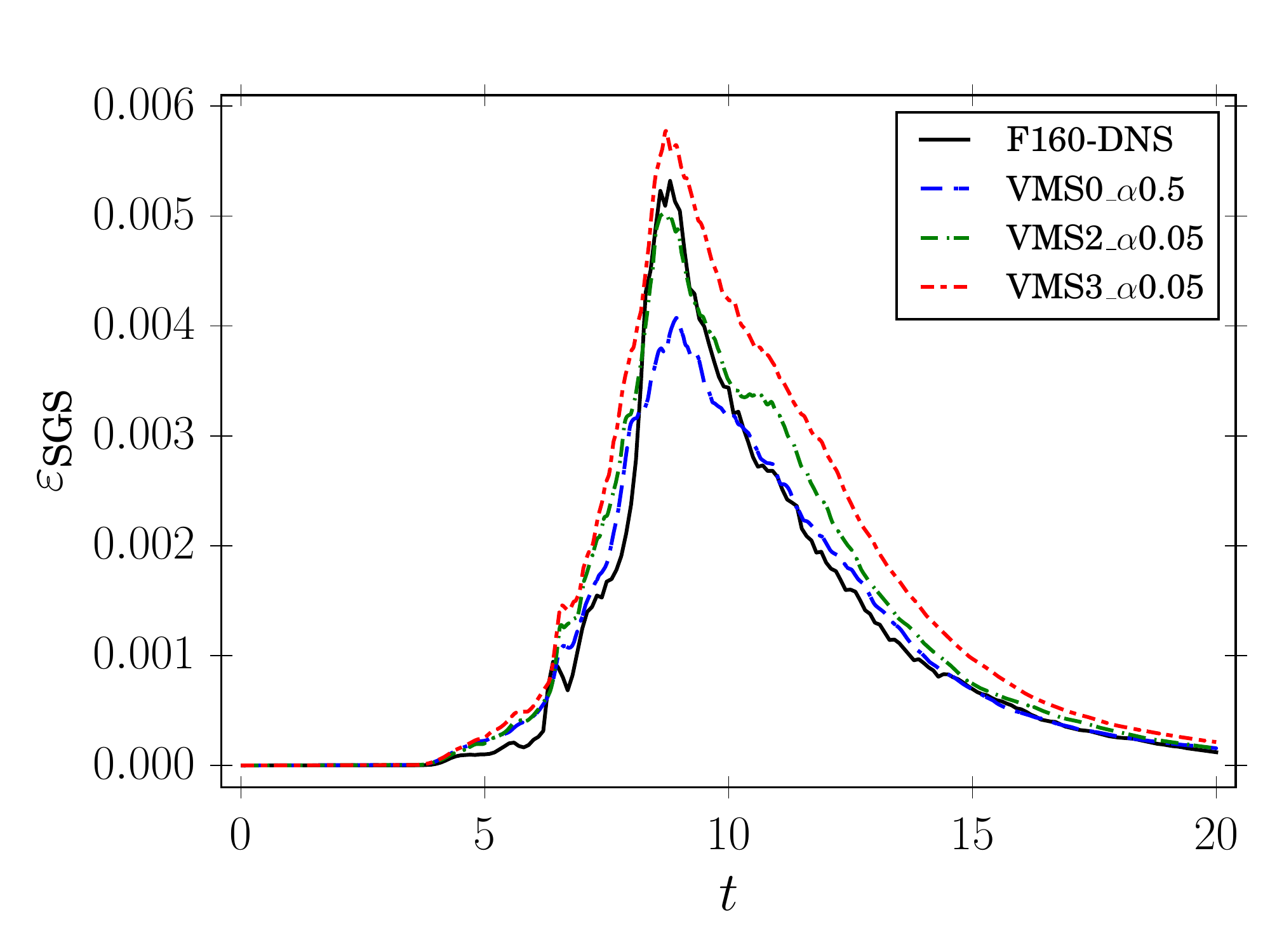}}
\caption{Effect of grid resolution; Kinetic energy dissipation components of the cases \textbf{P}6\_\textbf{R}161\_\textbf{VMS}{\color{blue}X}\_\textbf{DP}8\_$\bm{\alpha}${\color{blue}X}.}
\label{fig:R_diss}
\end{figure}

Comparing the high-resolution evolution of  $\varepsilon$ in Figure \ref{fig:R_diss} (a) to the low-resolution output in Figure \ref{fig:P_visc_diss}, one can appreciate that by increasing the number of degrees of freedom,  \textbf{VMS}2 still produces significantly better results than \textbf{VMS}0, although the difference between these two is smaller at higher resolution. Similar observations can be made by comparing the $\varepsilon_\text{SGS}$ evolution at high-resolution in Figure \ref{fig:R_diss} (b) to its lower-resolution counterpart in Figure \ref{fig:P_SGS_diss}. It is noteworthy saying that lowering the value of upwind dissipation to $\alpha=0.1$, for the implicit LES case (\textbf{P}6\_\textbf{R}161\_\textbf{VMS}0\_\textbf{DP}8\_$\bm{\alpha}$0.1) caused the divergence of the simulation.

\begin{figure}[!hbt]
\centering
\includegraphics[trim = 1mm 1mm 1mm 1mm, clip,width=0.6\linewidth]{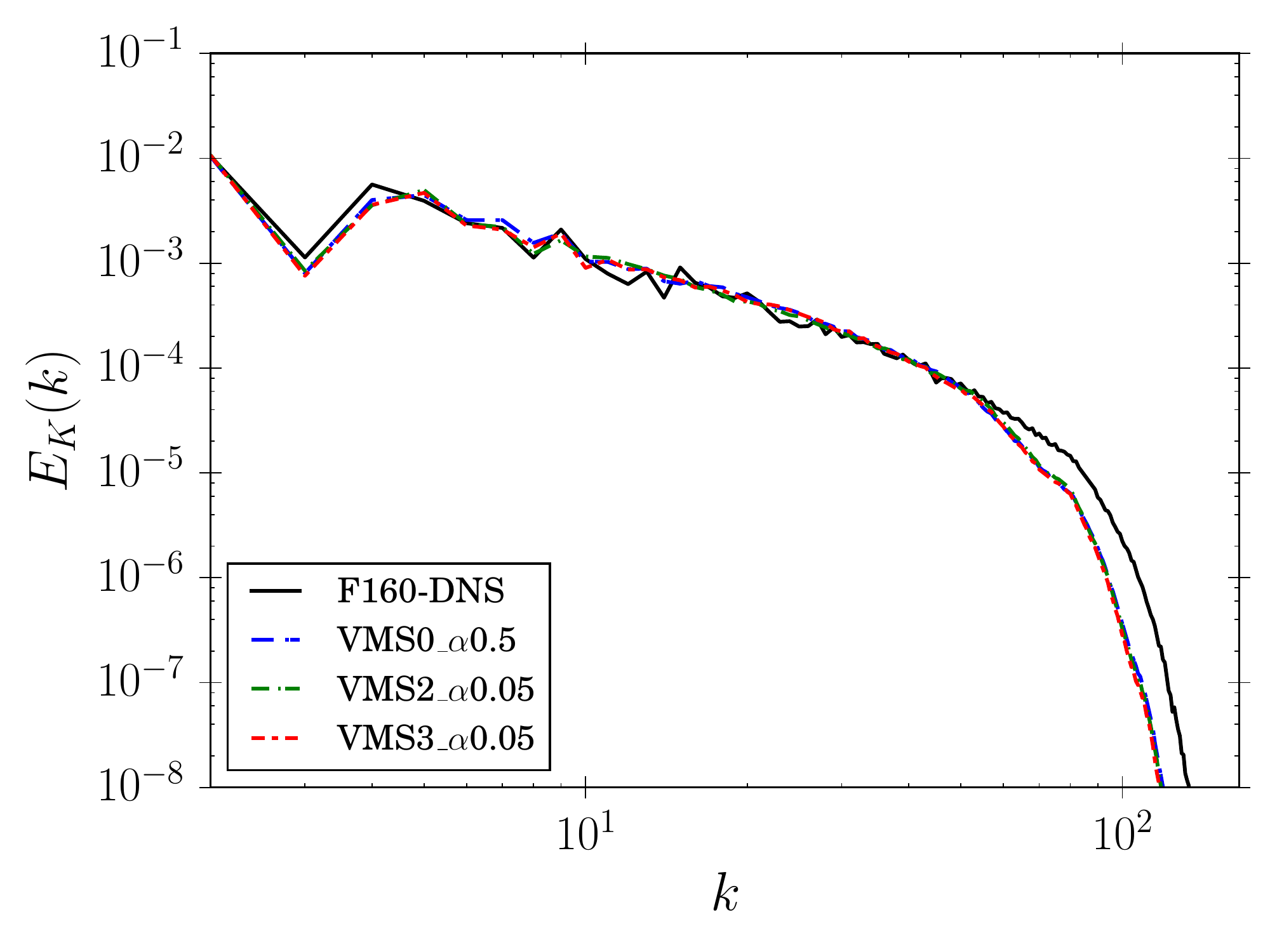}
\caption{Effect of grid resolution; Kinetic energy spectrum at $t=14$ of the cases \textbf{P}6\_\textbf{R}161\_\textbf{VMS}{\color{blue}X}\_\textbf{DP}8\_$\bm{\alpha}${\color{blue}X}. }
\label{fig:R_spectrum}
\end{figure}

In terms of kinetic energy spectrum, the high-resolution outputs in Figure \ref{fig:R_spectrum}, in contrast to the low-resolution ones in Figure \ref{fig:P_spectrum}, reveal again a reduction in the effect of scale partitioning and a better overall fidelity  for a larger resolution.

\begin{figure}[!hbt]
\centering
\subfloat[\textbf{P}8\_\textbf{R}81\_\textbf{VMS}3\_\textbf{DP}12\_$\bm{\alpha}0.05$]{
\includegraphics[trim = 1mm 1mm 1mm 1mm, clip,width=0.5\linewidth]{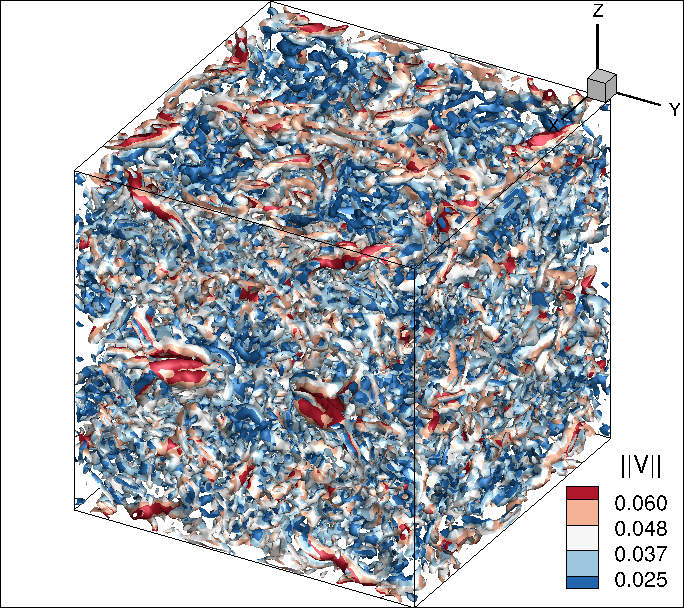}}~
\subfloat[\textbf{P}6\_\textbf{R}161\_\textbf{VMS}2\_\textbf{DP}8\_$\bm{\alpha}0.05$]{
\includegraphics[trim = 1mm 1mm 1mm 1mm, clip,width=0.5\linewidth]{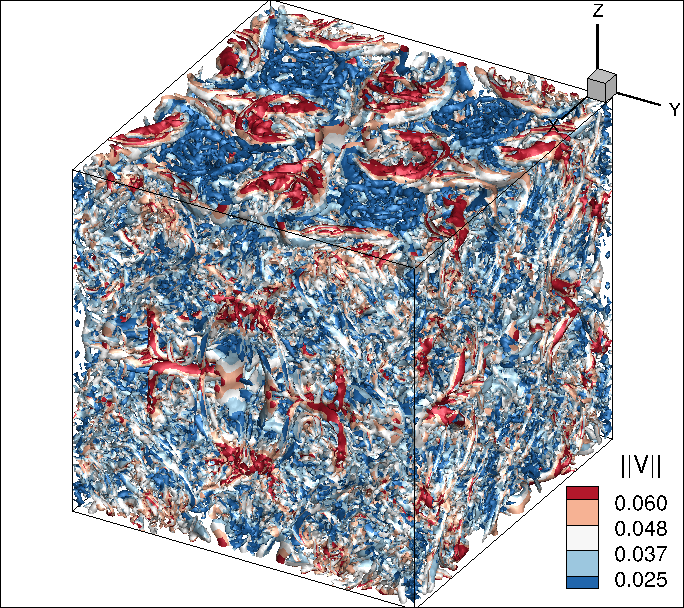}}
\caption{Effect of grid resolution; Iso-surfaces of  Q-criterion intensity (Q=0.01), coloured by velocity magnitude at $t=14$. }
\label{fig:R_Qcriterion}
\end{figure}

Finally, the iso-surfaces of Q-criterion in Figure \ref{fig:R_Qcriterion} let us appreciate the sharper definition of all resolved scales and the representation of a wider range of scales in the \textbf{R}161 case as compared to the \textbf{R}81.

\subsection{Alternative schemes : SD, $g_2$}
The major advantageous aspect of the FR/CPR method is the possibility of recovering different schemes, within the same formulation, by varying the flux correction functions at the basis of the method. In this section, we explore the effect of considering the correction functions of the SD and $g_2$ schemes in addition to those of the baseline DG approach.
 \begin{figure}[!hbt]
\centering
\subfloat[$\varepsilon$]{
\includegraphics[trim = 1mm 1mm 1mm 1mm, clip,width=0.5\linewidth]{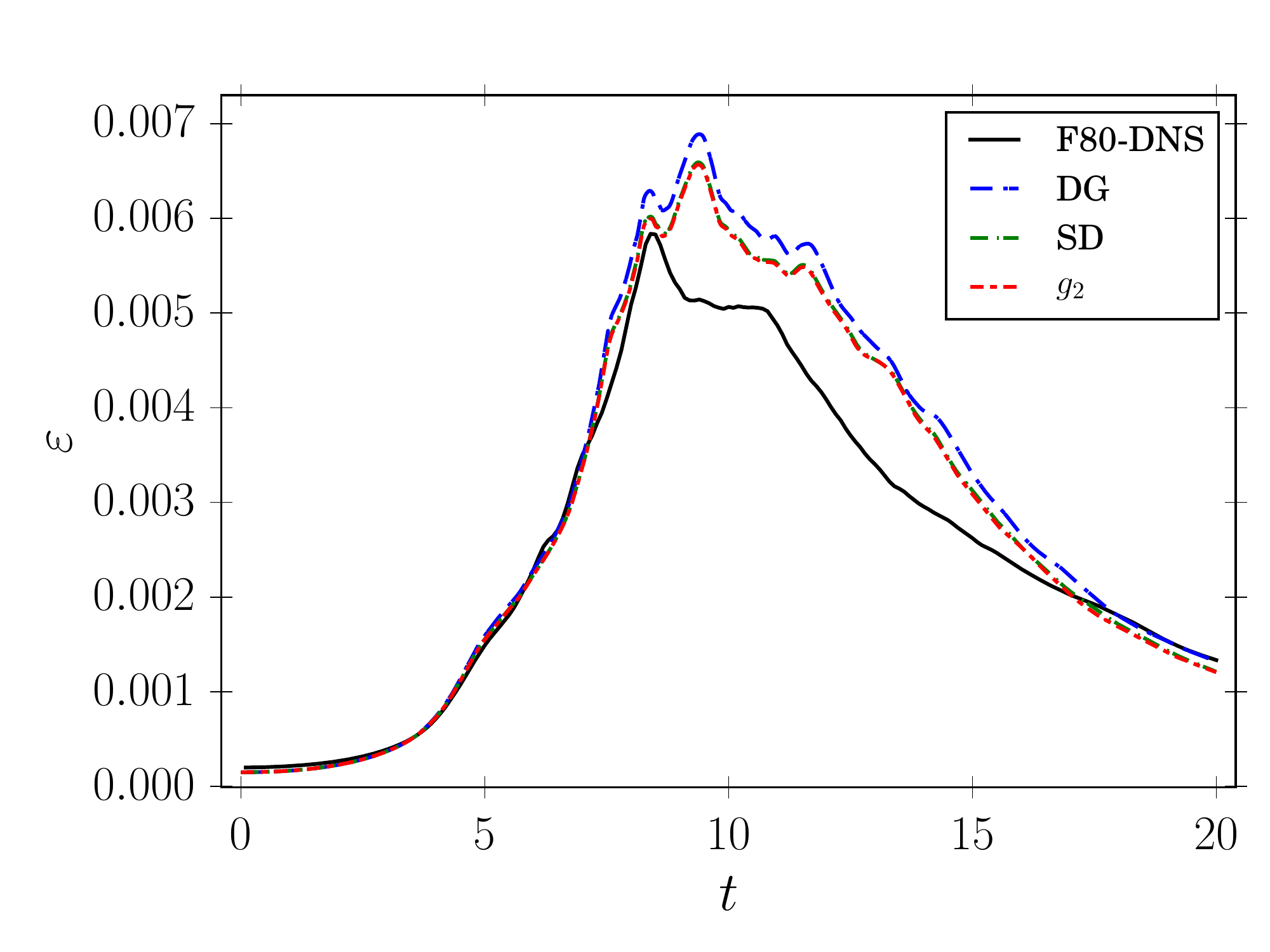}}~
\subfloat[$\varepsilon_\text{SGS}$]{
\includegraphics[trim = 1mm 1mm 1mm 1mm, clip,width=0.5\linewidth]{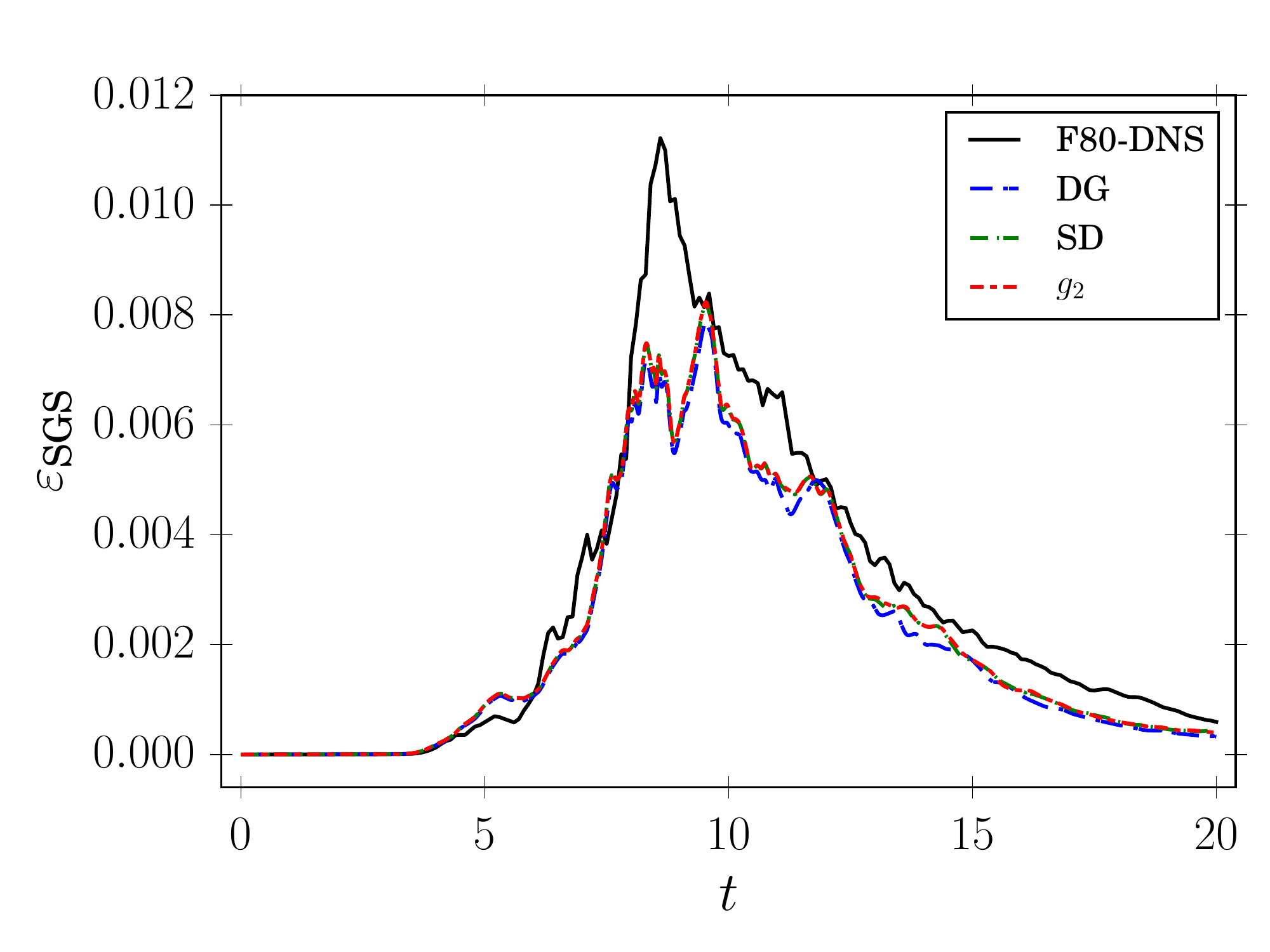}}
\caption{Effect of the FR/CPR schemes on the case \textbf{P}6\_\textbf{R}84\_\textbf{VMS}0\_\textbf{DP}8\_$\bm{\alpha}1.0$; Kinetic energy dissipation components.}
\label{fig:S1_diss}
\end{figure}

 \begin{figure}[!hbt]
\centering
\subfloat[$\varepsilon$]{
\includegraphics[trim = 1mm 1mm 1mm 1mm, clip,width=0.5\linewidth]{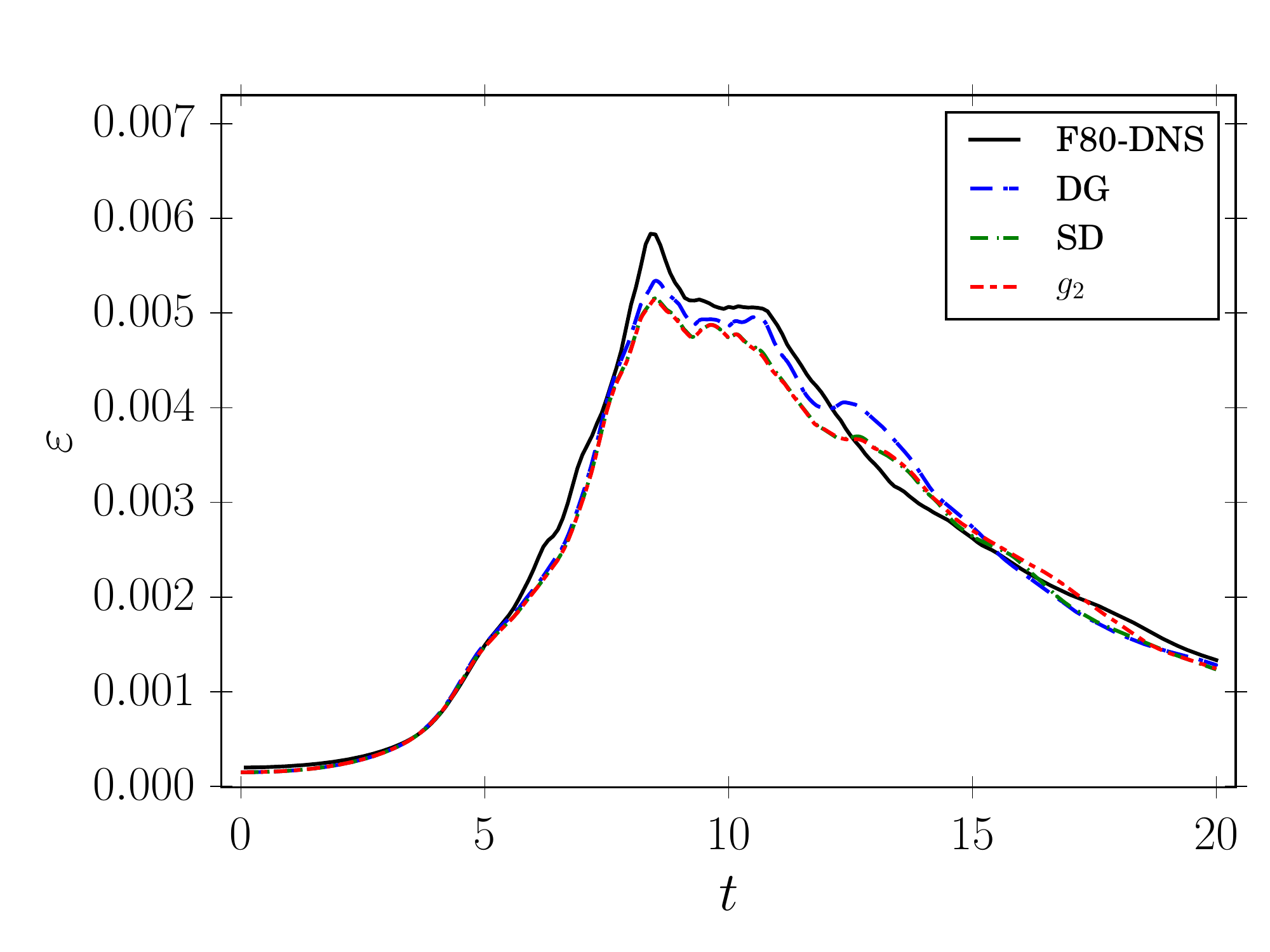}}~
\subfloat[$\varepsilon_\text{SGS}$]{
\includegraphics[trim = 1mm 1mm 1mm 1mm, clip,width=0.5\linewidth]{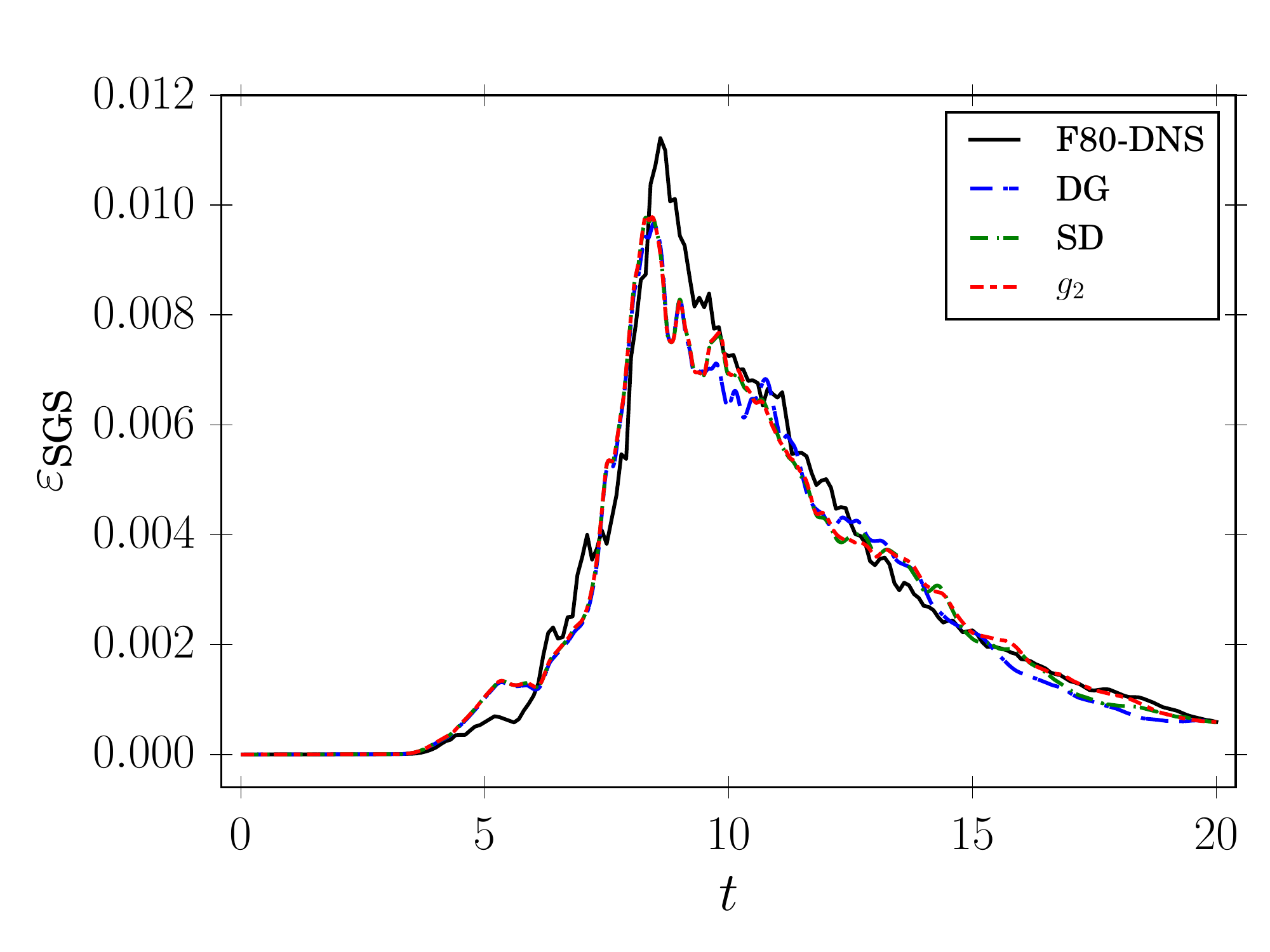}}
\caption{Effect of the FR/CPR schemes on the case \textbf{P}6\_\textbf{R}84\_\textbf{VMS}2\_\textbf{DP}8\_$\bm{\alpha}0.5$; Kinetic energy dissipation components.}
\label{fig:S2_diss}
\end{figure}

Figures \ref{fig:S1_diss} and \ref{fig:S2_diss} illustrate the impact of the mentioned correction functions on the resolved and the SGS dissipation components for \textbf{VMS}0  and  \textbf{VMS}2 cases as an example. The observation is that the SD and $g_2$  results are almost identical and are both very close to those of the FR-DG scheme. This corroborates the observations of \cite{Mengaldo2018} on ESFR schemes: a close affinity between SD and $g_2$ LES results, both comparable to DG.

We note that no significant difference in terms of maximum stable time step was detected between these three schemes.

\section{Conclusions}
\label{sec:conclus}
A variational multiscale approach for the simulation of compressible turbulent flows is presented that encompasses a large family of compact nodal discretization methods represented by the high-order flux reconstruction scheme. The mathematical aspects of the large-eddy simulation modelling are presented from a variational perspective by emphasizing the approximations made for each term of the equations. The resulting model based on a two-level partitioning of all the scales of turbulence into resolved and unresolved (subgrid), is then discretized via the flux reconstruction scheme. Another scale separation is subsequently introduced to isolate the large and small resolved scales, allowing thus for the application of the subgrid-scale model to the latter only. The proposed formulation is assessed on the benchmark problem of Taylor-Green vortex with a Reynolds number of 5000 for which the filtered DNS data is available. The results exhibited the role of the VMS modelling in improving the simulations agreement with reference data, compared to the baseline implicit LES, by adjusting the amplitude and the spectral spread of subgrid dissipation, especially on coarser grids. Furthermore, these numerical experiments enabled the identification of the isolated role of parameters such as de-aliasing and Roe's upwinding dissipation. For low-Mach flows, adjusting the magnitude of the latter to moderately lower values than the default unity, conjointly to the application of a VMS model, generated a noticeable improvement of outcomes with regards to physical metrics as well as to maximum stable time step.

\section{Acknowledgements}
We would like to acknowledge the financial support of FRQNT, NSERC, McGill Mechanical Engineering Department and the Grant number NIH R01 DC005788-13.
Computational resources and support were provided by Compute-Canada, Calcul-Qu\'ebec and McGill's HPC staff.
Finally, we would also like to thank Professor Siva Nadarajah for his kind support in the early stages of this study and Professor Eric Lamballais for providing the reference DNS data.

\appendix

  \setcounter{remark}{0}
 \renewcommand{\theremark}{\Alph{section}\arabic{remark}}

 \setcounter{lemma}{0}
 \renewcommand{\thelemma}{\Alph{section}\arabic{lemma}}
  \setcounter{proposition}{0}
 \renewcommand{\theproposition}{\Alph{section}\arabic{proposition}}
 \setcounter{theorem}{0}
 \renewcommand{\thetheorem}{\Alph{section}\arabic{theorem}}
 \setcounter{corollary}{0}
 \renewcommand{\thecorollary}{\Alph{section}\arabic{corollary}}
  \setcounter{definition}{0}
 \renewcommand{\thedefinition}{\Alph{section}\arabic{definition}}

{\section{Elemental geometrical mapping}
\label{app:mappings}
To compute the derivatives and the integrals efficiently via numerical methods, the variational residual on the \textit{physical} domain/element $\bm{\Omega}$,  is mapped to a \textit{computational} domain/element, $\bm{\Omega^r}$, defined in the reference Cartesian system of coordinates, $\bm{\xi}:= \xi_j\,\mathbf{{b}}^\mathbf{r}_j$, spanned by the orthonormal basis vectors, $\mathbf{{b}}^\mathbf{r}_j$. In this section, we present the mathematical developments which allow for such transformations. 

\begin{definition}
The mapping  (transformation)  function, $\bm{\mathfrak{M}}:=\mathfrak{M}_j(\bm{\xi})\,\mathbf{{b}}_j$, defined using $\mathbf{{b}}_j$, the basis vectors of the physical space, serves to retrieve the coordinates of the physical element, given those of the reference element, such that:
\begin{equation}
\bm{x}=\bm{\mathfrak{M}}(\bm{\xi}),
\label{eq:map}
\end{equation}
with the following components:
\begin{equation*}
x_1=\mathfrak{M}_{1}(\xi_1,\xi_2,\xi_3),\quad
x_2=\mathfrak{M}_{2}(\xi_1,\xi_2,\xi_3),\quad
x_3=\mathfrak{M}_{3}(\xi_1,\xi_2,\xi_3).
\end{equation*}
\end{definition}

\begin{lemma}
A volume integral on the physical element can be expressed as a volume integral on the reference element via the following relation:
\begin{equation*}
\int_{\bm{\Omega}} f(\bm{x}) \ d{\Omega} = \int_{\bm{\Omega^r}} f \left(\bm{\mathfrak{M}}(\bm{\xi})\right) \, J_{{\Omega}}(\bm{\xi}) \, d{\Omega^r},
\end{equation*}
where the physical and reference differential volumes\footnote{ $(\mathbf{a} \times \mathbf{b}) \cdot \mathbf{c}$ denotes the triple scalar product of vectors $\mathbf{a}$, $\mathbf{b}$ and $\mathbf{c}$.}  are respectively $  d{\Omega}:= \ (\bm{dx_1} \times \bm{dx_2}) \cdot \bm{dx_3}$ and $  d{\Omega^r}:= \ (\bm{d\xi_1} \times \bm{d\xi_2}) \cdot \bm{d\xi_3}$; and  $J_{{\Omega}}(\bm{\xi})$ designates the determinant of $\bm{J}_{{\Omega}}(\bm{\xi}):=\frac{\partial \mathfrak{M}_i (\bm{\xi})}{\partial \xi_j}$, the Jacobian matrix of the transformation.  
\begin{proof}
The spatial differentials of the physical element can be related to those of the reference element via the following relations:
\begin{align*}
\bm{dx_1} &:= dx_1 \,\mathbf{b}_1 = \frac{\partial \mathfrak{M}_1}{\partial \xi_1} \,d\xi_1\,\mathbf{{b}}^\mathbf{r}_1 + \frac{\partial \mathfrak{M}_1}{\partial {\xi_2}} \,d{\xi_2}\,\mathbf{{b}}^\mathbf{r}_2  + \frac{\partial \mathfrak{M}_1}{\partial {\xi_3}} \,d{\xi_3}\,\mathbf{{b}}^\mathbf{r}_3, \\
\bm{dx_2} &:= dx_2 \,\mathbf{b}_2  = \frac{\partial \mathfrak{M}_2}{\partial \xi_1} \,d\xi_1\,\mathbf{{b}}^\mathbf{r}_1 + \frac{\partial \mathfrak{M}_2}{\partial {\xi_2}} \,d{\xi_2}\,\mathbf{{b}}^\mathbf{r}_2  + \frac{\partial \mathfrak{M}_2}{\partial {\xi_3}} \,d{\xi_3}\,\mathbf{{b}}^\mathbf{r}_3, \\
\bm{dx_3} &:= dx_3 \,\mathbf{b}_3  = \frac{\partial \mathfrak{M}_3}{\partial \xi_1} \,d\xi_1\,\mathbf{{b}}^\mathbf{r}_1 + \frac{\partial \mathfrak{M}_3}{\partial {\xi_2}} \,d{\xi_2}\,\mathbf{{b}}^\mathbf{r}_2  + \frac{\partial \mathfrak{M}_3}{\partial {\xi_3}} \,d{\xi_3}\,\mathbf{{b}}^\mathbf{r}_3.
\end{align*}
Hence, the physical differential volume reads
\begin{align}
\begin{split}
 d{\Omega} &\equiv (\bm{dx_1} \times \bm{dx_2}) \cdot \bm{dx_3} = 
 \begin{vmatrix}
 \frac{\partial \mathfrak{M}_1}{\partial \xi_1} \,d\xi_1\, & \frac{\partial \mathfrak{M}_1}{\partial {\xi_2}} \,d{\xi_2}\,  &\frac{\partial \mathfrak{M}_1}{\partial {\xi_3}} \,d{\xi_3}\,\\
 \vspace{.1pt}\\
 \frac{\partial \mathfrak{M}_2}{\partial \xi_1} \,d\xi_1\, & \frac{\partial \mathfrak{M}_2}{\partial {\xi_2}} \,d{\xi_2}\,  &\frac{\partial \mathfrak{M}_2}{\partial {\xi_3}} \,d{\xi_3}\,\\
  \vspace{.1pt}\\
 \frac{\partial \mathfrak{M}_3}{\partial \xi_1} \,d\xi_1\, & \frac{\partial \mathfrak{M}_3}{\partial {\xi_2}} \,d{\xi_2}\,  &\frac{\partial \mathfrak{M}_3}{\partial {\xi_3}} \,d{\xi_3}\,   
\end{vmatrix}
  \vspace{.5pt}\\
&= J_{{\Omega}}(\bm{\xi}) \,d\xi_1 \, d{\xi_2} \, d{\xi_3} = J_{{\Omega}}(\bm{\xi}) \, d{\Omega^r}.
\end{split} 
\end{align}
Finally, Definition \ref{eq:map} yields $f(\bm{x}) \equiv f \left(\bm{\mathfrak{M}}(\bm{\xi})\right)$ which completes the proof.
\end{proof}
\label{app:lemma_volume}
\end{lemma}

\begin{definition}
The function $\bm{\mathcal{L}}(v_1,v_2)$ parametrizes $\bm{\Gamma}$, the surface of the physical element,  by independent variables $v_1$ and $v_2$ and provides the Cartesian coordinates of all points, $\mathbf{P}$, on the surface such that
\begin{equation*}
\mathbf{P}(x_1,x_2,x_3) = \bm{\mathcal{L}}(v_1,v_2).
\end{equation*}
For example, the sphere $x_1^2+x_2^2+x_3^2=r^2$ can be parametrized as
\begin{equation*}
\bm{\mathcal{L}}(v_1,v_2):= v_1\mathbf{b}_1+v_2\mathbf{b}_2+\sqrt{r^2-v_1^2-v_2^2}\,\mathbf{b}_3,
\end{equation*}
From which the followings ensue 
\begin{equation*}
x_1 = v_1, \quad x_2=v_2, \quad x_3=\sqrt{r^2-v_1^2-v_2^2}.
\end{equation*}
Another example is the plane $z=1$  parametrized as
\begin{equation*}
\bm{\mathcal{L}}(v_1,v_2):=v_1\mathbf{b}_1+v_2\mathbf{b}_2+\mathbf{b}_3.
\end{equation*}
\label{app:defsurfparam}
\end{definition}
\begin{definition}
Differential tangential vectors $\bm{dt}_1$ and $\bm{dt}_2$,  to the surface $\bm{\Gamma}$ at the point $\mathbf{P}$ are defined via Definition \ref{app:defsurfparam} and in terms of $\bm{x}$ coordinates as
\begin{equation}
\bm{dt}_1(\bm{x}) := \frac{\partial \bm{\mathcal{L}}}{\partial u} du, \qquad \bm{dt}_2(\bm{x}) := \frac{\partial \bm{\mathcal{L}}}{\partial v} dv.
\end{equation}
The tangential vector $\bm{dt}_1$ has the following components in  physical coordinates:
\begin{equation}
\bm{dt}_1(\bm{x}) = dx_1\mathbf{{b}}_1 + dx_2\mathbf{{b}}_2  + dx_3\mathbf{{b}}_3,
\end{equation}
and the followings in terms of the reference coordinates:
\begin{equation*}
\bm{dt}_1(\bm{\xi}) = d\xi_1\mathbf{{b}}^\mathbf{r}_1 + d\xi_2\mathbf{{b}}^\mathbf{r}_2  + d\xi_3\mathbf{{b}}^\mathbf{r}_3,
\end{equation*}
and similarly for $\bm{dt}_2$.
\label{app:surf_diffs}
\end{definition}

\begin{lemma}
The coordinates of the vector $\bm{dt}_1$ in the two systems are related via $\bm{dt}_1(\bm{x}) =  \bm{J}_{\Omega} \, \bm{dt}_1(\bm{\xi})$ and similarly for $\bm{dt}_2$.
\begin{proof}
The components of the $\bm{dt}_1$ in the physical system can be expressed via Definition \ref{app:surf_diffs} as
\begin{eqnarray}
\begin{split}
dx_1= &\frac{\partial \mathfrak{M}_1}{\partial \xi_1}d\xi_1 +\frac{\partial \mathfrak{M}_1}{\partial {\xi_2}}d\xi_2+ \frac{\partial \mathfrak{M}_1}{\partial {\xi_3}}d\xi_3,&\\
dx_2= &\frac{\partial \mathfrak{M}_2}{\partial \xi_1}d\xi_1 +\frac{\partial \mathfrak{M}_2}{\partial {\xi_2}}d\xi_2+ \frac{\partial \mathfrak{M}_2}{\partial {\xi_3}}d\xi_3,&\\
dx_3= &\frac{\partial \mathfrak{M}_3}{\partial \xi_1}d\xi_1 +\frac{\partial \mathfrak{M}_3}{\partial {\xi_2}}d\xi_2+ \frac{\partial \mathfrak{M}_3}{\partial {\xi_3}}d\xi_3,&
\end{split}
\label{eq:dt_map}
\end{eqnarray}
or in a concise form as
\begin{align*}
\bm{dt}_1(\bm{x})\equiv 
\begin{bmatrix}
	dx_1\\
	dx_2\\
	dx_3\\
\end{bmatrix} =
\bm{J}_{\Omega}
\begin{bmatrix}
	d\xi_1\\
	d\xi_2\\
	d\xi_3
\end{bmatrix}
\equiv \bm{J}_{\Omega} \, \bm{dt}_1(\bm{\xi}),
\end{align*}
and we similarly have $\bm{dt}_2(\bm{x}) =  \bm{J}_{\Omega} \, \bm{dt}_2(\bm{\xi})$.
\label{eq:app_dtmap}
\end{proof}
\end{lemma}

\begin{definition}
We define the differential surface on the boundaries of the physical domain/element, denoted as $d{\Gamma}$, to be the area of the parallelogram supported by the vectors $\bm{dt}_1$ and $\bm{dt}_2$, evaluated simply by taking the norm of the cross product of these two vectors:
\begin{equation}
d{\Gamma} :=  \ \| (\bm{dt}_1 \times \bm{dt}_2)\| =  \ \| \bm{dn}\|,
\label{eq:surf_dS}
\end{equation}
where $\bm{dn}$ is the local outward\footnote{With regards to the closed volume $\bm{\Omega}$.} normal vector to the surface $\bm{\Gamma}$ with magnitude corresponding to the differential surface size.
\label{eq:app_surfdiff}
\end{definition}

\begin{lemma}
A surface integral on the physical element boundary can be expressed as a surface integral on the reference element boundary, $\bm{\Gamma^r}$, via the following relation:
\begin{equation*}
\int_{\bm{\Gamma}} f(\bm{x}) \ d{\Gamma} = \int_{\bm{\Gamma^r}} f \left(\bm{\mathfrak{M}}(\bm{\xi})\right) \, J_{{\Gamma}}(\bm{\xi}) \, d{\Gamma^r},
\end{equation*}
where $J_\Gamma :=   \|J_{{\Omega}} \, \bm{J}_{\Omega}^{-\intercal} \, \bm{n}^{\bm{r}}\| $ with  $\bm{n}^{\bm{r}}:=n^r_j\mathbf{b}^{\bm{r}}_j$ designating the unit outward normal vector in the reference space.
\begin{proof}
Starting from Definition \ref{eq:app_surfdiff} and employing Lemma \ref{eq:app_dtmap}, the following relations hold:
\begin{align}
\begin{split}
d{\Gamma}&=\|\bm{dn}\| \\
&= \|\bm{dt}_1(\bm{x}) \times \bm{dt}_2(\bm{x}) \|\\
&= \|(\bm{J}_{\Omega} \,\bm{dt}_1(\bm{\xi})) \times (\bm{J}_{\Omega}\,\bm{dt}_2(\bm{\xi}))\| \\
&=\| J_{{\Omega}} \, \bm{J}_{\Omega}^{-\intercal} (\bm{dt}_1(\bm{\xi}) \times \bm{dt}_2(\bm{\xi}))\| \\
&\equiv \|J_{{\Omega}} \, \bm{J}_{\Omega}^{-\intercal} \, \bm{dn}^{\bm{r}}\|\\
&\equiv \|J_{{\Omega}} \, \bm{J}_{\Omega}^{-\intercal} \, \bm{n}^{\bm{r}} \,d{\Gamma}^{{r}}\|\\
&= \|J_{{\Omega}} \, \bm{J}_{\Omega}^{-\intercal} \, \bm{n}^{\bm{r}}\| \, d{\Gamma}^{{r}}\\
& \equiv J_{{\Gamma}} \, d{\Gamma^r},
\end{split}
\end{align}
where $\bm{dn}^{\bm{r}}:= \bm{dt}_1(\bm{\xi}) \times \bm{dt}_2(\bm{\xi}) \equiv\bm{n}^{\bm{r}} \,d{\Gamma}^{{r}}$ is  the differential normal vector in the reference space. Finally, Definition \ref{eq:map} yields $f(\bm{x}) \equiv f \left(\bm{\mathfrak{M}}(\bm{\xi})\right)$ which completes the proof.
\end{proof}
\label{lemma:appsurf_int}
\end{lemma}

\begin{lemma}
The local unit outward normal to the physical boundary, $\bm n$, is related to the local unit outward normal to the reference boundary, $\bm{n^r}$, via
\begin{equation}
\bm{n} =  \frac{J_{\Omega} \, \bm{J}_{\Omega}^{-T}}{J_{{\Gamma}}} \,\bm{n}^{\bm{r}}.
\label{eq:n_map}
\end{equation}
\begin{proof}
From the proof of Lemma \ref{lemma:appsurf_int}, the differential normal to the physical element can be expressed as $\bm{dn} = J_{{\Omega}} \, \bm{J}_{\Omega}^{-\intercal} \, \bm{n}^{\bm{r}} \,d{\Gamma}^{{r}}$. Hence the unit normal is obtained through dividing the differential normal by its norm:
\begin{equation}
\bm{n} = \frac{\bm{dn}}{\|\bm{dn}\|} = \frac{J_{{\Omega}} \, \bm{J}_{\Omega}^{-\intercal} \, \bm{n}^{\bm{r}} \,d{\Gamma}^{{r}}}{\|J_{{\Omega}} \, \bm{J}_{\Omega}^{-\intercal} \, \bm{n}^{\bm{r}} \,d{\Gamma}^{{r}} \|} = \frac{J_{{\Omega}} \, \bm{J}_{\Omega}^{-\intercal} \, \bm{n}^{\bm{r}} \,d{\Gamma}^{{r}}}{\|J_{{\Omega}} \, \bm{J}_{\Omega}^{-\intercal} \, \bm{n}^{\bm{r}} \|\, d{\Gamma}^{{r}}} =   \frac{J_{\Omega} \, \bm{J}_{\Omega}^{-T}}{J_{{\Gamma}}} \,\bm{n}^{\bm{r}}.
\end{equation}
\end{proof}
\label{lemma:appnormal}
\end{lemma}

\begin{remark}
It is interesting to note that depending on the choice of $\bm{\mathfrak{M}}$, it is possible that the test function on the reference element is not a polynomial of degree $p$ or less, i.e., $\phi^r(\bm{\xi})\notin  {\rm I\!P}^{p}(\bm{\Omega^r})$, even if $\phi(\bm{x})\in {\rm I\!P}^{p}(\bm{\Omega})$. Hence, most often, one explicitly defines the reference test function to ensure $\phi^r(\bm{\xi})\in  {\rm I\!P}^{p}(\bm{\Omega^r})$ (see \ref{app:bases} for useful definitions). This can be extended to the solution, such that
\begin{equation*}
Q^r(\bm{\xi}):= Q(\bm{\mathfrak{M}}(\bm{\xi})) \equiv Q(\bm{x})  ,
\end{equation*}
where $Q^r(\bm{\xi}) \in  {\rm I\!P}^{p}(\bm{\Omega^r})$ is rather ensured instead of $Q(\bm{x}) \in  {\rm I\!P}^{p}(\bm{\Omega})$.
\label{remark:app-space}
\end{remark}

\begin{theorem}
A sample variational formulation (such as the one in Eq. \ref{eq:greenform}) on the physical element can be expressed as a formulation on the reference element. In other words, the formulation
\begin{equation}
\int_{\bm{\Omega}}  \phi \,\frac{\partial Q}{\partial t} {d\Omega} - \int_{\bm{\Omega}} \frac{\partial \phi}{\partial x_i} \, F_{i} \,{d\Omega} + \int_{\bm{\Gamma}}  \phi \, F_i \,n_i \,{{d\Gamma}}= 0,
\end{equation}
is equivalent to the formulation
\begin{equation}
\int_{\bm{\Omega^r}}  \phi^r \,\frac{\partial Q^r}{\partial t} J_\Omega \,{d\Omega^r} - \int_{\bm{\Omega^r}} \frac{\partial \phi^r}{\partial \xi_j}  \frac{\partial \xi_j}{\partial x_i} \, F^r_{i} J_\Omega \,{d\Omega^r} + \int_{\bm{\Gamma^r}}  \phi^r \, F^r_i \, \frac{\partial\xi_j}{\partial x_i} \,n^r_j \, J_{\Omega}\,{{d\Gamma^r}}  = 0,
\end{equation}
where $F_i:=F_i(Q)$ and $F^r_i:=F_i(Q^r)$.
\begin{proof}
The equivalence of the first volume integral,
$$\int_{\bm{\Omega}}  \phi \,\frac{\partial Q}{\partial t} {d\Omega}  =
 \int_{\bm{\Omega^r}}  \phi^r \,\frac{\partial Q^r}{\partial t} J_\Omega \,{d\Omega^r},$$
is ensured via  Lemma \ref{app:lemma_volume} and Remark \ref{remark:app-space}. 
The transformation of the second volume integral,
$$\int_{\bm{\Omega}} \frac{\partial \phi}{\partial x_i} \, F_{i} \,{d\Omega}  =
 \int_{\bm{\Omega^r}} \frac{\partial \phi^r}{\partial \xi_j}  \frac{\partial \xi_j}{\partial x_i} \, F^r_{i} J_\Omega \,{d\Omega^r},$$
is immediate via  Lemma \ref{app:lemma_volume}, Remark \ref{remark:app-space} and considering $\frac{\partial \phi^r}{\partial x_i} = \frac{\partial \phi^r}{\partial \xi_j} \frac{\partial \xi_j}{\partial x_i}$. 
And finally, the surface integral equivalence reads
\begin{align}
&\int_{\bm{\Gamma}}  \phi \, \bm{F} \cdot \bm{n} \,{{d\Gamma}} =  \int_{\bm{\Gamma^r}}  \phi^r \, \bm{F} \cdot \bm{n}  \,J_\Gamma\,{{d\Gamma^r}} = \int_{\bm{\Gamma^r}}  \phi^r \, \bm{F^r} \cdot \left(  \frac{ \bm{J}_{\Omega}^{-T}\, J_{\Omega} }{J_{{\Gamma}}} \, \bm{n^r} \right)  \, J_\Gamma\,{{d\Gamma^r}}\\ 
= &\int_{\bm{\Gamma^r}}  \phi^r \, \bm{F^r} \cdot \left(  { \bm{J}_{\Omega}^{-T} } \, \bm{n^r} \right) \, J_{\Omega} \,{{d\Gamma^r}}
= \int_{\bm{\Gamma^r}}  \phi^r \, F^r_i \, \frac{\partial\xi_j}{\partial x_i} \,n^r_j \, J_{\Omega}\,{{d\Gamma^r}}, 
\end{align}
via Lemmas \ref{lemma:appsurf_int} and \ref{lemma:appnormal} as well as Remark \ref{remark:app-space}.
\end{proof}
\label{theorem:residual_map}
\end{theorem}

\begin{corollary}
The filtered Lagrange polynomials (Definition \ref{def:filtlag}) do not depend on $J_\Omega$ in affine elements\footnote{An affine element features a constant $J_\Omega$ such as parallelepipedic hexahedrals.}.
\begin{proof}
The only terms in Eq.  \eqref{eq:filtrd_Lag} that depend on $J_\Omega$  are $\widebartilde{{C}}_{mg}^{-1}$ and  $\widebartilde{{C}}_{hl}$. For affine elements, $J_\Omega$  is constant and can be respectively factorized from these matrices as $1/J_\Omega$ and $J_\Omega$  which cancel each other, thus leaving terms that do not depend on $J_\Omega$, assuming $ \widebartilde{{\phi}}_m^\mathcal{L}(\bm{x}) =  \widebartilde{{\phi}}_m^{\mathcal{L}}(\mathfrak{M}(\bm{\xi})) \equiv  \widebartilde{{\phi}}_m^{\mathcal{L}r}(\bm{\xi}) $. This result means that the filtered Lagrange polynomials can be computed once on the reference element and used on all affine elements.
\end{proof}
\end{corollary}
\begin{remark}
Alternatively to the approach presented here, one can start from the expression of the PDE in the generalized coordinates and derive the discrete form on the reference domain then. This is the approach adopted by Wang \textit{et al.} \cite[2.3.]{Wang-et-al_2011a} who presented two CPR/FR formulations: one on physical elements and the other on reference elements. It can be argued that the former  is free-stream preserving without being conservative and vice versa, unless the metric identities \cite{Abe2015} are satisfied, in which case both properties are satisfied by both formulations which become hence fully equivalent.
\end{remark}
\begin{remark}
The modal basis made of normalized Legendre polynomials (see \ref{app:bases}) might loose its normalized property due to the introduction of $J_\Omega$:
$$\int_{\bm{\Omega}} \phi_a^L(\bm{x}) \,\phi_a^L(\bm{x}) \,d\Omega = \int_{\bm{\Omega^r}} \phi_a^{Lr}(\bm{\xi}) \,\phi_a^{Lr}(\bm{\xi}) \,J_\Omega \,d\Omega^r \ne 1,$$
although
$\int_{\bm{\Omega^{r}}} \phi_a^{Lr}(\bm{\xi}) \,\phi_a^{Lr}(\bm{\xi}) \,d\Omega^r = 1,$
 in which case the application of an ortho-normalization process such as the  the Gram-Schmidt is required in the definition of a normalized set of modal basis polynomials.
\end{remark}
\begin{remark}
The reference tensor-product element is defined as $\bm{\Omega^r}:=[-1,1]^3$.
\end{remark}

\section{Modal and nodal basis functions}
\label{app:bases}
We present here the modal and nodal basis functions of the polynomial space ${\rm I\!P}^{p}$.
Normalized 1D Legendre polynomials are defined via the following recurrence formulas on the reference interval $\xi \in [-1,1]$:
\begin{align}
\label{eq:1dLegendreDef01}
\phi^{L\text{1D}}_0 (\xi) &:= C_0 \, 1,\\
\label{eq:1dLegendreDef02}
\phi^{L\text{1D}}_1 (\xi) &:= C_1 \, \xi,\\
\label{eq:1dLegendreDef03}
\phi^{L\text{1D}}_{a} (\xi) &:= C_a \, \frac{(2a-1) \,\xi \,\phi^{L\text{1D}}_{a-1} (\xi) \,- (a-1)\, \phi^{L\text{1D}}_{a-2} (\xi)}{a} \quad \forall a \in [2,...,p],
\end{align} 
where $C_b:=\frac{1}{\sqrt{2/(2b+1)}}$ is a normalization factor ensuring $\left<\phi^{L\text{1D}}_{b}, \phi^{L\text{1D}}_{b}\right>_\text{1D} := \int_{-1}^{+1}\phi^{L\text{1D}}_{b}(\xi)\,\phi^{L\text{1D}}_{b}(\xi) d\xi = 1$. Pre-normalized Legendre polynomials are retrieved by setting $C_b:=1$ in Eqs. \eqref{eq:1dLegendreDef01}-\eqref{eq:1dLegendreDef03}. Unless explicitly specified,  normalized Legendre polynomials are meant by $\phi^{L\text{1D}}_b(\xi)$.

For a nodal set $\xi_n$ with $n \in [0,...,p]$, the 1D Lagrange polynomials on the interval $\xi \in [-1,1]$ are expressed as:
$$\phi^\mathcal{L\text{1D}}_b (\xi):= \prod_{\substack{n\\ n\ne b}} \frac{\xi-\xi_n}{\xi_b-\xi_n}\equiv \frac{\xi - \xi_0}{\xi_b -\xi_0} \cdots \frac{\xi - \xi_{b-1}}{\xi_b -\xi_{b-1}}\frac{\xi - \xi_{b+1}}{\xi_b -\xi_{b+1}} \cdots \frac{\xi - \xi_p}{\xi_b -\xi_p},$$  $\forall b \in [0,...,p]$.

\section{DG, SD and $g_2$ correction functions}
\label{app:Fr_corrfuncs}
We present here the 1D FR correction functions for the three schemes discussed in this study, i.e. FR-DG, FR-SD and Huyhn's  FR-$g_2$ \cite{Huynh2007}. These are useful since the 3D corrections on tensor-product elements are performed as consecutive 1D corrections. We consider the interval $\xi \in [-1,1]$ and refer to its ends as $L$ and $R$. The correction function, $g_f$ of Eq. \eqref{eq:CPR3} associated to the left facette (the point $\xi=-1$) is thus denoted $g_L$ and vice versa. We recall that for a solution $Q \in {\rm I\!P}^{p}$,  all correction functions are polynomials of degree $p+1$.

Let us define the left and right Radau polynomials of degree $p$:
\begin{equation}
\psi_{L,p} := \frac{1}{2} (\phi^{L\text{1D}}_{p} +\phi^{L\text{1D}}_{p-1} ), \qquad \psi_{R,p} := \frac{(-1)^{p}}{2} (\phi^{L\text{1D}}_{p} -\phi^{L\text{1D}}_{p-1} ),
\end{equation}
where  $\phi^{L\text{1D}}_p$ is the 1D pre-normalized ($C_b=1$ in Eqs. \eqref{eq:1dLegendreDef01}-\eqref{eq:1dLegendreDef03}) Legendre polynomial of degree $p$ defined in  \ref{app:bases}.

The FR-DG correction functions are:
\begin{equation}
g_L := \psi_{R,p+1},  \qquad g_R := \psi_{L,p+1}.
\end{equation}

The FR-SD correction functions, noted as $g_{Ga}$ in the literature \cite{Huynh2007}, are:
\begin{equation}
g_L  := \frac{p+1}{2p+1} \psi_{R,p+1} + \frac{p}{2p+1} \psi_{R,p}, \qquad g_R  := \frac{p+1}{2p+1} \psi_{L,p+1} + \frac{p}{2p+1} \psi_{L,p}.
\end{equation}

Finally, the FR-$g_2$ correction functions are:
\begin{equation}
g_L  := \frac{p}{2p+1} \psi_{R,p+1} + \frac{p+1}{2p+1} \psi_{R,p}, \qquad g_R  := \frac{p}{2p+1} \psi_{L,p+1} + \frac{p+1}{2p+1} \psi_{L,p}.
\end{equation}

\section{De-aliasing}
\label{app:De-alaising}
In order to explain the mechanism used to reduce aliasing errors in this study, we start by  Eq. \eqref{eq:LES3} expressed in a nodal residual form (Lagrange ($\mathcal{L}$) coefficient at the solution node $q$):
\begin{equation}
\left( \frac{\partial \widebartilde{{Q}}_k}{\partial t}  \right)^\mathcal{L}_q=  -\left(\widebartilde{{{\mathscr{R}}}}_{}^{\text{inv/vis}}(\widebartilde{{\bm{Q}}})\right)_q^\mathcal{L} - { \widetilde{\phi}}^{\prime\mathcal{L}}_h(\bm{x}_q) \,\left({\widebartilde{{{\mathscr{R}}}}_{}^\text{mod}}(\widebartilde{{\bm{Q}}})\right)^\mathcal{L}_h,
\label{eq:app-deal1}
\end{equation}
where the combined inviscid and viscous residual is
\begin{equation*}
\left({\widebartilde{{{\mathscr{R}}}}_{}^\text{inv/vis}}(\widebartilde{{\bm{Q}}})\right)^\mathcal{L}_q :=\left(\frac{\partial  \widebartilde{{F}}_{ik}^{\text{inv}}}{\partial x_i} \right)^\mathcal{L}_q- \left(\frac{\partial \widebartilde{{F}}_{ik}^{\text{vis}}}{\partial x_i} \right)^\mathcal{L}_q  + \left( \widebartilde{{\mathscr{C}}}_k^{\text{inv}} \right)^\mathcal{L}_q - \left( \widebartilde{{\mathscr{C}}}_k^{\text{vis}} \right)^\mathcal{L}_q ,
\end{equation*}
 the residual of the SGS model reads
\begin{equation*}
\left({\widebartilde{{{\mathscr{R}}}}_{}^\text{mod}}(\widebartilde{{\bm{Q}}})\right)^\mathcal{L}_h :=-  \left( \frac{\partial \widebartilde{{F}}_{ik}^{\text{mod}}}{\partial x_i} 
+  \widebartilde{{\mathscr{C}}}_k^{\text{mod}}\right)^\mathcal{L}_h,
\end{equation*}
and the indices of the GLL solution nodes are $\{q,h\}\in[0,...,\widebartilde{{N}}_\text{DOFs}-1]$ with $\widebartilde{{N}}_\text{DOFs}:= (\widebartilde{{p}}+1)^3$.

As pointed out in  Section \ref{sec:proj},  this formulation is however prone to aliasing errors due to the primed, double-primed and other non-linear terms of the fluxes treated by the inherent collocation projection via interpolation  of the FR/CPR scheme.  Aliasing errors deteriorate  accuracy and might also  induce numerical instability. To alleviate this problem, we adopt the second method proposed in \cite{Spiegel2015a} which consists of first evaluating the residuals in an enriched polynomial space ${\rm I\!P}^{\dot{p}}(\bm{\Omega}_{e_i})$ with $\dot{p}>{\widebartilde{{p}}}$, and then projecting them onto the space ${\rm I\!P}^{\widebartilde{{p}}}(\bm{\Omega}_{e_i})$ to recover de-aliased residuals:
\begin{align}
\left(\widebartilde{{{\mathscr{R}}}}_{}^{\text{inv/vis}}(\widebartilde{{\bm{Q}}})\right)_q^\mathcal{L}&= \widebartilde{{\phi}}_l^{\mathcal{L}}(\bm{x}_q) \left(\widebartilde{{{\mathscr{R}}}}_{}^{\text{inv/vis}}(\widebartilde{{\bm{Q}}})\right)^\mathcal{L}_l = 
\widebartilde{{\mathcal{P}}}_q\left(\left(\dot{{\mathscr{R}}}^{\text{inv/vis}}(\dot{\bm{Q}}')\right)^\mathcal{L}_r \,\dot{\phi}^\mathcal{L}_r  \right), \\
\left({\widebartilde{{{\mathscr{R}}}}_{}^\text{mod}}(\widebartilde{{\bm{Q}}})\right)^\mathcal{L}_h &= \widebartilde{{\phi}}_l^{\mathcal{L}}(\bm{x}_h) \left({\widebartilde{{{\mathscr{R}}}}_{}^\text{mod}}(\widebartilde{{\bm{Q}}})\right)^\mathcal{L}_l = 
\widebartilde{{\mathcal{P}}}_h\left(\left({\dot{{\mathscr{R}}}_{}^\text{mod}}(\dot{\bm{Q}}')\right)_r^\mathcal{L} \,\dot{\phi}_r^\mathcal{L}  \right),
\label{eq:app-deal2}
\end{align}
where the index of the GLL solution nodes of the enriched space is $r\in[0,...,\dot{N}_\text{DOFs}-1]$ with $\dot{N}_\text{DOFs}:= (\dot{p}+1)^3$, $\dot{{Q}}'_r:= \widebartilde{{\phi}}_q^\mathcal{L}(\bm{x}_r)\,\widebartilde{{{Q}}}_q^\mathcal{L}$ is the solution interpolated from the resolved polynomial space to the enriched space nodes and $\widebartilde{{\mathcal{P}}}_q$ is the output of the projection (and recovery) operator, introduced in Section \ref{sec:proj}, evaluated at $\bm{x}_q$. The projection operator is more precisely defined as
\begin{equation}
\label{eq:app-proj1}
\widebartilde{{\mathcal{P}}}\left(\mathscr{S}\right) := \widebartilde{{\phi}}_h^\mathcal{L} \,\widebartilde{{\mathscr{S}}}_h^\mathcal{L} = \widebartilde{{\phi}}_h^\mathcal{L} \, \widebartilde{{{M}}}_{hq}^{-\mathcal{L}}\,\left<\widebartilde{{\phi}}_q^\mathcal{L} ,\mathscr{S}\right> \approx \widebartilde{{\phi}}_h^\mathcal{L} \, \widebartilde{{{M}}}_{hq}^{-\mathcal{L}}\,\left<\widebartilde{{\phi}}_q^\mathcal{L} ,I^{\dot{p}}(\mathscr{S})\right>
\,\equiv \widebartilde{{\phi}}_h ^\mathcal{L}\, \widebartilde{{{M}}}_{hq}^{-\mathcal{L}}\,\left<\widebartilde{{\phi}}_q^\mathcal{L} ,\dot{{\phi}}_r^\mathcal{L} \,\dot{\mathscr{S}}^\mathcal{L}_r\right>,
\end{equation}
where the integrals are computed by the collocated GLL quadrature of the enriched space.
\begin{proposition}
Employing either the  Lagrange or the Legendre basis for the projection and recovery in Eq. \eqref{eq:app-proj1} is equivalent, viz., $\widebartilde{{\phi}}_h^\mathcal{L} \, \widebartilde{{{M}}}_{hq}^{-\mathcal{L}}\,\left<\widebartilde{{\phi}}_q^\mathcal{L} ,\dot{{\phi}}_r^\mathcal{L} \,\dot{\mathscr{S}}_r^\mathcal{L}\right> \, =\, \widebartilde{{\phi}}_q^{L} \left<\widebartilde{{\phi}}_q^{L} ,\dot{{\phi}}_r^\mathcal{L} \,\dot{\mathscr{S}}^\mathcal{L}_r\right>$.
\begin{proof}
\begin{align*}
\widebartilde{{\phi}}_h^\mathcal{L} \, \widebartilde{{{M}}}_{hq}^{-\mathcal{L}}\,\left<\widebartilde{{\phi}}_q^\mathcal{L} ,\dot{{\phi}}_r^\mathcal{L} \, \dot{\mathscr{S}}_r^\mathcal{L} \right> &= \widebartilde{{\phi}}_{g}^{L} \,\widebartilde{{C}}_{gh}  \widebartilde{{C}}_{hl}^{-1}  \widebartilde{{{M}}}_{lm}^{-L}{\widebartilde{{C}}_{mq}^{-\intercal}}\,\left<\widebartilde{{\phi}}_q^\mathcal{L} ,\dot{{\phi}}_r^\mathcal{L} \, \dot{\mathscr{S}}^\mathcal{L}_r \right> \\
&=   \widebartilde{{\phi}}_{g}^{L} \,\delta_{gl} \,\delta_{lm}\, {\widebartilde{{C}}_{mq}^{-\intercal}}\,\left<\widebartilde{{\phi}}_q^\mathcal{L} ,\dot{{\phi}}_r^\mathcal{L} \,\dot{\mathscr{S}}^\mathcal{L}_r \right> = \widebartilde{{\phi}}_m^{L} \left<\widebartilde{{\phi}}_m^{L} ,\dot{{\phi}}_r^\mathcal{L} \,\dot{\mathscr{S}}^\mathcal{L}_r\right>,
\end{align*}
using Definition \ref{def:3dC}, the Lemma \ref{lem:3dCinv} and the proof of Lemma \ref{lem:3dnodMinv}.
\end{proof}
\end{proposition}
The outcome of this proposition is that for de-aliasing purposes, the construction of a modal basis is not mandatory and the intrinsic interpolatory basis of a nodal scheme can be directly employed.

\section{Derivation of kinetic energy dissipation components}
\label{app:disscomp}
In this section, we obtain the components of the volume-averaged kinetic energy dissipation \cite{Diosady2015,Plata2017}, $-D_t(E_K^{\bm{\Omega}})$ with $E_K^{\bm{\Omega}}:= \frac{1}{\Omega}\int_{\bm{\Omega}}  \frac{1}{2} \rho\,u_k u_k \,{{d\Omega}} $, where $\Omega:=\int_{\bm{\Omega}} \,{{d\Omega}}$ stands for the volume of the domain, to be monitored during the numerical simulation.  We proceed by deriving an equation of evolution employing the principles of conservation of mass and momentum for full scales of compressible turbulent flows, respectively:
\begin{equation}
\partial_t(\rho) + \partial_i(\rho u_i ) \equiv D_t(\rho)= 0,
\label{eq:app-cont1}
\end{equation}
and
\begin{equation}
\partial_t(\rho u_k) + \partial_i(\rho u_i \,u_k)  + \partial_i(P \,\delta_{ik}) - \partial_i\left(\tau_{ik}\right)  = 0,
\label{eq:app-mom1}
\end{equation}
for $\{i,k,l\} \in [1,2,3]$ and with $\tau_{ik} \equiv  \mu\left(  {\partial_i (u_k)} + {\partial_k (u_i)}  -\frac{2}{3} \, {\partial_l (u_l)}\,\delta_{ik} \right)$.

The first two terms of \eqref{eq:app-mom1} can be rewritten as
\begin{align*}
\partial_t(\rho u_k) + \partial_i(\rho u_i \,u_k)  &= \rho\,\partial_t( u_k) + \rho u_i\,\partial_i( u_k )  + \cancelto{0\,\text{via }\eqref{eq:app-cont1}}{u_k \left( \partial_t(\rho)  + \partial_i(\rho u_i) \right)}, \\
&\equiv \rho\, D_t(u_k),
\label{eq:app-mom2}
\end{align*}
which we substitute into Eq. \eqref{eq:app-mom1} that is then multiplied by $u_k$ to yield
\begin{equation}
u_k\,\left(\rho\, D_t(u_k)+\partial_k(P) - \partial_i\left(\tau_{ik}\right) \right)= 0.
\label{eq:app-mom3}
\end{equation}
The first term of \eqref{eq:app-mom3} can be reformulated as
\begin{equation*}
\rho\, D_t(u_k) \,u_k = \frac{1}{2} \rho\, D_t(u_k u_k)  = \frac{1}{2} D_t(\rho\,u_k u_k)  - \frac{1}{2} u_k u_k  \cancelto{0\,\text{via }\eqref{eq:app-cont1}}{D_t(\rho)},
\label{eq:app-mom4}
\end{equation*}
which is substituted to \eqref{eq:app-mom3} that we then integrate on a static spatial domain with periodic boundary conditions in all directions:
\begin{equation}
D_t\int_{\bm{\Omega}}  \frac{1}{2} (\rho\,u_k u_k) \,{{d\Omega}} \,+\,  \int_{\bm{\Omega}}  \partial_k(P) \,u_k \,{{d\Omega}}  \, - \, \int_{\bm{\Omega}} \partial_i \left(\tau_{ik} \right) u_k\, {{d\Omega}} = 0.
\label{eq:app-mom5}
\end{equation}

The following relations are obtained by integrating by parts and applying the divergence theorem:
\begin{equation} 
\label{eq:app-p}
\int_{\bm{\Omega}}  \partial_k(P) \,u_k \,{{d\Omega}} =  -\int_{\bm{\Omega}}  \partial_k(u_k) \,P \,{{d\Omega}} + \cancelto{0\,\text{(periodicity)}}{\int_{\bm{\Gamma}}  u_i n_i \,P \,{{d\Gamma}}},
\end{equation}
and
\begin{equation}
\label{eq:app-tau}
\int_{\bm{\Omega}} \partial_i ( \tau_{ik})\, u_k\, {{d\Omega}} = - \int_{\bm{\Omega}}  \tau_{ik} \,\partial_i(u_k)\, {{d\Omega}} +  \cancelto{0\,\text{(periodicity)}}{\int_{\bm{\Gamma}} u_k  \tau_{ik} \,n_i \, {{d\Gamma}}}.
\end{equation}

\begin{proposition}
The following relation holds:
\begin{equation}
\label{eq:app-lem}
\int_{\bm{\Omega}} \mu \, \left( {\partial_i (u_k)} + {\partial_k (u_i)} \right)  \,\partial_i(u_k) \,{{d\Omega}} = \int_{\bm{\Omega}}\frac{1}{2}\, \mu \, \left( {\partial_i (u_k)} + {\partial_k (u_i)} \right) \, \left( {\partial_i (u_k)} + {\partial_k (u_i)} \right) \,{{d\Omega}}.
\end{equation}
\begin{proof}
We have
\begin{equation*}
 {\partial_i (u_k)} \,\partial_i(u_k)= \frac{1}{2} \, \left( 2\,{\partial_i (u_k)} \, {\partial_i (u_k)} \right) \equiv \frac{1}{2} \, \left( {\partial_i (u_k)} \, {\partial_i (u_k)} + {\partial_k (u_i)} \, {\partial_k (u_i)} \right),
\end{equation*}
and hence
\begin{equation*}
 {\partial_i (u_k)} \,\partial_i(u_k) +  {\partial_k (u_i)} \,\partial_i(u_k) = \frac{1}{2} \, \left( {\partial_i (u_k)} \, {\partial_i (u_k)} + 2\,{\partial_k (u_i)} \,\partial_i(u_k)  + {\partial_k (u_i)} \, {\partial_k (u_i)} \right),
\end{equation*}
which can be reformulated as
\begin{equation*}
\left( {\partial_i (u_k)} + {\partial_k (u_i)} \right)  \,\partial_i(u_k)= \frac{1}{2}\, \left( {\partial_i (u_k)} + {\partial_k (u_i)} \right) \, \left( {\partial_i (u_k)} + {\partial_k (u_i)} \right).
\end{equation*}
\end{proof}
\end{proposition}

We also have:
\begin{equation}
\label{eq:app-rel23}
\int_{\bm{\Omega}} -\frac{2}{3} \, \mu \,{\partial_l (u_l)} \,\partial_i(u_k) \,\delta_{ik}  \,{{d\Omega}} = \int_{\bm{\Omega}} -\frac{2}{3} \, \mu \,{\partial_l (u_l)}  \,\partial_i(u_i) \,{{d\Omega}}.
\end{equation}

By substituting relations \eqref{eq:app-p}, \eqref{eq:app-tau}, \eqref{eq:app-lem} and \eqref{eq:app-rel23} into \eqref{eq:app-mom5} and multiplying by ($-1/\Omega$) we obtain:
\begin{equation}
-D_t(E_K^{\bm{\Omega}}) = \varepsilon(\mu,\bm{u}) + \varepsilon_d(\mu,\bm{u}) + \varepsilon_c(P,\bm{u}),
\label{eq:app-mom6}
\end{equation}
where the components are:
\begin{itemize}
\item viscous dissipation, $ \varepsilon(\mu,\bm{u}):= \frac{1}{\Omega} \int_{\bm{\Omega}}2\, \mu \, S_{ij}(\bm{u})\,S_{ij}(\bm{u}) \,{{d\Omega}}$ with   $S_{ij}(\bm{u}) \equiv \frac{1}{2} \, \left( {\partial_i (u_j)}+ {\partial_j (u_i)} \right)$, 
\item bulk viscosity dissipation, $\varepsilon_d(\mu,\bm{u}):= -\frac{1}{\Omega} \int_{\bm{\Omega}} \frac{2}{3} \, \mu \left(\partial_i(u_i)\right)^2 \,{{d\Omega}}$,
\item pressure-dilatation dissipation, $\varepsilon_c(P,\bm{u}):=-\frac{1}{\Omega}\int_{\bm{\Omega}}  \partial_k(u_k) \,P \,{{d\Omega}} $.
\end{itemize}

For under-resolved (large-eddy) simulations, only a range of the largest scales is solved for and hence Eq. \eqref{eq:app-mom6} needs to be adapted to accommodate this fact. To account for the effect of the unresolved scales on the total dissipation, we rather consider the following relation:
\begin{equation}
-D_t(\widebartilde{{E}}_K'^{\bm{\Omega}})  :=\varepsilon(\widebartilde{{\mu}}',\widebartilde{{\bm{u}}}') + \varepsilon_d(\widebartilde{{\mu}}',\widebartilde{{\bm{u}}}') + \varepsilon_c(\widebartilde{{P}}'',\widebartilde{{\bm{u}}}') + \varepsilon_\text{SGS},
\label{eq:app-mom7}
\end{equation}
where the prime exponent reminds that due to non-linearities the primed quantity in not necessarily within the resolved space and the double-prime designates a macro-quantity (macro-pressure here). The SGS dissipation is consequently estimated via
\begin{equation}
\varepsilon_\text{SGS} = -D_t(\widebartilde{{E}}_K'^{\bm{\Omega}})  - \varepsilon(\widebartilde{{\mu}}',\widebartilde{{\bm{u}}}') - \varepsilon_d(\widebartilde{{\mu}}',\widebartilde{{\bm{u}}}') - \varepsilon_c(\widebartilde{{P}}'',\widebartilde{{\bm{u}}}'),
\label{eq:app-sgs}
\end{equation}
with $-D_t(\widebartilde{{E}}_K'^{\bm{\Omega}})  $ evaluated by a proper numerical differentiation scheme such as finite differences. We will drop the accents and primes of the dissipation related quantities to lighten the notation but one should not omit the underlying fundamental differences.

\section{Energy spectrum computation}
\label{app:spec}
The script used in this work to compute the kinetic energy spectrum is available at \cite{Navah2018} and it is schematically presented in Algorithm \ref{alg:spec}.
\begin{algorithm}
\caption{Computation of the kinetic energy (KE) spectrum via radial averaging}
\label{alg:spec}
\begin{algorithmic}[1]
\Procedure{Spectrum}{$u_1,u_2,u_3$}
\Comment{$u_l:$ data on $i,j,k$ grid}
\State $Fu_l=  \text{FFT}(u_l)$ \Comment{fast Fourier transform}
\State $Au_l = \text{ABS}(Fu_l )/\text{SIZEOF}(u_l)$ \Comment{normalized amplitude of each mode}
\State $KEu_l = (Au_l)^2$ \Comment{contribution of $u_l$ to KE}
\State $SKEu_l = \text{FFTSHIFT}(KEu_l)$ \Comment{imaginary$\leftarrow$ $0^{th}$ mode $\rightarrow$real}
\State $\text{Box}_\text{side} = \sqrt[3]{\text{SIZEOF}(u_1)}$
\State $\text{Box}_\text{center} = \text{INT}(\text{Box}_\text{side} /2)$
\State $\text{Box}_\text{rad} = \sqrt{3(\text{Box}_\text{side})^2}/2$
\State $AvrgKEu_l = \text{ZEROS}(\text{Box}_\text{rad})$ \Comment{initialize the average KE}
\For{$i \in [1,...,\text{Box}_\text{side}]$}
\For{$j \in [1,...,\text{Box}_\text{side}]$}
\For{$k \in [1,...,\text{Box}_\text{side}]$}
\State $\text{rad}_{pos} = \text{INT}(\sqrt{(i-\text{Box}_\text{center})^2+(j-\text{Box}_\text{center})^2+(k-\text{Box}_\text{center})^2})$
\State $AvrgKEu_l[\text{rad}_{pos}] = AvrgKEu_l[\text{rad}_{pos}] + SKEu_l[i,j,k]$
\EndFor
\EndFor
\EndFor
\State $AvrgKE = 0.5(AvrgKEu_1+AvrgKEu_2+AvrgKEu_3)$ 
\EndProcedure
\end{algorithmic}
\end{algorithm}
\section*{References}
\bibliography{library}

\end{document}